\documentclass[12pt,letterpaper]{article}
\usepackage[usenames, dvipsnames]{xcolor}
\usepackage{jcapmod}

\usepackage{tocloft}
\usepackage[]{todonotes}
\usepackage{verbatim}
\usepackage{mathrsfs}
\usepackage{cleveref}
\usepackage{comment}
\usepackage[shortlabels]{enumitem}
\usepackage{pgf}
\usepackage{tikz-cd}
\usetikzlibrary{shapes,arrows}
\usetikzlibrary{calc}
\usepackage{pgfplots}
\pgfplotsset{compat=1.12}
\usepackage{tikz-3dplot}

\usepackage{amsthm}

\usepackage{tikz}
\usepackage{setspace,caption}
\usepackage{lipsum}
\usetikzlibrary{matrix,arrows,calc}
\makeatletter
\newsavebox\myboxA
\newsavebox\myboxB
\newlength\mylenA

\definecolor{cornellRed}{HTML}{B31B1B}
\definecolor{cornellBlue}{HTML}{0068AC}
\definecolor{cornellGreen}{HTML}{6EB43F}

\usepackage{bbm} 					%%% Blackboard Bold math symbols
\usepackage{slashed} 				%%% Feynman slash
\usepackage{graphicx}				%%% Graphics
\usepackage{subcaption}			%%% Subcaptions for subfigures
\usepackage{psfrag}				%%% Editing EPS files in TeX
\usepackage{tensor}				%%% The \indices command
\usepackage{fouridx}				%%% Arbitrary subscripts/superscripts
\usepackage{bm}					%%% Bold Math symbols
\usepackage{mdframed}				%%% Box for \aside
\usepackage{multirow}				%%% More flexible tables
\usepackage{soul}					%%% Strike-through text
\usepackage{bbold}				%%% Blackboard Bold
\usepackage{multicol}				%%% Pages with multiple columns
\usepackage{tikz-cd}
\usepackage{rotating}

\usetikzlibrary{arrows}

\tikzset{
commutative diagrams/.cd,
arrow style=tikz,
diagrams={>=latex}}

\usepackage{amsmath}
\usepackage{amssymb}
\usepackage{amsfonts}
\usepackage{mathtools}

\usepackage{feynmf}
\usepackage{marvosym}

\usepackage{import}

\newcommand*\xoverline[2][0.75]{%
    \sbox{\myboxA}{$\m@th#2$}%
    \setbox\myboxB\null% Phantom box
    \ht\myboxB=\ht\myboxA%
    \dp\myboxB=\dp\myboxA%
    \wd\myboxB=#1\wd\myboxA% Scale phantom
    \sbox\myboxB{$\m@th\overline{\copy\myboxB}$}%  Overlined phantom
    \setlength\mylenA{\the\wd\myboxA}%   calc width diff
    \addtolength\mylenA{-\the\wd\myboxB}%
    \ifdim\wd\myboxB<\wd\myboxA%
       \rlap{\hskip 0.5\mylenA\usebox\myboxB}{\usebox\myboxA}%
    \else
        \hskip -0.5\mylenA\rlap{\usebox\myboxA}{\hskip 0.5\mylenA\usebox\myboxB}%
    \fi}
\makeatother

\newcommand{\loc}{\mathrm{loc}}

\newcommand{\im}{\,\mathrm{Im}\,}
\newcommand{\re}{\,\mathrm{Re}\,}

\definecolor{cobalt}{RGB}{44, 98, 120}
\definecolor{celadon}{rgb}{0.67, 0.88, 0.69}
\definecolor{dm}{cmyk}{.20, 0, .30, 0}
\definecolor{burgundy}{rgb}{0.5, 0.0, 0.13}
\definecolor{plotBlue}{RGB}{94, 130, 181}
\definecolor{bisque}{rgb}{1.0, 0.89, 0.77}

\hypersetup{
  colorlinks,
  citecolor=Violet,
  linkcolor=cobalt,
  urlcolor=Blue}

\DeclareSymbolFontAlphabet{\mathbb}{AMSb}
\NewDocumentCommand{\xrightarrows}{ O{}O{} }{%
\mathrel{%
\vcenter{\hbox{%
\begin{tikzpicture}
  \node[minimum width=1cm,minimum height=1ex,anchor=south,align=center] (a){\text{\vphantom{hg}#1}\\[0.5ex] \vphantom{hg}#2};
  \draw[<-] ([yshift=0.35ex]a.west) -- ([yshift=0.35ex]a.east);
  \draw[->] ([yshift=-0.35ex]a.west) -- ([yshift=-0.35ex]a.east);
\end{tikzpicture}
}}%
}%
}

\newif\iffastcompile

\fastcompilefalse
\iffastcompile
\newcommand{\mk}[1]{}
\else
\newcommand{\mk}[1]{\todo[color=burgundy!30, size=\scriptsize, bordercolor=burgundy!30]{MK: #1}}
\fi

\iffastcompile
\newcommand{\mc}[1]{}
\else
\newcommand{\mc}[1]{\todo[color=bisque!30, size=\scriptsize, bordercolor=bisque!30]{MC: #1}}
\fi

\ProvideTextCommandDefault{\Dbar}{%
\leavevmode\lower.5ex\rlap{\hskip-.07em\accent"16}D%
}

\usepackage{environ}
\usepackage{changepage}

\begin{document}
	\newcommand{\main}{.}
\begin{titlepage}

\setcounter{page}{1} \baselineskip=15.5pt \thispagestyle{empty}
\setcounter{tocdepth}{2}
	{\hfill \small MIT-CTP/5623}
\bigskip\

\vspace{1cm}
\begin{center}
{\large \bfseries A Worldsheet Description of Flux Compactifications}
\end{center}

\vspace{0.55cm}

\begin{center}
\scalebox{0.95}[0.95]{{\fontsize{14}{30}\selectfont Minjae Cho$^{a}$,~ Manki Kim$^{b}$\vspace{0.25cm}}}

\end{center}

\begin{center}

\vspace{0.15 cm}
{\fontsize{11}{30}
\textsl{$^{a}$Princeton Center for Theoretical Science, Princeton University, Princeton, NJ 08544}\\
\textsl{$^{b}$Center for Theoretical Physics, Department of Physics, Massachusetts Institute of Technology, Cambridge, MA 02139}}\\
\vspace{0.25cm}

\vskip .5cm
\end{center}

\vspace{0.8cm}
\noindent

\vspace{1.1cm}
We demonstrate how recent developments in string field theory provide a framework to systematically study type II flux compactifications with non-trivial Ramond-Ramond profiles. We present an explicit example where physical observables can be computed order by order in a small parameter which can be effectively viewed as string coupling constant. We obtain the corresponding background solution of the string field equations of motions up to the second order in the expansion. Along the way, we show how the tadpole cancellations of the string field equations lead to the minimization of the F-term potential of the low energy supergravity description. String field action expanded around the obtained background solution furnishes a “worldsheet” description of the flux compactifications.

\vspace{3.1cm}

\noindent\today

\end{titlepage}
\tableofcontents\newpage

\section{Introduction}
One of the most urgent questions in string theory is the existence of string backgrounds closer to our four-dimensional real world, with no supersymmetry and small positive cosmological constant in the infrared \cite{SupernovaSearchTeam:1998fmf,SupernovaCosmologyProject:1998vns,Matorras:2022xel}. The main challenge in describing more realistic string vacua with no or minimal supersymmetry is that theoretical tools to analyze, both perturbative and non-perturbative, $\alpha'$ and $g_s$ corrections to such string backgrounds are scarce. At the center of this challenge, there lies a common feature of type II string compactifications: non-trivial Ramond-Ramond (RR) profiles.\footnote{There are numerous intriguing string backgrounds with non-trivial RR fluxes. They notably include the majority of scenarios related to holography and flux compactifications \cite{Maldacena:1997re,Aharony:1999ti,Kachru:2003aw,Balasubramanian:2005zx,DeWolfe:2005uu,Silverstein:2004id,Grana:2005jc,Douglas:2006es}.} The standard on-shell worldsheet approach based on the Ramond-Neveu-Schwarz (RNS) formalism \cite{Friedan:1985ge,Friedan:1985ey,Witten:2012bh,Witten:2013cia,Sen:2015hia} is well-known for its limitations in describing the RR profiles due to their half-integer picture numbers.\footnote{For an attempt to describe the RR profiles within the RNS formalism, see for example, \cite{Berenstein:1999jq,Berenstein:1999ip}. } Although there are alternative formalisms that may be useful in describing specific RR flux backgrounds,\footnote{Such alternatives include the pure-spinor formalism and the hybrid formalism \cite{Berkovits:1999im,Berkovits:1999in,Berkovits:2000fe,Berkovits:2000ph,Berkovits:2000nn,Berkovits:2001us,Berkovits:1994wr,Berkovits:2001tg,Linch:2006ig,Kappeli:2006fj}, where spacetime supersymmetries are made manifest at the level of the worldsheet. Whether such an approach may still apply for less supersymmetric backgrounds is an interesting question beyond the scope of this work. For an interesting early attempt to understand flux compactifications in the hybrid formalism, see \cite{Linch:2008rw}.} it is essential to establish a more universal and systematic framework for the study of flux compactifications.

It is worth emphasizing that the question of vacuum structure is inherently a low-energy problem. Hence, it is reasonable to expect that constructing solutions to the low-energy approximation of string theory, the low-energy supergravity, is equivalent to constructing string backgrounds, as long as the involved approximation is judiciously performed. And this is why most of the attempts to construct more interesting string backgrounds usually involve finding solutions to the low-energy supergravity equations of motion that can also accommodate RR fluxes. However, the low-energy approximation necessarily comes with a list of limitations. It is not clear how to compute $\alpha'$ and $g_s$ corrections purely within the low-energy supergravity, and physical quantities that are not geometrical or not protected by supersymmetry are extremely difficult to access.\footnote{Sometimes, special structures such as integrability provide access to more physical data about the theory. See, e.g. \cite{Beisert:2010jr} for a review in the context of holography.} Therefore, a first-principle worldsheet description of string backgrounds with RR fluxes is necessary to systematically compute observables of stringy nature.

It is rather surprising that there is such a contrast between the low-energy supergravity and the string worldsheet theory when it comes to the description of a background. The main difference is that the general covariance which makes Einstein field equation so universal and powerful is not manifest in the current formulation of string worldsheet theories. Within the RNS formalism, the closest to the Einstein field equations is provided by the string field equations of motions (SFEOM). Just as solutions to the former describe a supergravity background, those to the latter describe a superstring background. Nonetheless, as of today, string field theory (SFT) necessitates an exact worldsheet theory as its starting point to even formulate the SFEOM, which are then expressed in terms of the degrees of freedom of that starting point. However, this leads us to an intriguing possibility: certain interesting backgrounds, for which an exact worldsheet description is currently elusive, may be realized as solutions to SFEOM formulated around an exact worldsheet theory. In such cases, SFT offers a ``worldsheet" description of these string backgrounds, in the sense that it facilitates a systematic computation of genuinely string-theoretic physical observables.

The possibility that a background for which we do not know of an exact worldsheet CFT description arises as a solution to SFEOM can be realized, for example, when the solution has a different spacetime asymptotics than the starting point that was used to formulate SFEOM. Another example, not necessarily independent of the first, involves a string field solution featuring non-zero RR flux components while the starting point represents a pure Neveu-Schwarz-Neveu-Schwarz (NSNS) background. Such a solution can be described in SFT because RR fields are merely components of the string fields comprising the theory's field content. Therefore, there is no inherent impediment to working with nontrivial RR field configurations in SFT. In such cases, the solution does not necessarily correspond to a worldsheet CFT belonging to the CFT moduli space of the starting point. In fact, it is unclear if this solution can ever accommodate a local worldsheet CFT description in the conventional sense.\footnote{A related concept in SFT is the background independence \cite{SEN1990551,Sen:1993kb,Bergman:1994qq,Belopolsky:1995vi,Sen:2017szq}, which asserts that different SFTs described by different string couplings or worldsheet CFTs sharing the moduli space are related to each other by string field redefinitions. Solutions with RR fluxes suggest how such a notion may be generalized beyond the worldsheet CFT moduli space.}

The aforementioned possibilities exemplify the conceptual innovation inherent in SFT. Once a field theory description of strings is obtained from a given starting point, the original worldsheet CFT can be set aside in favor of the field theory description. Solutions to the field equations correspond to string backgrounds, and expanding the string field action around these solutions establishes the Feynman rules needed to compute physical observables. This field theory essentially serves as the ``worldsheet" formulation of the string background corresponding to the solutions. Of course, the limitation that such a field theory description is given by the degrees of freedom of the starting point is still present. Moreover, since the current formulation of any superstring field theories involving dynamical closed strings is non-polynomial in the closed string fields, achieving a consistent expansion scheme that aligns with the string coupling expansion is imperative for systematic computations of background solutions and their associated physical observables.

In this work, we study a simple class of flux compactifications that arises as solutions to SFEOM of type IIB superstring theory with D3-branes and O3-planes.\footnote{As long as the D7-brane tadpole is canceled locally, we don't expect any obstruction to generalize our construction to the backgrounds with D7-branes and O7-planes. But, we did not pursue this generalization in this work.} To do so, we shall make heavy use of the type IIB open-closed-unoriented SFT in RNS formalism for which a consistent BV-formalism was recently completed \cite{Sen:2014dqa,Sen:2015hha,Sen:2015uaa,deLacroix:2017lif,FarooghMoosavian:2019yke}. At the level of low-energy supergravity, these backgrounds describe non-compact four-dimensional spacetimes warped by compact six-dimensional spaces (which we take to be a six-torus in this work) featuring NSNS 3-form, RR 3-form, and RR 5-form fluxes alongside D3-branes and O3-planes.

As alluded to in previous paragraphs, to obtain SFEOM, we must start with a consistent string background for which a worldsheet CFT description is available. Because the closest string background is the toroidal compactifications without non-trivial fluxes, this implies that the non-trivial fluxes shall be treated perturbatively, at best, at least within the current formulation of string field theory. This idea was explored in \cite{Cho:2018nfn} where pp-wave background and $AdS_3$ background with nontrivial RR fluxes were described as solutions to type IIB SFEOM. These solutions were expressed as a perturbative expansion in the RR flux around a pure NSNS background where an exact worldsheet description is available.\footnote{Since the quantization of the RR flux is a non-perturbative effect, such an expansion is still consistent as long as string perturbation theory is concerned, even though the convergence of the expansion to the quantized result was not addressed in detail.}

One immediate worry follows. In the context of conventional holographic $AdS$ backgrounds, which can be attained as a near-horizon limit of D-branes, changing the Ramond-Ramond flux quanta does not necessarily destabilize moduli per se. This is in part due to the fact that near-horizon $AdS$ solutions come in infinite families. So, treating the RR flux quanta as a small parameter may not cause a severe problem. On the other hand, in the context of flux compactifications, the stability of vacua sensitively depends on which flux quanta one chooses, and it is not clear if it even makes sense to consider the RR flux as a small parameter while ensuring that the NSNS flux is indeed quantized.\footnote{Note that the worldsheet description is sensitive to the quantization of the NSNS flux. Hence, it is important to treat the NSNS flux quanta as O(1) quantities.} This raises a serious obstacle to studying the flux compactifications in string field theory.

As we will demonstrate in the main text, the situation proves more favorable for the flux compactification under consideration in this work. In the F-term minima, the vacuum expectation value of complex structure moduli is correlated with the vacuum expectation value of the dilaton. Furthermore, the source terms that describe the fluxes depend not only on the flux quanta but also on complex structure moduli. Essentially, this correlation between the complex structure moduli and the dilaton will enable us to work with fully quantized fluxes in SFT, with the only effective expansion parameter being $g_s^{1/2}$, which can be kept small in string perturbation theory. A recursive method to write down the string field solution and physical quantities in such an expansion has been well-established in SFT literature- see e.g. \cite{deLacroix:2017lif}. We will obtain the solution to order $g_s$ where all tadpoles in the SFT action vanish when expanded around the solution.\footnote{In \cite{Sen:2015uoa}, a similar perturbative string field solution where the massless tadpoles are absent was obtained in the context of SO(32) heterotic string theory on Calabi-Yau threefolds.} Even though our string field solution will be a direct translation of a supergravity solution, the SFT action around this solution provides a genuinely string-theoretic framework for computing physical observables that transcend the scope of low-energy effective descriptions.

\subsection{Summary of the draft}
The main goal of this paper is to construct a class of perturbative background solutions in string field theory that can be used for systematic investigations of flux compactifications in type IIB string theory. As such, the results of this draft involve the ingredients from string compactifications, string perturbation theory, and string field theory. This section serves as both the summary and the reading guide.

In \S\ref{sec:supergravity}, we review flux compactifications in type IIB string theory and provide a simple explicit example of such. \S\ref{sec:supergravity} is the only section where we will use the \emph{Einstein-frame} metric. In the rest of the draft, we shall use the \emph{string-frame} metric. We start with the low energy approximation of type IIB string theory. In \S\ref{sec:gkp}, we review type IIB flux compactifications of Giddings-Kachru-Polchinski (GKP) \cite{Giddings:2001yu}. In \S\ref{sec:gkp}, we shall retain terms in the supergravity action up to $\mathcal{O}(g_s\alpha'^2)$ compared to the leading order terms. In particular, we shall ignore the $\mathcal{O}(\alpha'^3)$ corrections to the low energy supergravity. We review how solving equations of motion up to $\mathcal{O}(g_s\alpha'^2)$ leads to the imaginary-self-dual (ISD) condition of the complexified threeform flux $G_3:=F_3-\tau H_3$
\begin{equation}
    \star_{6}G_3=i G_3\,.
\end{equation}

In \S\ref{sec:low energy supergravity}, we review the four dimensional low energy effective theory of type IIB flux compactifications. In \S\ref{sec:low energy supergravity}, we use the flux superpotential \cite{Gukov:1999ya}
\begin{equation}
    \int G_3\wedge\Omega\,,
\end{equation}
and the string-tree-level K\"ahler potential to analyze the F-term potential in the 4d. We review that up to $\mathcal{O}(g_s \alpha'^2),$ the low energy supergravity enjoys the no-scale structure, which is broken by the famous Becker-Becker-Haack-Louis correction at order $\mathcal{O}(\alpha'^3)$ \cite{Becker:2002nn}. We review how the F-term conditions for complex structure moduli and the axio-dilaton lead to the ISD condition for the complexified flux $G_3.$ We then explain that due to the no-scale structure, even if the F-term condition for the K\"ahler moduli is not solved, the F-term potential is still minimized. This implies that without considering further corrections, e.g. the non-perturbative superpotential, supersymmetry is broken perturbatively for \emph{generic} choices of quantized fluxes. We then explain that to preserve supersymmetry, the complexified flux $G_3$ should be of $(2,1)$ Hodge type, which implies that the vacuum expectation value of the flux superpotential should vanish. 

For generic flux vacua, it is unclear if there can be a sensible perturbative expansion scheme that is amenable to perturbative string field theory analysis, because string coupling is not expected to be small in general \cite{Denef:2004ze}. Therefore we shall make a judicious choice of candidate flux vacua to study. We then deploy a strategy to find supersymmetric flux vacua in toroidal compactifications based on the idea of \cite{Demirtas:2019sip}. We explain that in the class of flux vacua we study there is a linear relation between complex structure moduli $u^i$ and the axio-dilaton $\tau$
\begin{equation}
    u^i=p^i\tau\,,
\end{equation}
which is expected to be perturbatively exact. Furthermore, we shall explain that $g_s$ can be essentially treated as a free parameter, because the following direction in the moduli space
\begin{equation}
    (u^i(\lambda),\tau(\lambda))=(p^i\lambda,\lambda)\
\end{equation}
parametrized by $\lambda$ remains massless, which makes the perfect candidate for the perturbative analysis. In \S\ref{subsection:explicitExample}, we provide an explicit example of toroidal supersymmetric flux compactifications.

In \S\ref{sec:WS}, we review the worldsheet CFT for the $\Bbb{R}^{1,9}$ and $\Bbb{R}^{1,3}\times T^6$ target spacetimes. In \S\ref{subsection:non compact CFT}, we set up the worldsheet CFT convention for $\Bbb{R}^{1,9}$ target space. In \S\ref{sec:closed states}, we collect vertex operators for the massless states, and compute three point functions involving the massless states. In \S\ref{sec:bdrystate}, we review the boundary states and fix the conventions. In \S\ref{subsection:toruscft}, we set up the worldsheet CFT convention for $\Bbb{R}^{1,3}\times T^6$ target space. Furthermore, we introduce vielbeins that will prove useful in \S\ref{sec:SFT}.

In \S\ref{sec:SFT}, we apply the type II open-closed-unoriented string field theory to obtain perturbative background solutions in SFT that correspond to the flux vacua introduced in \S\ref{sec:supergravity}. In \S\ref{subsection:sft review}, we give a brief review of type II open-closed-unoriented string field theory. String field theory, as currently formulated, requires as a good starting point a well defined worldsheet CFT. And at the same time, it is not known how to directly construct the worldsheet CFT for flux vacua in the RNS formalism. Hence, we shall start with a purely geometric string background that can be deformed into a string background that corresponds to a flux vacuum. Because quantized fluxes are not small, it may na\"ively seem impossible to deform the background by quantized fluxes in a controlled manner. In \S\ref{subsection:epsilon expansion}, we point out that the vertex operator for the quantized fluxes not only depend on the flux quanta but also on complex structure moduli. This allows us to set up an expansion scheme that we call the $\epsilon$ expansion. To do so, we introduce a small parameter $\epsilon,$ take a large complex structure limit, and treat $g_s$ and $\im u_i^{-1}$ as small parameters of order $\mathcal{O}(\epsilon).$ This allows us to treat the deformation by fluxes in string field theory as small perturbations of order $\mathcal{O}(\epsilon^{1/2})$, which will then allow us to find string background order by order in $\epsilon.$ As $\epsilon$ expansion scheme is based on the smallness of $g_s$ and $\im u_i^{-1}\,,$ and the moduli stabilization will impose constraints on the vacuum expectation value of moduli, for the consistency of the expansion scheme it is imperative to check that the scaling of moduli are respected by moduli stabilization. We shall check this in \S\ref{sec:2ndbgd} by solving SFEOM. Beyond order $\mathcal{O}(\epsilon),$ the $\epsilon$ expansion can be treated as $g_s$ expansion.

In \S\ref{sec:1stbgd} and \S\ref{sec:2ndbgd}, we solve the equations of motion at order $\mathcal{O}(\epsilon^{1/2})$ and $\mathcal{O}(\epsilon)$ respectively. One notable outcome of \S\ref{sec:2ndbgd} is that in SFT the tadpole due to localized sources, such as D-branes, are smeared. We then discuss the subtlety in the absence of massless tadpoles in SFT in \S\ref{section:tadpole}. Once correctly implemented, it leads to the ISD condition and the integrated Bianchi identity, in agreement with the expectation from supergravity.

In \S\ref{sec:conclusions}, we conclude with a list of interesting future directions. In \S\ref{app:tor}, we summarize the supergravity conventions for toroidal compactifications. In \S\ref{app:spin}, we summarize the spinor conventions for the toroidal compactifications.

\section{Supergravity background}\label{sec:supergravity}
In this section, we give a supergravity description of the flux compactification background studied in this work. It arises as a solution to the IIB supergravity with local sources given by D3-branes and O3-planes. D7-branes and O7-planes may also be added, even though we will focus on the cases without them. We shall mostly follow the conventions of \cite{Kachru:2019dvo}. The ten-dimensional type IIB supergravity action in \emph{Einstein}-frame for the bosonic components is given by
\begin{equation}
S^{(E)}=\frac{1}{2\kappa_{10}^2}\int d^{10}X \sqrt{-G^{(E)}} \left(\mathcal{R}^{(E)}_{10}-\frac{\partial_A \tau\partial^A\bar{\tau}}{2(\im\tau)^2}-\frac{G_3\cdot \overline{G}_3}{2\im\tau}-\frac{\tilde{F}_5^2}{4}\right)+\frac{1}{8i\kappa_{10}^2}\int \frac{C_4\wedge G_3\wedge\overline{G}_3}{\im\tau}+S_{loc}\,,
\end{equation}
where $\mathcal{R}_{10}$ is the Ricci scalar computed from $G^{(E)},$ $\tau$ is the axio-dilaton, $G_3:=F_3-\tau H_3,$ and $\tilde{F}_5=F_5-\frac{1}{2} C_2\wedge H_3+\frac{1}{2} B_2 \wedge F_3.$ The local action $S_{loc}$ contains contributions from D-branes and O-planes. Note that throughout this draft, we shall use the following conventions
\begin{equation}
F^{(2k+1)}\cdot F^{(2k+1)}:=\frac{1}{(2k+1)!}G^{a_1b_1}\dots G^{a_{2k+1}b_{2k+1}}F^{(2k+1)}_{a_1\dots a_{2k+1}}F^{(2k+1)}_{b_1\dots b_{2k+1}}\,,
\end{equation}
and
\begin{equation}
|F^{(2k+1)}|^2:=(2k+1)!F^{(2k+1)}\cdot F^{(2k+1)}\,.
\end{equation}
The same action in \emph{string}-frame is written as
\begin{equation}
S^{(st)}=S_{NS}^{(st)}+S_{R}^{(st)}+S_{CS}+S_{loc}^{(st)}\,,
\end{equation} 
where
\begin{equation}
S_{NS}^{(st)}=\frac{1}{2\kappa_{10}^2}\int d^{10}X\sqrt{-G^{(st)}} e^{-2\Phi}\left(\mathcal{R}_{10}^{(st)}+4\partial^{\mu}\Phi\partial_{\mu}\Phi-\frac{1}{2}H_3\cdot H_3\right)\,,
\end{equation}
\begin{equation}
S_{R}^{(st)}=-\frac{1}{4\kappa_{10}^2}\int d^{10}X\sqrt{-G^{(st)}}\left(F_1\cdot F_1+\tilde{F}_3\cdot \tilde{F}_3+\frac{1}{2}\tilde{F}_5\cdot \tilde{F}_5\right)\,,
\end{equation}
\begin{equation}
S_{CS}=-\frac{1}{4\kappa_{10}^2}\int C_4\wedge H_3\wedge F_3\,.
\end{equation}
The string-frame metric is related to the Einstein-frame metric via
\begin{equation}
G^{(E)}=e^{-\Phi/2}G^{(st)}\,.
\end{equation}
Note that we defined $\tilde{F}_3:= F_3-C_0\wedge H_3.$ We remark that the RR fields are normalized differently in supergravity than the NSNS fields.

Now, we shall consider Calabi-Yau orientifold compactifications in the presence of fluxes. A few comments are in order. The orientifold compactification is consistent iff the Ramond-Ramond tadpole is canceled. This inevitably induces warping, which mixes the compact and the non-compact directions. In the limit where the warping is small, the topology of the spacetime will be of $\Bbb{R}^{3,1}\times X/\mathcal{I},$ where $X$ is a Calabi-Yau and $\mathcal{I}$ is the orientifold involution $\mathcal{I}^2=I.$ But, at a finite warping, the topology of the spacetime should be understood as a $\Bbb{R}^{3,1}$ fibration over $X/\mathcal{I}.$

A cautionary remark should follow. In generic compactifications, in which the compact manifold has a non-trivial Euler characteristic, it is expected that a non-trivial vacuum expectation value of the flux superpotential will cause a perturbative runaway behavior \cite{Sethi:2017phn}. This runaway behavior is induced by $\mathcal{O}(\alpha'^3)$ corrections to the effective action, which manifests itself as the famous $\mathcal{O}(\alpha'^3)$ no-scale structure breaking correction to the K\"{a}hler potential known as the BBHL correction \cite{Becker:2002nn}. As such a background cannot be regarded as a vacuum, due to its time-dependent nature, one should be careful to study such backgrounds. The common wisdom is that one shall find flux compactification in which the slow roll parameters and the perturbative supersymmetry breaking scale are small, which allows one to treat the background as an approximate vacuum \cite{Kachru:2018aqn}. Then, one can reliably approximate the non-perturbative superpotential to balance the energy from the supersymmetry breaking flux to find a stable $AdS$ vacuum. In this work, however, we shall consider a simpler setting in which the compact manifold is a simple toroidal orientifold, and the vacuum expectation of the flux superpotential vanishes exactly. In toroidal orientifolds, we wouldn't see such a runaway behavior at the leading order in $g_s$, but one should carefully treat the runaway behavior in generic Calabi-Yau compactifications when adopting the worldsheet description of flux compactifications.\footnote{This will require carefully including $\alpha'$ corrections to the tree-level actions and equations of motions of string field theory.}

\subsection{Ansatz and assumptions for the ten-dimensional supergravity}\label{sec:gkp}
In this section, we shall review the ten-dimensional supergravity solutions of \cite{Giddings:2001yu}.

We will take the following metric ansatz in \emph{Einstein}-frame\footnote{From now on, we shall denote the Einstein-frame metric by $G$ unless we are required to specify the frame for clarity.}
\begin{equation}
ds^2=G_{AB}dX^AdX^B=e^{2A(y)} g_{\mu\nu}(x)dx^\mu dx^\nu+ e^{-2A(y)}g_{ab}(y) dy^a dy^b\,,
\end{equation}
with $x$ denoting coordinates in the four non-compact dimensions ($\mu=0,1,2,3$) and $y$ denoting coordinates in the six compact dimensions ($a=4,5,6,7,8,9$). Moreover, we take the ansatz for the RR 5-form flux as
\begin{equation}
\tilde{F}_5=(1+\star_{10})\sqrt{-g_4}d\alpha(y)\wedge dx^0\wedge dx^1\wedge dx^2\wedge dx^3\,,
\end{equation}
which is manifestly self-dual. Note that $\star_{10}$ is the ten-dimensional Hodge star operator.  The Ricci tensors can be written as
\begin{equation}
\mathcal{R}_{\mu\nu}[G]=\mathcal{R}_{4,\mu\nu}[g]-e^{4A}g_{\mu\nu}\nabla^2 A\,,
\end{equation}
\begin{equation}
\mathcal{R}_{ab}[G]=\mathcal{R}_{6,ab}[g]+\nabla^2 Ag_{ab}-8\partial_a A\partial_b A\,,
\end{equation}
where $\mathcal{R}_4[g]$ and $\mathcal{R}_6[g]$ are the Ricci tensors of $g_{\mu\nu}$ and $g_{ab},$ respectively. We define the D3-brane charge density as
\begin{equation}
\rho_{D3}=\rho_{D3}^{loc}+\frac{1}{2\mu_3\kappa_{10}^2}H\wedge F(\epsilon \partial y_4\wedge \dots \wedge \partial y_9)\,,
\end{equation}
where $\epsilon=\pm1$ determines the orientation of the internal manifold and $\rho_{D3}^{\loc}$ is the local D3-brane charge density that is normalized for a spacetime filling D3-brane at $y_0^i$ as
\begin{equation}
\rho_{D3}^{loc}(y^i)=\delta^{(6)}(y^i-y_0^i)\,.
\end{equation}
Note that we normalized the Dirac delta function such that
\begin{equation}
\int d^6 y\sqrt{g_6}\delta^{(6)}(y-y_0)=1\,. 
\end{equation}
Combining the Bianchi identity 
\begin{equation}\label{eqn:Bianchi}
d\tilde{F}_5= 2\mu_3\kappa_{10}^2 \rho_{D3} d\text{Vol}_{X/\mathcal{I}} =H\wedge F+2\mu_3\kappa_{10}^2\rho_{D3}^{loc}d\text{Vol}_{X/\mathcal{I}}\,,
\end{equation}
and the Einsteins equations, one can find the following equations \cite{Giddings:2001yu}
\begin{equation}\label{eqn:phi p}
\nabla^2 \Phi_-=e^{-4A}\partial_a \Phi_-\partial^a\Phi_-+\frac{e^{8A}}{3! \im\tau} |G_3+i\star_6 G_3|^2+2\mu_3\kappa_{10}^2e^{8A}( \mathcal{J}^{loc}-\rho_{D3}^{loc})+\mathcal{R}_4[g]\,,
\end{equation}
\begin{equation}\label{eqn:phi n}
\nabla^2 \Phi_+=e^{-4A}\partial_a \Phi_+\partial^a\Phi_++\frac{e^{8A}}{3! \im\tau} |G_3-i\star_6 G_3|^2+2\mu_3\kappa_{10}^2e^{8A}( \mathcal{J}^{loc}+\rho_{D3}^{loc})+\mathcal{R}_4[g]\,,
\end{equation}
where we defined 
\begin{equation}
\mathcal{J}^{loc}:= \frac{1}{4\mu_3} \left[T_{ab}^{loc}G^{ab}-T_{\mu\nu}^{loc}G^{\mu\nu}\right]\,,
\end{equation}
for the localized energy-momentum tensor $T^{loc}_{AB}$ and
\begin{equation}
\Phi_\pm=e^{4A}\pm \alpha\,.
\end{equation}
It is important to note that D3-branes and O3-planes satisfy\footnote{When D7-branes and O7-planes are present, one can place 4 D7-branes on every O7-plane to cancel the D7-brane tadpole locally. This will ensure that the axio-dilaton remains a modulus in string perturbation theory. We will assume that D7-branes and O7-planes are absent in the main text as our explicit example does not contain seven-branes.}
\begin{equation}\label{eqn:D3O3}
\mathcal{J}^{loc}-\rho_{D3}^{loc}=0\,.
\end{equation}

We shall choose $F_3$ and $H_3$ such that they are integral elements of $H_3(X,\Bbb{Z}),$ and we shall furthermore assume that the following equality holds
\begin{equation}\label{eqn:isd}
G_3=-i\star_6 G_3\,.
\end{equation}
The reason for imposing the aforementioned equality will be clear. We will call the condition \eqref{eqn:isd} the Imaginary-Self-Dual (ISD) condition. To understand the consequence of \eqref{eqn:isd}, let us examine the real and the imaginary parts of \eqref{eqn:isd}
\begin{equation}
F_3-C_0 H_3=-\frac{1}{g_s}\star_6 H_3\,,
\end{equation}
\begin{equation}
-\frac{1}{g_s}H_3=- \star_6(F_3-C_0H_3)\,.
\end{equation}
If $C_0=0,$ we have an identity
\begin{equation}
F_3=-\frac{1}{g_s}\star_6 H_3\,,
\end{equation}
therefore
\begin{equation}
F_3\wedge\star_6 F_3= \frac{1}{g_s^2} H_3\wedge \star_6 H_3.
\end{equation}

At the leading order in $\alpha'$ expansion, the non-compact components of the Einstein equation are solved by $g_{\mu\nu}=\eta_{\mu\nu},$ as the Calabi-Yau manifold is Ricci flat. Then, equations (\ref{eqn:phi p}) and (\ref{eqn:phi n}) become
\begin{equation}\label{eqn:redPhi m}
\nabla^2 \Phi_-=e^{-4A}\partial_a \Phi_-\partial^a\Phi_-\,,
\end{equation}
\begin{equation}\label{eqn:redPhi p}
\nabla^2 \Phi_+=e^{-4A}\partial_a \Phi_+\partial^a\Phi_++\frac{e^{8A}}{3! \im\tau} |G_3-i\star_6 G_3|^2+4\mu_3\kappa_{10}^2e^{8A}\rho_{D3}^{loc}\,,
\end{equation}
with the further condition (\ref{eqn:D3O3}). We introduce a bracket for a 6-form
\begin{equation}
-\left[d y^1\wedge dy^2\wedge dy^3 \wedge dy^4 \wedge dy^5\wedge dy^6\right]=1\,.
\end{equation}
Furthermore, we shall assume
\begin{equation}
\rho_{D3}^{loc}=\sum_{y_{D3}}\delta^{(6)}(y-y_{D3})-\frac{1}{4}\sum_{y_{O3}}\delta^{(6)}(y-y_{O3})\,.
\end{equation}

The equation \eqref{eqn:redPhi m} can be solved by the following ansatz
\begin{equation}\label{eqn:supergravitysolm}
\Phi_-=0\,.
\end{equation}
Using the identity
\begin{equation}
e^{-4A}=\frac{2}{\Phi_++\Phi_-}\,,
\end{equation}
we can rewrite \eqref{eqn:redPhi p} as
\begin{equation}\label{eqn:redPhi p2}
\nabla^2 (\Phi_+)^{-1}=\frac{1}{24\im\tau}|G_3-i\star_6 G_3|^2+\mu_3\kappa_{10}^2\rho_{D3}^{\loc}\,,
\end{equation}
whose solution is simply given as
\begin{equation}\label{eqn:supergravitysolp}
(\Phi_+)^{-1}=\mu_3\kappa_{10}^2 \int d^6 Y \sqrt{g}\mathcal{G}^{(6)}(Y-Y_0) \rho_{D3}(Y_0)\,.
\end{equation}
where $\mathcal{G}^{(6)}(Y-Y_0)$ is the six-dimensional Green's function that implicitly depends on the complex structure moduli of $X$ which is defined as
\begin{equation}\label{eq:green}
\nabla^2 \mathcal{G}^{(6)}(Y-Y_0)=\delta^{(6)}(Y-Y_0)-\frac{1}{\text{Vol}(X/\mathcal{I})}\,.
\end{equation} 

To facilitate $g_s$ factor counting, let us convert our solution to that of the string-frame. In the string-frame, the last two terms on the RHS of \eqref{eqn:redPhi p2} contain a factor of $g_s^{3/2}$, while $\nabla^2$ on the LHS contains a factor of $g_s^{1/2}$. Therefore, at the leading order, $\Phi_+$ should be of order $g_s$, and we can study the solutions systematically in the $g_s$ expansion. We thus take the expansion
\begin{equation}\label{eqn:Aalphaexpansion}
A=\sum_{n=1,2,...}g_s^nA^{(n)},~~~\alpha=\sum_{n=1,2,...}g_s^n\alpha^{(n)},
\end{equation}
and correspondingly
\begin{equation}
\Phi_-=\sum_{n=1,2,...}g_s^n\Phi_-^{(n)},~~~\Phi_+^{-1}=\sum_{n=1,2,...}g_s^n(\Phi_+^{-1})^{(n)}.
\end{equation}
Then, (\ref{eqn:redPhi m}) implies
\begin{equation}
\Phi_-^{(1)}=0\,,
\end{equation}
while \eqref{eqn:redPhi p} at order $g_s$ becomes
\begin{equation}
g_s\nabla^2(\Phi_+^{-1})^{(1)}=\frac{g_s^{1/2}}{24}|G_3-i\star_6 G_3|^2+\mu_3\kappa_{10}^2(g_s^{-3/2}\rho_{D3}^{\loc})\,.
\end{equation}
As a solution to the above equation, we find
\begin{equation}
\Phi_+^{(1)}=\mu_3\kappa_{10}^2 \int d^6 Y \sqrt{g^{(st)}}\mathcal{G}^{(6)}(Y-Y_0) (g_s^{-3/2}\rho_{D3}(Y_0))\,.
\end{equation}
As one can see from the definition, $\mathcal{G}^{(6)}$ is proportional to $g_s$.

A few remarks concerning the supergravity solution a l\'{a} GKP are in order. For a generic flux choice, the ISD condition
\begin{equation}
G_3=-i\star_6 G_3\,,
\end{equation}
is expected to stabilize all complex structure moduli and the axio-dilaton. In particular, it is not guaranteed that at the point where moduli are stabilized, $g_s$ will be small enough to ensure that the corrections in $g_s$ are subleading. For such backgrounds, the perturbative worldsheet description won't be available. Furthermore, as we will discuss in detail in the next section, if the complexified threeform flux has a non-vanishing $(0,3)$ component, supersymmetry will be broken perturbatively. And this broken supersymmetry can in general cause a perturbative runaway behavior for generic Calabi-Yau orientifold compactifications. Although the supersymmetry can be restored non-perturbatively via Euclidean D3-branes and D7-brane gaugino condensations, studying the non-perturbative effects adds an extra layer of technical complication for string field theoretic analysis. Hence, as we will explain in the next section, in this work, we shall carefully choose a special class of flux vacua that allows us a perturbative treatment of string worldsheet and retains supersymmetry even perturbatively.

\subsection{Low energy 4d supergravity analysis}\label{sec:low energy supergravity}
In this section, we shall discuss in detail the structure of the flux vacua that we will focus on. Most of the discussions presented in this section can be recast in ten-dimensional supergravity. However, the description of the flux vacua in four-dimensional compactification will be more economical. Hence, in this section, we shall adopt the four-dimensional supergravity description. 

The two-derivative action of 4d $\mathcal{N}=1$ supergravity is determined by the K\"{a}hler potential $K(\phi,\bar{\phi})$, the superpotential $W(\phi)$, and the holomorphic gauge coupling $f(\phi)$. The effective supergravity action involving the K\"{a}hler potential and the superpotential reads 
\begin{equation}
S=-\frac{1}{2}\int d^4x \left[ R-K_{\phi\bar{\phi}}\partial^\mu \phi\partial_\mu\bar{\phi}-V_F\right]\,,
\end{equation}
where we defined
\begin{equation}
K_{\phi\bar{\phi}}:=\partial_\phi\bar{\partial}_{\bar{\phi}}K(\phi,\bar{\phi})\,,
\end{equation}
\begin{equation}
V_F= e^{ K} \left(K^{a\bar{b}}D_aW \bar{D}_{\bar{b}}\overline{W}-3|W|^2\right)\,,
\end{equation}
and
\begin{equation}
D_a W=(\partial_a+\partial_a K)W\,.
\end{equation}

At the leading order in $g_s,$ the perturbative K\"{a}hler potential is determined as \cite{Grimm:2004uq}
\begin{equation}
K=-2\log\left(\mathcal{V}-g_s^{3/2}\frac{\zeta(3)\chi}{4(2\pi)^3}\right)-\log\left(\tau-\bar{\tau}\right)-\log\left(i\int_X \Omega\wedge\overline{\Omega}\right)\,,
\end{equation}
where $\mathcal{V}$ is Einstein-frame Calabi-Yau volume, and we included $\alpha'^3$ corrections to the K\"{a}hler potential \cite{Becker:2002nn}. In principle, one can also take into account the non-perturbative corrections to the K\"{a}hler potential by employing mirror symmetry \cite{Hori:2003ic}, at the leading order in $g_s,$ see for example \cite{Demirtas:2021nlu}. But, we will not need to consider the non-perturbative corrections to the K\"{a}hler potential in this work. 

The superpotential at the string tree-level is given as \cite{Gukov:1999ya}
\begin{equation}
W_{tree}=\int_X \left(F_3-\tau H_3\right)\wedge\Omega(z)\,.
\end{equation}
An important feature of the superpotential is that due to the holomorphy and the axionic shift symmetry, the flux superpotential is not renormalized perturbatively \cite{Dine:1986vd,Burgess:2005jx}. As a result, the tree-level superpotential is perturbatively exact
\begin{equation}
W_{pert}=W_{tree}\,,
\end{equation}
and the only corrections that can enter are of the non-perturbative nature. 

In the later part of the draft, we shall study SFEOM and compare the SFEOM to the effective potential of the low-energy supergravity descriptions. For this, it would be useful to review how the F-term potential can be derived via dimensional reduction. To simplify the discussion, we shall assume that the warping is extremely weak. The energy density is stored in largely two parts of the ten-dimensional effective supergravity action: the threeform field strength
\begin{equation}\label{eqn:pot1}
-\frac{1}{4\kappa_{10}^2}\int_{\Bbb{R}^{1,3}\times X}d^{10}X\sqrt{-G^{(E)}}\frac{G_3\cdot\overline{G}_3}{\im\tau}\,,
\end{equation}
and the DBI action of the spacetime filling D3-branes and O3-planes
\begin{equation}\label{eqn:pot2}
-\mu_3\int_{\Bbb{R}^{1,3}\times X} d^{10}X\sqrt{-G^{(E)}}\rho_{D3}^{loc}\,.
\end{equation}
One can rewrite $G_3$ as
\begin{equation}
G_3=\frac{1}{2}(G_++G_-)\,,
\end{equation}
where we define
\begin{equation}
G_\pm:=G_3\mp i\star_6 G_3\,.
\end{equation}
Then, we can rewrite \eqref{eqn:pot1} as
\begin{equation}
-\frac{1}{2\kappa_{10}^2\im\tau}\int_{\Bbb{R}^{1,3}\times X} d^{10}X\sqrt{-G^{(E)}} G_-\cdot \overline{G}_--\frac{1}{2\kappa_{10}^2\im\tau}\int _{\Bbb{R}^{1,3}\times X}d\text{Vol}_{\Bbb{R}^{1,3}}\wedge H\wedge F\,.
\end{equation}
Then, we collect \eqref{eqn:pot1} and \eqref{eqn:pot2} to wrtie the effective potential
\begin{equation}
\mathcal{V}_{eff}=\mathcal{V}_1+\mathcal{V}_2\,,
\end{equation}
where
\begin{equation}
\mathcal{V}_1:=\frac{1}{2\kappa_{10}^2\im\tau}\int_{ X} d^{6}X\sqrt{G^{(E)}} G_-\cdot \overline{G}_-\,,
\end{equation}
and
\begin{equation}
\mathcal{V}_2:=\frac{1}{2\kappa_{10}^2}\int_{X}H\wedge F+\mu_3\int_{ X} d^{6}X\sqrt{G^{(E)}}\rho_{D3}^{loc}\,.
\end{equation}
As was shown in \cite{Giddings:2001yu},
\begin{equation}
V_F=\mathcal{V}_1\,.
\end{equation}
As a result, variation of the F-term potential with respect to a variation of the internal manifold $\varphi,$ e.g. Kahler moduli or complex structure moduli, 
\begin{equation}
\partial_{\varphi} V_F \delta\varphi
\end{equation}
is equivalent to
\begin{equation}\label{eqn:der eff pot}
\frac{1}{2\kappa_{10}^2\im\tau}\int_Xd^6X \sqrt{G^{(E)}}\left(-\frac{1}{2} (G_-)_{abc}(\overline{G}_-)_{def}\delta G^{ad}G^{be}G^{cf} +\frac{\delta G_{ab} G^{ab}}{2}G_-\cdot \overline{G}_-\right)\,.
\end{equation}
This form of the variation of the effective potential is exactly the one we will find in the string field theory analysis. Note in the literature, $\mathcal{V}_2$ is often omitted because the cancellation of the RR tadpole automatically guarantees that $\mathcal{V}_2$ vanishes. In the string field theory analysis, however, we shall find that $\mathcal{V}_2=0$ arises as an independent consistency condition.

We shall now briefly review the supersymmetry conditions given the superpotential and the tree-level approximation of the K\"{a}hler potential. As the goal of this draft is to understand the perturbative flux vacua, we shall ignore the non-perturbative terms in the superpotential and focus on the perturbative superpotential. The supersymmetric minima can be found by solving the F-term equations
\begin{equation}
D_a W_{pert}=0\,.
\end{equation}

The F-term conditions for complex structure and the axio-dilaton are, respectively,
\begin{equation}
D_{z_a}W_{pert}=\int_X G_3\wedge\chi^a=0\,,
\end{equation}
and
\begin{equation}
D_\tau W_{pert}=-\frac{2i}{\im\tau}\int_X \overline{G}_3\wedge\Omega=0\,,
\end{equation}
where $\chi^a$ is a $(2,1)$ form defined by
\begin{equation}
\partial_{z_a}\Omega=-K_{z_a}\Omega+\chi^a\,.
\end{equation}
Therefore, to solve the F-term equations for the complex structure moduli and the axio-dilaton, the moduli should be adjusted such that the complexified threeform is ISD, meaning $G_3$ is decomposed to $(2,1)\oplus (0,3)$ Hodge type.

Now let us study the effective potential when the F-term conditions for the complex structure moduli and the axio-dilaton modulus are solved, without assuming that the F-term conditions for the K\"{a}hler moduli are solved. Let us recall that the F-term potential is given by
\begin{equation}
V_F=e^K ( K^{a\bar{b}} D_aW_{pert}D_{\bar{b}}\overline{W}_{pert}-3|W_{pert}|^2)\,.
\end{equation}
Because the F-terms for the complex structure moduli and the axio-dilaton modulus are solved and the perturbative superpotential does not depend on the K\"{a}hler moduli, the F-term potential reduces to
\begin{equation}
V_F = e^K (K^{T_a \bar{T}_{\bar{b}}}K_{T_a} K_{\bar{T}_b}-3)|W_{pert}|^2\,.
\end{equation}
At the leading order in $g_s$ and $\alpha',$ the F-term potential enjoys the no-scale structure
\begin{equation}
K^{T_a \bar{T}_{\bar{b}}}K_{T_a} K_{\bar{T}_b}-3=0\,,
\end{equation}
and the F-term potential vanishes even if the F-term equations for the K\"{a}hler moduli are not solved. Hence, at the leading order in $\alpha'$ and $g_s,$ the F-term potential is minimized if the F-term equations for the complex structure and the axio-dilaton are solved. Of course, in general, even at the leading order in $g_s,$ there is the famous $\alpha'^3$ corrections to the  K\"{a}hler potential \cite{Becker:2002nn}. And, this $\alpha'^3$ corrections will cause the perturbative runaway if the F-term equations for the K\"{a}hler moduli are not solved. Note, however, for the explicit example we consider in this paper, such an $\alpha'^3$ correction is absent.

The F-term conditions for the K\"{a}hler moduli are given as
\begin{equation}
D_{T_a}W_{pert}=K_{T_a} W_{pert}=0\,.
\end{equation}
Therefore, to find perturbative supersymmetric vacua away from the infinite distance points, one needs to choose the quantized fluxes $F_3$ and $H_3$ such that a point in the moduli space at which the following equations can be solved
\begin{equation}\label{eqn:perturbative flux vacua}
G_3\wedge\Omega=G_3\wedge\chi^a=G_3\wedge\overline{\Omega}=0\,,
\end{equation}
meaning that at the F-term minimum the complexified threeform flux must be of $(2,1)$ Hodge type. Because the non-trivial F-term conditions away from the infinite distance limits only depend on complex structure moduli and the axio-dilaton, there is precisely one more equation to solve than the number of complex structure moduli and the axio-dilaton. As such, generically, the F-term conditions do not have solutions. So, to find perturbative supersymmetric vacua, we shall look for special choices of fluxes.

An important remark on the perturbative supersymmetric flux vacua \eqref{eqn:perturbative flux vacua} should follow. Because the perturbative superpotential vanishes at the F-term minimum, the F-term conditions can be rewritten as
\begin{equation}
\partial_a W_{pert}=W_{pert}=0\,.
\end{equation}
Therefore, the F-term conditions do not depend on the K\"{a}hler potential. Because of this, despite the fact that the K\"{a}hler potential is expected to be renormalized, the perturbative corrections to the effective action cannot destabilize the vacuum because the tree-level superpotential is perturbatively exact. As a result, the perturbative supersymmetric flux vacua of \eqref{eqn:perturbative flux vacua} should furnish a good background for string perturbation theory.

Before we discuss the construction of supersymmetric vacua, we shall make a few remarks about the non-perturbative corrections to the superpotential. Supersymmetry constrains the classes of the non-perturbative corrections that can enter the superpotential. In total, there are four different types of possible non-perturbative corrections: D3-brane gaugino condensation, D7-brane gaugino condensation, D(-1)-instantons, and D3-instantons. The D7-brane gaugino condensation and the D3-instanton contributions to the superpotential take the following form \cite{Witten:1996bn}
\begin{equation}
\mathcal{A} e^{-2\pi c T}\,,
\end{equation}
where $T$ is an Einstein-frame four-cycle volume. In general, $\mathcal{A}$ is expected to depend on all but K\"{a}hler moduli. On the other hand, the D3-brane gaugino condensation and D(-1)-instanton contributions take the following form
\begin{equation}
\mathcal{A} e^{-2\pi c\tau}\,.
\end{equation}
Of those, because we will work with the configurations in which D7-brane is absent, the D7-brane gaugino condensation cannot happen. Similarly, as was shown in \cite{Kim:2022jvv}, the D(-1)-instanton contribution to the superpotential is absent if the D7-brane tadpole is canceled locally. Therefore, the D(-1)-instanton contribution is also absent in the examples we shall consider. On the other hand, D3-instantons and D3-brane gaugino condensations can be, in principle, generated.\footnote{Because the D3-branes wrapped on four cycles in toroidal orbifolds, in general, have too many fermion zero modes, to study the generation of the D3-instanton superpotential one must carefully take into account the effects of the flux couplings \cite{Kallosh:2005gs,Grimm:2011dj,Bianchi:2011qh,Bianchi:2012kt}.}

With this understanding, we shall proceed to the cases where the flux terms and the local source terms on the RHS of \eqref{eqn:phi p} and \eqref{eqn:phi n} can be systematically treated in string perturbation theory. For this purpose, we will adopt a very special type of flux vacua recently introduced in \cite{Demirtas:2019sip}, which was studied in the attempt to realize small vacuum expectation value (VEV) of the flux superpotential in Calabi-Yau orientifold compactifications. Because the period integral of the toroidal compactification has an analogous structure to that of large complex structure limits of the Calabi-Yau compactifications, we can borrow the construction of \cite{Demirtas:2019sip} with a simple modification. The special structure employed in \cite{Demirtas:2019sip} will allow us to identify the axio-dilaton VEV as an expansion parameter, and understanding this structure will help us better understand the expansion scheme for the worldsheet theory in flux compactifications. We will, therefore, quickly review the construction of \cite{Demirtas:2019sip}. 

We shall focus on a large complex structure limit of Calabi-Yau compactifications. In the large complex structure limit, the period integral $\int_\gamma\Omega$ for a three cycle $\gamma\in H_3(X,\Bbb{Z})$ enjoys the following exact expansion 
\begin{equation}
\int_\gamma \Omega(z)= f_{poly}(\gamma,z)+f_{np}(\gamma,z)\,,
\end{equation}
where $z$ is the flat coordinate for complex structure moduli, $f_{poly}$ is a degree 3 polynomial in $z$
\begin{equation}
f_{poly}(\gamma,z)=\frac{1}{3!}A_{\gamma,i,j,k}z^iz^jz^k+\frac{1}{2}B_{\gamma,i,j}z^iz^j+C_{\gamma,i}z^i+D_{\gamma}\,,
\end{equation}
and $f_{np}(\gamma,z)$ is a collection of exponentially suppressed terms
\begin{equation}
f_{np}(\gamma,z)=\sum_{\vec{d}}\mathcal{N}_{\vec{d},\gamma}e^{2\pi i \vec{d}\cdot \vec{z}}\,,
\end{equation}
which vanishes in toroidal compactifications.

The central idea of \cite{Demirtas:2019sip} is that there is a choice of quantized fluxes $H_3$ and $F_3$ such that
\begin{equation}\label{eq:quadratic}
\int_{F_3}\Omega(z)=\frac{1}{2}B_{F_3,i,j}z^iz^j+f_{np}(F_3,z)\,,
\end{equation}
\begin{equation}
\int_{H_3}\Omega(z)=C_{H_3,i}z^i\,,
\end{equation}
and the flux superpotential is therefore given as
\begin{equation}
W=\int_{X} G_3\wedge\Omega= \frac{1}{2}B_{F_3,i,j}z^iz^j-C_{H_3,i}\tau z^i+f_{np}(F_3,z)\,.
\end{equation}
Provided that $B_{F_3,i,j}$ has a non-trivial determinant and
\begin{equation}
C_{H_3,i}(B_{F_3,i,j})^{-1}C_{H_3,j}=0\,,
\end{equation}
one can find a solution to the approximate F-term equation ignoring $f_{np}$
\begin{equation}\label{eqn:pfv cond}
z^i=C_{H_3,j} (B_{F_3,i,j})^{-1}\tau=p^i\tau\,,
\end{equation}
and the solution is valid as long as $z^i$ is in a K\"{a}hler cone of the mirror Calabi-Yau and $\im(z^i)\gg1.$ By reinstating $f_{np}(F_3,z),$ one can then also stabilize $\tau$ at small $g_s$ and small vacuum expectation value of $W_0:=\langle \int G\wedge\Omega\rangle.$ %
To adopt the flux vacua of \cite{Demirtas:2019sip} to toroidal compactifications, we must take into account that $f_{np}$ vanishes exactly. As a result, the F-term condition for complex structure moduli and the axio-dilaton \eqref{eqn:pfv cond} become perturbatively exact, and the tree-level superpotential vanishes exactly. Therefore, the F-term condition for K\"{a}hler moduli is also solved simultaneously. It is important to note that because there are no exponentially suppressed terms in $W_{tree},$ there is a perturbative moduli space parametrized by $z^i=p^i\tau.$ This will allow us to tune the $g_s$ to a small value without going off-shell. Furthermore, as previously emphasized, the F-term conditions for the perturbative supersymmetric flux vacua are not sensitive to the K\"{a}hler potential. As a result, the class of flux vacua that will be studied in this draft is resilient to the unknown perturbative $g_s$ and $\alpha'$ corrections.

\subsection{An explicit example}\label{subsection:explicitExample}
We now introduce a simple example that fits into the ansatz and assumptions discussed in the previous subsection. We take $X=T^6/\Bbb{Z}_2,$ and thus consider the toroidal orientifold, where the orientifold action $\mathcal{I}$ acts on the complex coordinates of $T^6$ as\footnote{For the conventions for $T^6$, see \S\ref{app:tor}.}
\begin{equation}
\mathcal{I}:(Z_1,Z_2,Z_3)\mapsto -(Z_1,Z_2,Z_3)\,.
\end{equation}
Before the orientifolding, the type IIB compactification has $\mathcal{N}=8$ supersymmetry. The orientifolding removes half of the supersymmetries, and in the fluxless cases, the target space has $\mathcal{N}=4$ supersymmetry. Turning on quantized threeform fluxes will completely break supersymmetry at generic points in the moduli space. In contrast, some of the supersymmetries will be restored at special subloci of the moduli space.

We shall now find the orientifold planes. Orientifold planes are identified as fixed loci of the orientifold actions. Noting that the complex coordinates are periodic under the translations
\begin{equation}
Z^i\sim Z^i+1\sim Z^i+u_i\,,
\end{equation}
we find that there are, in total, 64 O3-planes at
\begin{equation}
\frac{1}{2} (\vec{m}\cdot I+\vec{n}\cdot \vec{u})=\left(\frac{m_1+n_1 u_1}{2},\frac{m_2+n_2 u_2}{2},\frac{m_3+n_3u_3}{2}\right)\,,
\end{equation}
where $m_i$ and $n_i$ are integers. As a result, the D3-brane tadpole induced by the O3-planes is $-16.$ We shall choose the appropriate 3-form fluxes and place a proper number of D3-branes to cancel the tadpole
\begin{equation}
\frac{1}{2(2\pi)^4\alpha'^2}\int_{T^6} H_3\wedge F_3+N_{D3}=16\,.
\end{equation}
A simplifying feature of this orientifold is that there are no O7-planes and hence no D7-branes.

We will turn on the following quantized 3-form fluxes $F_3$ and $H_3$ 
\begin{equation}\label{eqn:F3explicit}
\frac{1}{(2\pi)^2\alpha'}F_3=4 dY^2\wedge dY^3 \wedge dY^5-2 dY^1\wedge dY^4\wedge dY^5-2 dY^1\wedge dY^3\wedge dY^6\,,
\end{equation}
\begin{equation}\label{eqn:H3explicit}
\frac{1}{(2\pi)^2\alpha'}H_3=4 dY^1\wedge dY^4\wedge dY^6-2dY^2\wedge dY^3\wedge dY^6-2 dY^2\wedge dY^4\wedge dY^5\,.
\end{equation}
This flux choice was found in \cite{Cicoli:2022vny}. The D3-brane tadpole stored in the fluxes is then
\begin{equation}
\frac{1}{2(2\pi)^4\alpha'^2}\int_{T^6} H_3\wedge F_3=12\,.
\end{equation}
Therefore, to cancel the tadpole, we shall include four spacetime-filling D3-branes in the compactification, located at $Y=Y_{D3}^i$ for $i=1\,,\dots,4.$ As a result, we have
\begin{equation}
\rho_{D3}^{loc}=\sum_{i=1}^4\delta^{(6)}(Y-Y_{D3}^i)-\frac{1}{4}\sum_{n,m}\delta^{(6)} \left(Y-\frac{1}{2} (\vec{m}\cdot I+\vec{n}\cdot \vec{u})\right)\,.
\end{equation}

We are now ready to compute the flux superpotential. We have
\begin{equation}
\frac{1}{(2\pi)^2\alpha'} \int_{T^6} F_3 \wedge \Omega=-4 u_2 u_3+2 u_1u_3+2u_1u_2\,,
\end{equation}
and 
\begin{equation}
-\frac{1}{(2\pi)^2\alpha'}\int_{T^6}\tau  H_3\wedge \Omega=-\tau(4 u_1-2u_2-2u_3)\,.
\end{equation}
Therefore, we find
\begin{equation}
\frac{1}{(2\pi)^2\alpha'}\int_{T^6} G_3\wedge\Omega=-4u_2u_3+2u_1u_3+2u_1u_2-\tau(4u_1-2u_2-2u_3)\,.
\end{equation}
Note that the non-perturbative term $f_{np}$ is absent in this specific example. The F-term equations for the complex structure moduli and the axio-dilaton are given as
\begin{equation}
2u_3+2u_2-4\tau=-4u_3+2u_1+2\tau=-4u_2+2u_1+2\tau=-4u_1+2u_2+2u_3=0\,.
\end{equation}
Therefore, we find a solution to the F-term equation with one flat direction parametrized by $\tau$
\begin{equation}
u_1=u_2=u_3=\tau\,.
\end{equation}
As advertised in the previous subsection, the VEVs of complex structure moduli are proportional to that of $\tau$. At the F-term minimum, we find that $G_3=F_3-\tau H_3$ can be written as
\begin{align}
G_3=&\frac{i}{\im\tau} \left(2d\bar{Z}^1\wedge dZ^2\wedge dZ^3-dZ^1\wedge d\bar{Z}^2\wedge dZ^3-dZ^1\wedge dZ^2\wedge d\bar{Z}^3\right)\,.
\end{align}
Because $G_3$ does not contain $(0,3)$ form, we find that at the F-term minimum, the perturbative superpotential vanishes exactly 
\begin{equation}
\left\langle\int_{T^6} G_3\wedge\Omega\right\rangle=0\,,
\end{equation}
and the F-term equations for K\"{a}hler moduli are automatically solved as well. Note also that 3-form fluxes scale in terms of $\tau$ as $|F_3|^2\sim\tau,$ $|H_3|^2\sim\tau^{-1}$, and $[H_3\wedge F_3]\sim\tau^0$.

\section{Worldsheet conventions}\label{sec:WS}
In this section, we collect relevant worldsheet conventions and check the BRST exactness of the vertex operators for the NSNS field strength. In this section, we shall work in the \emph{string-frame}.

\subsection{$\Bbb{R}^{1,9}$}\label{subsection:non compact CFT}

For the free field CFT with $(\mathcal{N},\bar{\mathcal{N}})=(1,1)$ worldsheet supersymmetry, we write the matter part of the action as
\begin{equation}
S_{matter}=\frac{1}{2\pi}\int d^2z \left( \frac{1}{\alpha'}\partial X^\mu \bar{\partial}X_\mu+\frac{1}{2} \psi^\mu\bar{\partial}\psi_\mu+\frac{1}{2}\tilde{\psi}^\mu\partial \tilde{\psi}_\mu\right)\,.
\end{equation}
In the ghost system, we have the usual $b,~c,~\beta,~\gamma$ system, whose action is given as
\begin{equation}
S_{ghost}=\frac{1}{2\pi}\int d^2z \left[b\bar{\partial} c+\bar{b}\partial \bar{c}+\beta\bar{\partial}\gamma+\bar{\beta}\partial \bar{\gamma} \right]
\end{equation}
To ``bosonize" the $\beta,~\gamma$ system and introduce the scalar fields $\phi,~\bar{\phi}$ and fermionic fields $\xi,~\eta,~\bar{\xi},~\bar{\eta}$
\begin{equation}
\beta=\partial \xi e^{-\phi}\,,\quad \gamma=\eta e^\phi\,,\quad \bar{\beta}=\bar{\partial}\bar{\xi}e^{-\bar{\phi}}\,,\quad \bar{\gamma}=\bar{\eta}e^{\bar{\phi}}\,.
\end{equation}
Note that $e^{\pm \phi}$ and $e^{\pm\bar{\phi}}$ behave as worldsheet fermions.

The OPEs of the aforementioned fields are written as
\begin{align}
X^\mu (z,\bar{z})X^\nu (0,0)\sim& -\frac{\alpha'}{2}\eta^{\mu\nu}\ln |z|^2\,,\qquad\psi^\mu(z)\psi^\nu(0)\sim \frac{\eta^{\mu\nu}}{z}\,,\nonumber\\
c(z)b(0)\sim& \frac{1}{z}\,,\qquad \xi(z)\eta(0)\sim \frac{1}{z}\,,\nonumber\\
\partial \phi(z)\partial \phi(0)\sim&-\frac{1}{z^2}\,,\qquad e^{q_1\phi(z)}e^{q_2\phi(0)}\sim z^{-q_1q_2} e^{(q_1+q_2)\phi(0)}\,.
\end{align}
We write the mode expansions of the worldsheet fields
\begin{equation}
\partial X^\mu(z)=-i\sqrt{\frac{\alpha'}{2}}\sum_{m=-\infty}^\infty \frac{\alpha_m^\mu}{z^{m+1}}\,,\quad \bar{\partial}X^\mu =-i\sqrt{\frac{\alpha'}{2}}\sum_{m=-\infty}^\infty \frac{\bar{\alpha}_m^\mu}{\bar{z}^{m+1}}\,,
\end{equation}
\begin{equation}
\psi^\mu(z)=\sum_{n\in \Bbb{Z}+\nu} \frac{\psi_n^\mu}{z^{n+\frac{1}{2}}}\,,\quad\bar{\psi}^\mu(\bar{z})=\sum_{n\in\Bbb{Z}+\nu}\frac{\bar{\psi}^\mu_n}{\bar{z}^{n+\frac{1}{2}}}\,,
\end{equation}
\begin{equation}
b(z)=\sum_{m=-\infty}^\infty \frac{b_m}{z^{m+2}}\,,\quad c(z)=\sum_{m=-\infty}^{\infty} \frac{c_m}{z^{m-1}}\,,
\end{equation}
\begin{equation}
\beta(z)=\sum_{n\in\Bbb{Z}+\nu}\frac{\beta_n}{z^{n+\frac{3}{2}}}\,,\quad \gamma(z)=\sum_{n\in\Bbb{Z}+\nu}\frac{\gamma_n}{z^{n-\frac{1}{2}}}\,,
\end{equation}
where $\nu$ is 0 in the Ramond sector and $1/2$ in the Neveu-Schwarz sector.

Therefore, we have the following (anti) commutation relations
\begin{equation}
[\alpha_m^\mu,\alpha_n^\nu]=[\bar{\alpha}_m^\mu,\bar{\alpha}_n^\nu]=m \eta^{\mu\nu}\delta_{m,-n}\,,
\end{equation}
\begin{equation}
\{ \psi_m^\mu,\psi_n^\nu\}=\{\bar{\psi}_m^\mu,\bar{\psi}_n^\nu\}=\eta^{\mu\nu}\delta_{m,-n}\,,
\end{equation}
\begin{equation}
[\gamma_m,\beta_n]=\delta_{m,-n}\,,\quad \{b_m,c_n\}=\delta_{m,-n}\,.
\end{equation}

The matter stress energy tensor and its supersymmetric partner take the form
\begin{equation}
T_m=-\frac{1}{\alpha'} \partial X^\mu \partial X_\mu-\frac{1}{2}\psi_\mu \partial \psi^\mu\,,
\end{equation}
and
\begin{equation}
G_m=i\sqrt{\frac{2}{\alpha'}}\psi^\mu\partial X_\mu\,.
\end{equation}
The ghost stress energy tensor is given as
\begin{equation}
T_g=(\partial b) c-2\partial (bc)+(\partial \beta)\gamma -\frac{3}{2}\partial (\beta\gamma)\,,
\end{equation}
and its supersymmetric partner is written as
\begin{equation}
T_F=-\frac{1}{2}(\partial \beta) c+\frac{3}{2}\partial (\beta c)-2b\gamma\,.
\end{equation}

We write the BRST current as
\begin{equation}
j_B=c\left( T_m-\frac{1}{2}(\partial \phi)^2-\partial^2\phi-\eta\partial\xi\right)+\eta e^\phi G_m +bc\partial c-\eta\partial \eta b e^{2\phi}\,,
\end{equation}
and the corresponding BRST charge as
\begin{equation}
Q_B=\frac{1}{2\pi i}\oint j_B\,.
\end{equation}
The PCO is normalized as
\begin{equation}
\chi:=\{ Q_B,\xi\}=c\partial \xi+e^{\phi}G_m-\partial \eta be^{2\phi}-\partial (\eta b e^{2\phi})\,.
\end{equation}

Finally, let us turn to the spin fields. Our spin field convention will mostly follow that of \cite{Alexandrov:2021shf,Alexandrov:2021dyl}. We shall denote the 16-component spin fields in the holomorphic matter sector by $\Sigma^\alpha$ and $\Sigma_\alpha,$ which carry opposite chirality. Similarly, in the anti-holomorphic sector, we have $\bar{\Sigma}^\alpha$ and $\bar{\Sigma}_\alpha.$ We shall choose the convention such that $e^{-\phi/2}\Sigma_\alpha,$ $e^{-3\phi/2}\Sigma^\alpha,$ $e^{-\bar{\phi}/2}\bar{\Sigma}_\alpha,$ and $e^{-3\bar{\phi}/2}\bar{\Sigma}^\alpha$ are GSO even. The reason for choosing the same chirality for the GSO even spin fields in the holomorphic and the anti-holomorphic sectors is because, in type IIB string theory, the holomorphic and the anti-holomorphic spacetime fermions carry the same chirality. We write some important OPEs involving the spin fields
\begin{align}
\psi^\mu(z) e^{-\phi/2}\Sigma_\alpha(w)=&-\frac{1}{\sqrt{2}}(z-w)^{-1/2} (\Gamma^\mu)_{\alpha\beta}e^{-\phi/2}\Sigma^\beta(w)+\cdots\,,\\
\psi^\mu(z)e^{-\phi/2}\Sigma^\alpha(w)=&-\frac{1}{\sqrt{2}}(z-w)^{-1/2}(\Gamma^\mu)^{\alpha\beta}e^{-\phi/2}\Sigma_\beta+\cdots\,,\\
e^{-3\phi/2}\Sigma^\alpha e^{-\phi/2}\Sigma_\beta=&(z-w)^{-2} \delta_\beta^\alpha e^{-2\phi}(w)+\cdots\,,\\
e^{-\phi/2}\Sigma_\alpha(z) e^{-\phi/2}\Sigma_\beta(w)=&\frac{1}{\sqrt{2}}(z-w)^{-1}(\Gamma^\mu)_{\alpha\beta}e^{-\phi}\psi_\mu(w)+\dots\,.
\end{align}
Note that the $16\times 16$ gamma matrices $\Gamma^\mu$ satisfy the following identities
\begin{equation}
\{\Gamma^\mu,\Gamma^\nu\}=2\eta^{\mu\nu} I_{16}\,,\quad (\Gamma^\mu)_{\alpha\beta}=(\Gamma^\mu)_{\beta\alpha}\,,\quad (\Gamma^\mu)^{\alpha\beta}=(\Gamma^\mu)^{\beta\alpha}\,,
\end{equation}
\begin{equation}
(\Gamma^\mu)^{\alpha\beta}=(\Gamma^\mu)_{\alpha\beta}\quad\text{for }\mu\neq0\,,\quad (\Gamma^0)^{\alpha\beta}=\delta^{\alpha\beta}\,,\quad (\Gamma^0)_{\alpha\beta}=-\delta_{\alpha\beta}\,.
\end{equation}
We define $\Gamma_{10}$ as
\begin{equation}
(\Gamma^{10})^\alpha_{~\beta}:=(\Gamma^0\cdots \Gamma^9)^\alpha_{~\beta}=\delta^{\alpha}_{~\beta}\,,
\end{equation}
and
\begin{equation}
(\Gamma^{10})_\alpha^{~\beta}:= (\Gamma^0\cdots \Gamma^9)_\alpha^{~\beta}=-\delta_\alpha^{~\beta}\,.
\end{equation}
We then have the usual identities
\begin{equation}
\{\Gamma^\mu,\Gamma^{10}\}=0\,,
\end{equation}
\begin{equation}
(\Gamma^{10})^2=I_{16}\,.
\end{equation}

\subsubsection{Closed string states}\label{sec:closed states}
We now collect vertex operators for the massless states. Let us start with the NS sectors following the conventions of \cite{Polchinski:1998rq}. In the (0,0) picture, we have
\begin{equation}
V_{NSNS}^{(0,0)}(z,\bar{z})=-\frac{2g_c}{\alpha'}\epsilon_{\mu\nu}\left(i\partial X^\mu+\frac{\alpha'}{2}p\cdot \psi\psi^\mu\right)\left(i\bar{\partial}X^\nu+\frac{\alpha'}{2}p\cdot\bar{\psi}\bar{\psi}^\nu\right)e^{ip\cdot X}\,+...,
\end{equation}
where $...$ include terms with different ghost structures, and the polarization tensors $\epsilon^{(h)},$ $\epsilon^{(B)},$ and $\epsilon^D$ for the graviton state, the antisymmetric two form, and the dilaton, respectively, satisfy 
\begin{align}
\epsilon^{(h)}_{\mu\nu}=&\epsilon^{(h)}_{\nu\mu}\,,\qquad \epsilon_{\mu\nu}^{(h)}\eta^{\mu\nu}=p^\mu\epsilon_{\mu\nu}^{(h)}=0\,,\\
\epsilon^{(B)}_{\mu\nu}=&-\epsilon^{(B)}_{\nu\mu}\,,\qquad p^\mu\epsilon_{\mu\nu}^{(B)}=0\,,\\
\epsilon^{(D)}_{\mu\nu}=&\frac{1}{\sqrt{8}}(\eta_{\mu\nu}-p_\mu\bar{p}_\nu-\bar{p}_\mu p_\nu)\,,\qquad p^\mu \epsilon_{\mu\nu}^{(D)}=0\,,
\end{align}
where $\bar{p}_\mu$ is an auxiliary vector that is defined as $\bar{p}_\mu:= n_\mu/(n\cdot p)\,,$ where $n$ is a generic null vector. We record the NSNS vertex operators in a few more pictures
\begin{align}
V_{NSNS}^{(-1,0)}=&g_c\sqrt{\frac{2}{\alpha'}}\epsilon_{\mu\nu}e^{-\phi} \psi^\mu\left(i\bar{\partial} X^\nu+\frac{\alpha'}{2}p\cdot \bar{\psi}\bar{\psi}^\nu\right)e^{ip\cdot X}\,+...,\\
V_{NSNS}^{(0,-1)}=&g_c\sqrt{\frac{2}{\alpha'}}\epsilon_{\mu\nu} \left(i\partial X^\mu+\frac{\alpha'}{2}p\cdot \psi\psi^\mu\right)e^{-\bar{\phi}}\bar{\psi}^\nu e^{ip\cdot X}\,+...,\\
V_{NSNS}^{(-1,-1)}=&-g_c \epsilon_{\mu\nu}e^{-\phi}\psi^\mu e^{-\bar{\phi}}\bar{\psi}^\nu e^{ip\cdot X}\,.
\end{align}
Note that following \cite{Polchinski:1998rq}, we identify $\kappa=\kappa_{10}e^{\Phi_0}=2\pi g_c.$ The aforementioned vertex operators correspond to the linearized fields $h_{\mu\nu},$ $a_{\mu\nu},$ and $D$ defined as follows \cite{Blumenhagen:2013fgp}
\begin{align}\label{eqn:NSNS fluctuations}
g_{\mu\nu}=&\eta_{\mu\nu}-2\kappa h_{\mu\nu}-\frac{2\kappa}{\sqrt{d-2}}\eta_{\mu\nu} D\,,\\
B_{\mu\nu}=&-2\kappa a_{\mu\nu}\,,\\
\Phi=&\Phi_0-\frac{1}{2}\kappa\sqrt{d-2} D\,,
\end{align}
where $d=10.$ A few important remarks are in order. $g_{\mu\nu}$ as written above is metric in string-frame. The corresponding spacetime action up to two derivatives in the string-frame is then given as
\begin{align}\label{eqn:action 2der NSNS}
S\supset&\frac{1}{2\kappa_{10}^2}\int d^{10}X \sqrt{-g} e^{-2\Phi} \left[ R+4(\partial \Phi)^2-\frac{1}{12}|H_3|^2\right]\,,\\
=&\int d^{10}X \sqrt{-g}\frac{1}{2}  \left( \frac{1}{2}h^{\mu\nu} \square h_{\mu\nu}-(\partial D)^2-\frac{1}{2} |da|^2\right)+\text{gauge fixing terms}+\mathcal{L}_{3pt}+\dots\,.
\end{align}
For convenience, we write the three-point interaction terms
\begin{align}
\int d^{10}X \mathcal{L}_{3pt}=& \int d^{10}X  \kappa_{10}e^{\Phi_0}\left[h^{\mu\nu} h^{\rho \sigma}\partial_\mu \partial_\nu h_{\rho\sigma}+2\partial^\sigma h_{\mu\nu}\partial^\mu h^{\nu\rho}h_{\rho\sigma}-h^{\mu\nu}\partial_\mu D\partial_\nu D\right]\nonumber\\
&\qquad\qquad\qquad\quad-\frac{e^{-\Phi_0}}{6\kappa_{10}}\frac{D}{\sqrt{8}} |H_3|^2-\frac{1}{4\kappa_{10}} e^{-\Phi_0}h^{\mu\nu} H_{\mu ab}H_\nu^{~ab}\,.
\end{align}

Now, we collect the conventions for the Ramond-Ramond sector. We first write the vertex operator for the massless RR fields in $(-1/2,-1/2)$ picture
\begin{equation}
V_{RR}^{(-\frac{1}{2},-\frac{1}{2})}=\frac{g_c\sqrt{\alpha'}}{8\sqrt{2}\kappa_{10}} F^{\alpha\beta}c\bar{c}e^{-\phi/2}\Sigma_\alpha e^{-\bar{\phi}/2}\bar{\Sigma}_\beta e^{ip\cdot X}\,,
\end{equation}
where we define
\begin{equation}
F^{\alpha\beta}=\sum_{k} \frac{i}{(2k+1)!}F_{M_1\dots M_{2k+1}}^{(2k+1)}( \Gamma^{M_1\dots M_{2k+1}})^{\alpha\beta}\,,
\end{equation}
and
\begin{equation}
F_{M_1\dots M_{2k+1}}^{(2k+1)}=i p_{M_1} C^{(2k)}_{M_2\dots M_{2k+1}}+\text{permutations}\,.
\end{equation}
This normalization of the vertex operator corresponds to the action \cite{Blumenhagen:2013fgp}
\begin{equation}
S=-\frac{1}{2\kappa_{10}^2}\int d^{10}X \sqrt{-g} \frac{1}{2(2k+1)!} F_{M_1\dots M_{2k+1}}F^{M_1\dots M_{2k+1}}\,.
\end{equation}
By varying the NSNS fields \eqref{eqn:NSNS fluctuations}, we obtain the following interaction vertices
\begin{equation}
S\supset -\frac{1}{2\kappa_{10}^2}\int d^{10}X \left[\frac{\kappa}{(2k)!} h^{\mu\nu} F_{\mu M_2\dots M_{2k+1}}F_{\nu}^{~M_2\dots M_{2k+1}}-\kappa\frac{4-2k}{(2k+1)!\sqrt{8}} D|F^{(2k+1)}|^2\right]\,. 
\end{equation}
We also record the vertex operators for the massless RR fields in $(-1/2,-3/2)$ and $(-3/2,-3/2)$ pictures \cite{Billo:1998vr,Alexandrov:2021shf,Alexandrov:2021dyl}
\begin{align}
V_{RR}^{(-1/2,-3/2)}=&\frac{g_c}{4\sqrt{2}\kappa_{10}}\left[A_{\alpha\beta}(\slashed{p})^{\beta\gamma} c\bar{c}e^{-3\phi/2}\Sigma^\alpha e^{-\bar{\phi}/2}\bar{\Sigma}_\gamma -E_\alpha^{~\beta}c\bar{c} e^{-3\phi/2}\Sigma^\alpha e^{-\bar{\phi}/2}\bar{\Sigma}_\beta\right.\nonumber\\
&\left.+\sqrt{\alpha'}D^\alpha_{~\beta}(\slashed{p})^{\beta\gamma}(\partial c+\bar{\partial}\bar{c}) c\bar{c}\partial\xi e^{-5\phi/2}\Sigma_\alpha e^{-\bar{\phi}/2}\bar{\Sigma}_\gamma+D^\alpha_{~\beta}c\bar{c}\partial\xi e^{-5\phi/2}\Sigma_\alpha \bar{\eta}e^{\bar{\phi}/2}\bar{\Sigma}^\beta\right]e^{ip\cdot X}\,,
\end{align}
and
\begin{align}
V_{RR}^{(-3/2,-3/2)}=&\frac{g_c}{2\sqrt{2}\kappa_{10}} \left[\frac{1}{\sqrt{\alpha'}}A_{\alpha\beta}c\bar{c}e^{-3\phi/2}\Sigma^\alpha e^{-3\bar{\phi}/2}\bar{\Sigma}^\beta+E_{\alpha}^{~\beta}(\partial c+\bar{\partial}\bar{c}) c\bar{c} e^{-3\phi/2}\Sigma^\alpha\bar{\partial}\bar{\xi} e^{-5\bar{\phi}/2}\bar{\Sigma}_\beta \right.\nonumber\\
&\qquad\quad\left.+D^{\alpha}_{~\beta}(\partial c+\bar{\partial}\bar{c})c\bar{c}\partial\xi e^{-5\phi/2}\Sigma_\alpha e^{-3\bar{\phi}/2}\bar{\Sigma}^\beta\right]e^{ip\cdot X}\,,
\end{align}
where we identify
\begin{equation}
F^{\alpha\beta}=\left(\slashed{p}A\slashed{p}-\slashed{p}E+D\slashed{p}\right)^{\alpha\beta}\,.
\end{equation}
The description in terms of $A,~E,$ and $D$ is redundant due to the gauge symmetry
\begin{equation}
A\mapsto A+\Lambda\,,\quad E\mapsto E+\Lambda' \slashed{p}\,,\quad D\mapsto D+\slashed{p} (\Lambda'-\Lambda)\,.
\end{equation}
Following \cite{Alexandrov:2021shf,Alexandrov:2021dyl}, we choose a gauge in which $A=0.$ In this gauge, we have
\begin{equation}
E_\alpha^{~\beta}=\frac{1}{2}\sum_k \frac{1}{(2k)!}C^{(2k)}_{M_1\dots M_{2k}} (\Gamma^{M_1\dots M_{2k}})_\alpha^{~\beta}\,,\quad D^\alpha_{~\beta}=-\frac{1}{2}\sum_k \frac{1}{(2k)!} C^{(2k)}_{M_1\dots M_{2k}} (\Gamma^{M_1\dots M_{2k}})^\alpha_{~\beta}\,.
\end{equation} 

To check the numerical factors of the vertex operators, we will perform a few sample calculations, which will prove useful later. Let us first compute a three-point function of the NSNS vertex operators. We shall put one vertex operator in the (0,0) picture and the other two in the (-1,-1) picture. We compute
\begin{align}
A_{NS}:=&\langle V_{NSNS}^{(0,0)}(p_1,\epsilon_1)V_{NSNS}^{(-1,-1)}(p_2,\epsilon_2)V_{NSNS}^{(-1,-1)}(p_3,\epsilon_3)\rangle_{S^2}\,,\\
=&\frac{2g_c^3}{\alpha'}\biggl\langle\epsilon_{\alpha\beta}^1\epsilon_{\gamma\delta}^2\epsilon_{\epsilon\eta}^3 \left(i\partial X^\alpha+\frac{\alpha'}{2}p_1\cdot \psi\psi^\alpha\right)\left(i\bar{\partial}X^\beta +\frac{\alpha'}{2}p\cdot \bar{\psi}\bar{\psi}^\beta\right) \nonumber\\
&\quad\quad \times e^{-\phi}\psi^\gamma e^{-\bar{\phi}}\bar{\psi}^\delta e^{-\phi}\psi^\epsilon e^{-\bar{\phi}}\bar{\psi}^\eta c cc\bar{c}\bar{c}\bar{c} e^{ip_1\cdot X}e^{ip_2\cdot X}e^{ip_3\cdot X}\biggr\rangle_{S^2}\,,\\
=&-\frac{\alpha'g_c^3 C_{S^2}}{2}\delta^{(10)}(p_1+p_2+p_3)\epsilon_{\alpha\beta}^1\epsilon_{\gamma\delta}^2\epsilon_{\epsilon\eta}^3\nonumber\\
&\times(p_3^\alpha \eta^{\gamma\epsilon}+p_2^\epsilon \eta^{\alpha\gamma}+p_1^\gamma\eta^{\epsilon\alpha})(p_3^\beta \eta^{\delta\eta}+p_2^\eta\eta^{\beta\delta}+p_1^\delta \eta^{\eta\beta})\,.
\end{align}
By using the relation
\begin{equation}
C_{S^2}=\frac{8\pi}{\alpha' g_c^2}\,,
\end{equation}
and taking the terms of order $\mathcal{O}(p^2),$ one can check that we reproduce the action \eqref{eqn:action 2der NSNS}. Let us compute an interaction between one NSNS vertex and two RR vertices. We will put the NSNS vertex in the (-1,-1) picture and the RR vertices in the (-1/2,-1/2) picture. We compute
\begin{align}
A_{RR}:=&\langle V_{NSNS}^{(-1,-1)}(p_1,\epsilon_1)V_{RR}^{(-\frac{1}{2},-\frac{1}{2})}(p_2,F_2)V_{RR}^{(-\frac{1}{2},-\frac{1}{2})}(p_3,F_3)\rangle_{S^2}\,,\\
=&\frac{\alpha' g_c^3}{128\kappa_{10}^2}\epsilon^1_{\mu\nu}F_1^{\alpha\beta}F_2^{\gamma\delta}\biggl\langle e^{-\phi}\psi^\mu e^{-\bar{\phi}}\bar{\psi}^\nu e^{-\phi/2}\Sigma_\alpha e^{-\bar{\phi}/2}\bar{\Sigma}_\beta e^{-\phi/2}S_\gamma e^{-\bar{\phi}/2}\bar{S}_\delta\nonumber\\
&\qquad\qquad\qquad\qquad\quad\times ccc\bar{c}\bar{c}\bar{c} e^{ip_1\cdot X}e^{ip_2\cdot X}e^{ip_3\cdot X}\biggr\rangle\,,\\
=&\frac{\alpha'g_c^3C_{S^2}}{256\kappa_{10}^2}\delta^{(10)}(p_1+p_2+p_3)\epsilon_{\mu\nu}^1 F_1^{\alpha\beta}F_2^{\gamma\delta} (\Gamma^\mu)_{\alpha\gamma}(\Gamma^\nu)_{\beta\delta}\,.
\end{align}
By plugging in 
\begin{equation}
F^{\alpha\beta}=\sum_k \frac{i}{(2k+1)!}F_{M_1\dots M_{2k+1}}^{(2k+1)} (\Gamma^{M_1\dots M_{2k+1}})^{\alpha\beta}\,,
\end{equation}
we compute
\begin{align}\label{eqn:RR correlation}
A_{RR}=&-\frac{\alpha'g_c^3 C_{S^2}}{256 \kappa_{10}^2}\delta^{(10)}(p_1+p_2+p_3)\epsilon_{\mu\nu}^1 \sum_{k,k'} \frac{F^1_{M_1\dots M_{2k+1}} F^2_{N_1\dots N_{2k'+1}}}{(2k+1)!(2k'+1)!}\text{Tr}\left(\Gamma^\mu \Gamma^{M_1\dots M_{2k+1}} \Gamma^\nu \Gamma^{N_1 \dots N_{2k'+1}}\right)\,.
\end{align}
For $k=k',$ we have
\begin{align}
A_{RR}=&-\frac{g_s}{4\kappa_{10}} \delta^{(10)}(p_1+p_2+p_3)\epsilon_{\mu\nu}^1 \sum_{k} \frac{F^1_{M_1\dots M_{2k+1}} F^2_{N_1\dots N_{2k+1}}}{(2k+1)!(2k+1)!}\nonumber\\
&\times \biggl[- \eta^{\mu\nu} \delta^{M_1\dots M_{2k+1},N_1\dots N_{2k+1}}+2 \left(\eta^{\mu M_1}\eta^{\nu N_1} \delta^{M_2\dots M_{2k+1}, N_2\dots N_{2k+1}}+\text{cyclic perm}\right)\biggr]\,,\\
=&-\delta^{(10)}(p_1+p_2+p_3)\frac{g_s}{4\kappa_{10}(2k)!}\left[ 2\epsilon^{\mu\nu}_1 F^{1,M_2\dots M_{2k+1}}_\mu F^{2}_{\nu M_2\dots M_{2k+1}} -\frac{\epsilon_1^{\mu\nu}\eta_{\mu\nu}}{2k+1} F_{M_1\dots M_{2k+1}}^1 F^{2, M_1\dots M_{2k+1}} \right]
\end{align}

\subsubsection{Boundary states}\label{sec:bdrystate}
We shall now collect the boundary states describing D-branes and O-planes \cite{Callan:1987px,Polchinski:1987tu,Billo:1997eg,DiVecchia:1997vef,DiVecchia:1999mal}. Our treatment will closely follow that of \cite{DiVecchia:1999mal}. 

For pedagogy, we shall first start with the boundary states in bosonic string theory, and then we will move on to the boundary states in superstring theories. In the parameterization
\begin{equation}
z=e^{t+i\sigma}\,,
\end{equation}
we will take $t=0$ or equivalently $|z|=1$ as the boundary of the string worldsheet. We shall assume that the string ends on a Dp-brane, which we shall assume to be static in the frame we choose. We will denote the directions parallel to the Dp-brane by $X^\alpha,$ where $\alpha=0,\dots,p,$ and the directions transverse to the Dp-brane by $X^i,$ where $i=p+1,\dots,d-1.$ We will assume that the Dp-brane is located at $y^i,$ for $i=p+1,\dots,d-1.$ Then, we have the following boundary conditions for the bosonic coordinates
\begin{equation}
\partial_\tau X^\alpha|_{\tau=0} |B_X\rangle =0\,,
\end{equation}
and
\begin{equation}
X^i|_{\tau=0}|B_X\rangle =y^i\,,
\end{equation}
where $|B_X\rangle$ is the boundary state for the matter CFT. In terms of the lowering and raising operators, we have the following boundary conditions
\begin{equation}
(\alpha_n^\mu+S^\mu_{~\nu} \bar{\alpha}_{-n}^\nu)|B_X\rangle =0\,,\quad\forall n\neq0\,,
\end{equation}
\begin{equation}
p^\alpha |B_X\rangle=0\,,\quad (q^i-y^i)|B_X\rangle=0\,,
\end{equation}
where
\begin{equation}
S^{\mu\nu}=(\eta^{\alpha\beta},-\eta^{ij})\,.
\end{equation}
The boundary $|B_X\rangle$ that satisfies the aforementioned boundary conditions is determined as
\begin{equation}
|B_X\rangle= \delta^{(d-p-1)}(q^i-y^i) \exp\left(\sum_{n=1}^\infty -\frac{1}{n} \alpha_{-n}^\mu S_{\mu\nu}\bar{\alpha}_{-n}^\nu\right)|0\rangle\,,
\end{equation}
where $|0\rangle$ is the SL(2;R) vacuum.

The full boundary state must also include the ghost boundary state
\begin{equation}
|B\rangle= \mathcal{N}|B_X\rangle \otimes |B_{gh}\rangle\,,
\end{equation}
where $\mathcal{N}$ is a normalization constant. The BRST invariance will fix the ghost boundary state for us, but before showing how the BRST invariance fixes $|B_{gh}\rangle,$ we shall first study the boundary condition for the matter Virasoro generators. Let us first study the boundary condition for $L_0$
\begin{equation}
L_0^{(m)}=\frac{\alpha' p^2}{4}+\sum_{n=1}^\infty \alpha_{-n}^\mu \alpha_{\mu n}\,.
\end{equation}
Let us act $\sum_{n=1}^\infty \alpha_{-n}^\mu \alpha_{\mu n}$ on the boundary state $|B_X\rangle,$ 
\begin{align}
\sum_{n=1}^\infty \alpha_{-n}^\mu \alpha_{\mu n}|B_X\rangle=&-\sum_{n=1}^\infty \alpha^\mu_{-n} S_{\mu\nu}\bar{\alpha}^\nu_{-n}|B_X\rangle =- \sum_{n=1}^\infty S_{\mu\nu} \bar{\alpha}^\nu_{-n}\alpha^\mu_{-n}|B_X\rangle =\sum_{n=1}^\infty \bar{\alpha}_{-n}^\mu \bar{\alpha}_{\mu n}|B_X\rangle\,,
\end{align}
where we used
\begin{equation}
S_{\mu\nu}S^{\nu\rho}=\delta^\rho_\mu\,,
\end{equation}
in the last step. As a result, we find
\begin{equation}
L_0^{(m)}|B_X\rangle= \bar{L}_0^{(m)}|B_X\rangle\,,
\end{equation}
Similarly, one can also easily conclude
\begin{equation}
L_n^{(m)}|B_X\rangle= \bar{L}_{-n}^{(m)}|B_X\rangle\,,
\end{equation}
for all $n.$

Now we are ready to fix the ghost boundary state by imposing the BRST invariance
\begin{equation}
(Q_B+\bar{Q}_B)|B\rangle=0\,.
\end{equation}
The BRST invariance then requires the following boundary condition for the ghost sector
\begin{equation}
(c_n+\bar{c}_{-n})|B_{gh}\rangle=0\,,\quad (b_n-\bar{b}_{-n})|B_{gh}\rangle=0\,.
\end{equation}
The solution to the boundary condition is given as
\begin{equation}
|B_{gh}\rangle=\exp\left(\sum_{n=1}^\infty c_{-n}\bar{b}_{-n}-b_{-n}\bar{c}_{-n}\right) \frac{c_0+\bar{c}_0}{2}c_1\bar{c}_1|0\rangle\,.
\end{equation}
As a result, we obtain the Dp-brane boundary state in bosonic string theory
\begin{equation}
|B\rangle =\mathcal{N}\delta^{(d-p-1)}(q^i-y^i) \exp\left(\sum_{n=1}^\infty( -\frac{1}{n} \alpha_{-n}^\mu S_{\mu\nu}\bar{\alpha}_{-n}^\nu)+\sum_{n=1}^\infty( c_{-n}\bar{b}_{-n}-b_{-n}\bar{c}_{-n})\right)\frac{c_0+\bar{c}_0}{2}c_1\bar{c}_1|0\rangle\,.
\end{equation}
One can determine the normalization constant $\mathcal{N}$ by requiring the factorization limit of open string at one-loop is consistent with the closed string channel \cite{Callan:1987px,Polchinski:1987tu}. But, we shall not write the normalization constant for bosonic string theory here.

Let us now study a boundary state for an Op-plane. This will be represented by the cross cap. Suppose that an orientifold action $\Omega$ acts on the target space coordinates as
\begin{equation}
\Omega: X^\alpha\mapsto X^\alpha\,,\quad \Omega: X^i\mapsto -X^i\,.
\end{equation}
Then, the orientifold plane is located at the fixed locus $X^i=y^i,$ for all $i.$ In the flat spacetime, $y^i$ is simply the origin. To find the crosscap state, we impose the following boundary conditions
\begin{equation}
X^\alpha(0,\sigma)|C\rangle=X^\alpha(0,\sigma+\pi)|C\rangle\,,\quad \partial_\tau X(0,\sigma)|C\rangle=-\partial_\tau X(0,\sigma+\pi)|C\rangle\,,
\end{equation}
\begin{equation}
X^i(0,\sigma)|C\rangle=-X^i(0,\sigma+\pi)|C\rangle\,,\quad \partial_\tau X^i(0,\sigma)|C\rangle=\partial_\tau X^i(0,\sigma+\pi)|C\rangle\,.
\end{equation}
The boundary conditions for the oscillators are then given as
\begin{equation}
(\alpha_n^\alpha+(-1)^n\bar{\alpha}_{-n}^\alpha)|C\rangle=0\,,\quad p^\alpha |C\rangle=0\,,
\end{equation}
\begin{equation}
(\alpha_n^i-(-1)^n\bar{\alpha}_{-n}^i|C\rangle=0\,,\quad (X^i-y^i) |C\rangle=0\,.
\end{equation}
We can combine the first column of the boundary conditions into
\begin{equation}
(\alpha^\mu_n +S^{\mu}_{~\nu} (-1)^n\bar{\alpha}^{\nu}_{-n})|C\rangle=0\,.
\end{equation}

We split the crosscap state into the matter part and the ghost part
\begin{equation}
|C\rangle=\mathcal{N}_C|C_X\rangle\otimes |C_{gh}\rangle\,,
\end{equation}
then we find
\begin{equation}
|C_X\rangle= \delta^{(d-p-1)}(q^i-y^i) \exp\left(\sum_{n=1}^\infty -\frac{(-1)^n}{n}\alpha^\mu_{-n}S_{\mu\nu}\bar{\alpha}^\nu_{-n}\right) |0\rangle\,,
\end{equation}
and
\begin{equation}
|C_{gh}\rangle=\exp\left(\sum_{n=1}^\infty (-1)^n(c_{-n}\bar{b}_{-n}-b_{-n}\bar{c}_{-n})\right) \frac{c_0+\bar{c}_0}{2}c_1\bar{c}_1|0\rangle\,.
\end{equation}
Note that the following identity holds
\begin{equation}
|C_X\rangle\otimes |C_{gh}\rangle= e^{i\pi L_0} |B_X\rangle\otimes |B_{gh}\rangle\,.
\end{equation}

Now, we shall review the boundary states in superstring theories. In superstring theories, the bosonic contributions and the $b,~c$ ghost part of the boundary states are the same as the boundary states in bosonic string theory. The difference between superstring and bosonic string theories originates from the worldsheet fermion fields and the $\beta,~\gamma$ ghosts. We shall first study the boundary states for the worldsheet fermions and then the boundary states for the $\beta,~\gamma$ system.

The boundary conditions for the worldsheet fermions are summarized as
\begin{equation}
(\psi(0,\sigma)^\mu-\eta S^\mu_{~\nu}\bar{\psi}^\nu(0,\sigma))|B_{\psi},\eta\rangle=0\,,
\end{equation}
where $\eta=\pm1.$ In terms of the oscillators, we have
\begin{equation}
(\psi_n^\mu-\eta S^{\mu}_{~\nu}\bar{\psi}_{-n}^\nu)|B_\psi,\eta\rangle=0\,.
\end{equation}
Note $n\in \Bbb{Z}$ in the Ramond sector and $n\in\Bbb{Z}+1/2$ in the Neveu-Schwarz sector. We now consider the $\beta,~\gamma$ ghost system. Requiring the BRST invariance again constraints the boundary states in the $\beta,~\gamma$ ghost system
\begin{equation}
(\gamma_n+\eta \bar{\gamma}_{-n})|B_{sgh},\eta\rangle=0\,,\quad (\beta_n+\eta\bar{\beta}_{-n})|B_{sgh},\eta\rangle=0\,.
\end{equation}

The boundary state in the matter sector that satisfies the above boundary condition in the NS sector is given as
\begin{equation}
|B_\psi,\eta\rangle_{NS}= \exp\left(\sum_{n=1/2}^\infty \eta \psi_{-n}^\mu S_{\mu\nu}\bar{\psi}_{-n}^\nu \right)|0\rangle\,.
\end{equation}
Similarly, the boundary state in the ghost sector is given as
\begin{equation}
|B_{sgh},\eta\rangle_{NS}=\exp\left(\eta \sum_{n=1/2}^\infty (\gamma_{-n}\bar{\beta}_{-n}-\beta_{-n}\bar{\gamma}_{-n})\right)|-1,-1\rangle\,,
\end{equation}
where we define
\begin{equation}
|-1,-1\rangle :=e^{-\phi}(0)e^{-\bar{\phi}}(0)|0\rangle\,,
\end{equation}
a vacuum with $(-1,-1)$ picture. Note that the boundary state has picture number $-2$ and ghost number $3$ as it should.

The boundary state in the Ramond sector is a little more complicated. Let us first study the zero mode sector of the fermions in the Ramond sector. We shall make a gauge choice such that the boundary state in the Ramond-Ramond sector is given as a linear combination of $(-1/2,-3/2)$ picture and $(-3/2,-1/2)$ picture.  The fermionic zero modes satisfy the following boundary conditions
\begin{equation}
(\psi_0^\mu-\eta S^{\mu}_{~\nu}\bar{\psi}_{0}^\nu)|B_\psi,\eta\rangle_{R}=0\,,
\end{equation}
\begin{equation}
(\gamma_0+\eta\bar{\gamma}_0)|B_{sgh},\eta\rangle_R=0\,,\quad (\beta_0+\eta\bar{\beta}_0)|B_{sgh},\eta\rangle_R=0\,.
\end{equation}
To study the zero mode sector of the boundary state, we shall consider a linear combination of the most zero mode state
\begin{align}
|B_{0},\eta\rangle_R=&\exp\left(\eta\gamma_0\bar{\beta}_0\right)\left(M^{\alpha}_{~\beta}e^{-\phi/2}\Sigma_\alpha e^{-3\bar{\phi}/2}\bar{\Sigma}^\beta+\tilde{M}_{\alpha}^{~\beta} e^{-\phi/2}\Sigma^\alpha e^{-3\bar{\phi}/2}\bar{\Sigma}_\beta \right)|0\rangle\nonumber\\
&+\exp(-\eta\bar{\gamma}_0\beta_0)\left(N_\alpha^{~\beta} e^{-3\phi/2} \Sigma^\alpha e^{-\bar{\phi}/2}\overline{\Sigma}_\beta+\tilde{N}^\alpha_{~\beta} e^{-3\phi/2}\Sigma_\alpha e^{-\bar{\phi}/2}\overline{\Sigma}^\beta\right)|0\rangle\,.
\end{align}
Then, the boundary conditions are translated into
\begin{align}
\left(M^\alpha_{~\beta} e^{-\phi/2} (\Gamma^\mu)_{\alpha\gamma}\Sigma^\gamma e^{-3\bar{\phi}/2}\bar{\Sigma}^\beta-\eta S^\mu_{~\nu} \tilde{M}_\alpha^{~\beta} e^{-\phi/2}\Sigma^\alpha (\Gamma^\nu)_{\beta\delta}e^{-3\bar{\phi}/2}\bar{\Sigma}^\delta\right)|0\rangle=0\,,
\end{align}
\begin{equation}
\left(-\eta S^\mu_{~\nu}M^\alpha_{~\beta} e^{-\phi/2}\Sigma_\alpha e^{-3\bar{\phi}/2} (\Gamma^\nu)^{\beta\gamma}\overline{\Sigma}_\gamma+\tilde{M}_{\alpha}^{~\beta} e^{-\phi/2}(\Gamma^\mu)^{\alpha\gamma} \Sigma_\gamma e^{-3\bar{\phi}/2}\overline{\Sigma}_\beta\right)|0\rangle=0\,,
\end{equation}
\begin{equation}
\left(N_{\alpha}^{~\beta} e^{-3\phi/2}(\Gamma^\mu)^{\alpha\gamma} \Sigma_\gamma e^{-\bar{\phi}/2}\overline{\Sigma}_\beta-\eta S^\mu_{~\nu} \tilde{N}^\alpha_{~\beta} e^{-3\phi/2}\Sigma_\alpha e^{-\bar{\phi}/2} (\Gamma^\nu)^{\beta\gamma}\overline{\Sigma}_\gamma\right)|0\rangle=0\,,
\end{equation}
and
\begin{equation}
\left(-\eta S^\mu_{~\nu}N_{\alpha}^{~\beta}e^{-3\phi/2}\Sigma^\alpha e^{-\bar{\phi}/2}(\Gamma^\nu)_{\beta\gamma}\overline{\Sigma}^\gamma+\tilde{N}^\alpha_{~\beta} e^{-3\phi/2} (\Gamma^\mu)_{\alpha\gamma} \Sigma^\gamma e^{-\bar{\phi}/2}\overline{\Sigma}^\beta\right)|0\rangle=0\,.
\end{equation}
As a result, we find
\begin{equation}
(\Gamma^\mu)_{\alpha\gamma} M^\gamma_{~\beta}=\eta S^\mu_{~\nu} \tilde{M}_\alpha^{~\gamma}(\Gamma^\nu)_{\gamma\beta}\,,
\end{equation}
\begin{equation}
\eta S^{\mu}_{~\nu} M^{\alpha}_{~\gamma} (\Gamma^\nu)^{\gamma\beta}=\tilde{M}_{\gamma}^{~\beta} (\Gamma^\mu)^{\alpha\gamma}\,,
\end{equation}
\begin{equation}
N_{\gamma}^{~\beta} (\Gamma^\mu)^{\alpha\gamma}=\eta S^{\mu}_{~\nu} \tilde{N}^{\alpha}_{~\gamma} (\Gamma^\nu)^{\gamma\beta}\,,
\end{equation}
and
\begin{equation}
\eta S^\mu_{~\nu} N_\alpha^{~\gamma} (\Gamma^\nu)_{\gamma\beta}=\tilde{N}^\gamma_{~\beta} (\Gamma^\mu)_{\alpha\gamma}\,.
\end{equation}

The solutions to the above equations are given as 
\begin{equation}
M^{\alpha}_{~\beta}= (\Gamma^{0\dots p})^{\alpha}_{~\beta}\,,\quad \tilde{M}_\alpha^{~\beta} =\eta (\Gamma^{0\dots p})_\alpha^{~\beta}\,,
\end{equation}
\begin{equation}
N_{\alpha}^{~\beta}=(\Gamma^{0\dots p})_{\alpha}^{~\beta}\,,\quad \tilde{N}^\alpha_{~\beta}=\eta(\Gamma^{0\dots p})^\alpha_{~\beta}\,.
\end{equation}
Note that the overall signs of the $M,\tilde{M},N,\tilde{N}$ are chosen in agreement with the GSO projection. 

Including the non-zero mode contributions, we find
\begin{align}
|B_{\psi,sgh},\eta\rangle_R=&\exp\left(\sum_{n=1}^\infty\eta \psi_{-n}^\mu S_{\mu\nu}\bar{\psi}^\nu_{-n}+ \sum_{n=1}^\infty\eta (\gamma_{-n}\bar{\beta}_{-n}-\beta_{-n}\bar{\gamma}_{-n})\right)|B_0,\eta\rangle\,.
\end{align}

The boundary states we constructed for superstring theories contain GSO odd states. To project out such unphysical modes, we shall GSO project the boundary states
\begin{equation}
|B\rangle_{NS}=\frac{1}{2} \left(|B,+\rangle_{NS}-|B,-\rangle_{NS}\right)\,,
\end{equation}
\begin{equation}
|B\rangle_{R}=\frac{1}{2} \left(|B,+\rangle_{R}+|B,-\rangle_{R}\right)\,.
\end{equation}

The crosscap state can be simply obtained by inserting appropriate phases $e^{i\pi n}$ 
\begin{equation}
|C_\psi,\eta\rangle_{NS}= \exp\left(\sum_{n=1/2}^\infty e^{i\pi n}\eta \psi_{-n}^\mu S_{\mu\nu}\bar{\psi}_{-n}^\nu \right)|0\rangle\,,
\end{equation}
\begin{equation}
|C_{sgh},\eta\rangle_{NS}=\exp\left(\eta \sum_{n=1/2}^\infty e^{i\pi n} (\gamma_{-n}\bar{\beta}_{-n}-\beta_{-n}\bar{\gamma}_{-n})\right)|-1,-1\rangle\,,
\end{equation}
and
\begin{align}
|C_{\psi,sgh},\eta\rangle_R=&\exp\left(\sum_{n=1}^\infty e^{i\pi n}\eta \psi_{-n}^\mu S_{\mu\nu}\bar{\psi}^\nu_{-n}+ \sum_{n=1}^\infty e^{i\pi n}\eta (\gamma_{-n}\bar{\beta}_{-n}-\beta_{-n}\bar{\gamma}_{-n})\right)|B_0,\eta\rangle_R\,.
\end{align}
Note that the zero mode sector of the boundary states is identical to both the cross cap and the disk. Hence we recycled $|B_0,\eta\rangle_R$ for convenience. 

We shall now fix the overall normalization of the boundary states by comparing various one-point functions on the disk to the linearized DBI and CS actions. As the massless vertex operator insertion is the simplest to consider, we will focus on the massless parts of the boundary states. 

In the NS sector, the boundary state has the following massless state contribution where $\Bbb{P}$ stands for the projection to $L_0+\bar{L}_0=0$ sector
\begin{align}
\Bbb{P}|B\rangle_{NS}=&\mathcal{N}_B (\Bbb{P}\delta^{(d-p-1)}(q^i-y^i))\frac{c_0+\bar{c}_0}{2} c_1\bar{c}_1\left(\psi_{-1/2}^\mu S_{\mu\nu}\bar{\psi}_{-1/2}^\nu+\gamma_{-1/2}\bar{\beta}_{-1/2}-\beta_{-1/2}\bar{\gamma}_{-1/2}\right)\nonumber\\
&\times e^{-\phi}(0)e^{-\bar{\phi}}(0)|0\rangle\\
=&-\mathcal{N}_B(\Bbb{P}\delta^{(d-p-1)}(q^i-y^i))  \left[\frac{\partial c+\bar{\partial}\bar{c}}{2}c\bar{c}\left( e^{-\phi} \psi^\mu S_{\mu\nu}e^{-\bar{\phi}}\bar{\psi}^\nu - (\eta\bar{\partial}\bar{\xi} e^{-2\bar{\phi}}-\partial\xi\bar{\eta}e^{-2\phi}) \right)\right]|0\rangle\,.
\end{align}
Here, $\Bbb{P}\delta^{(d-p-1)}(q^i-y^i))=\frac{1}{\text{Vol}(Y)}$ is the inverse volume of the free bosons along which Dirichlet boundary conditions are imposed. To fix the normalization of the boundary states, we shall use the zero-momentum graviton trace operator and the 10d dilaton operator which are written as, respectively,
\begin{equation}
    V_{g^{ab}}:=\frac{1}{4\pi} c\bar{c}\left( e^{-\phi}\psi^a e^{-\bar{\phi}}\bar{\psi}^b+\frac{\eta^{ab}}{2} (\eta\bar{\partial}\bar{\xi}e^{-2\bar{\phi}}-\partial \xi \bar{\eta}e^{-2\phi})\right)\,,
\end{equation}
and
\begin{equation}
    V_{\phi}=\frac{1}{8\pi} c\bar{c}\left(\eta_{ab}e^{-\phi}\psi^a e^{-\bar{\phi}}\bar{\psi}^b +(\eta\bar{\partial}\bar{\xi} e^{-2\bar{\phi}}-\partial \xi \bar{\eta}e^{-2\phi}) \right)\,.
\end{equation}
Note that the graviton trace operator corresponds to the variation of the metric in Einstein-frame \cite{Belopolsky:1995vi}. Typically, the ghost term is omitted as the polarization of the graviton vertex operator is taken to be traceless. But, for the polarization with a non-trivial trace, the inclusion of the ghost term is necessary to ensure that the string coupling is not varying \cite{Belopolsky:1995vi}. Also, the dilaton operator $V_{\phi}$ corresponds to the dilaton $g_s=e^{\phi}$ in Einstein-frame. 

Let us recall an identity,
\begin{equation}
    \{d_{gh}\}_{D^2}= g_s^{-1}\mathcal{N}_B\,,
\end{equation}
for the ghost-dilaton operator
\begin{equation}
    d_{gh}:=c\bar{c}(\eta\bar{\partial}\bar{\xi} e^{-2\bar{\phi}}-\partial \xi \bar{\eta}e^{-2\phi})\,.
\end{equation}
We then compute
\begin{equation}
    \{V_{g^{ab}}\}_{D^2}=\frac{\mathcal{N}_B}{8\pi g_s}(S^{ab}+\eta^{ab})\,,
\end{equation}
and
\begin{equation}
    \{V_{\phi}\}_{D^2}=\frac{\mathcal{N}_B}{8\pi g_s}(p-3)\,.
\end{equation}

We shall fix the normalization $\mathcal{N}_B$ by comparing to the DBI action and its variations with respect to dilaton and metric in Einstein-frame. Let us start with the DBI action in string-frame
\begin{equation}
    S_{DBI}=-T_p\int d^{p+1}x e^{-\phi} \sqrt{-G^{(st)}}\,.
\end{equation}
In Einstein-frame, we have
\begin{equation}
    S_{DBI}=-T_p\int d^{p+1}x e^{\phi(p-3)/4 }\sqrt{-G^{(E)}}\,.
\end{equation}
By varying the dilaton $\phi=\phi_0+\delta\phi$ and the metric $G^{(E)}=\eta+h,$ we obtain
\begin{equation}\label{eqn:dbi variation}
    S_{DBI}=-T_p\int d^{p+1} x e^{\phi(p-3)/4} \sqrt{-\eta} \left(1+\frac{p-3}{4}\delta\phi +\frac{1}{2}h^\mu_\mu\right)\,,
\end{equation}
where $\mu$ ranges from $0$ to $p.$ The variation \eqref{eqn:dbi variation} is reproduced if we fix the normalization to be
\begin{equation}
    \mathcal{N}_B=-2\pi T_p\,.
\end{equation}

% The corresponding graviton disk one-point function is given by
% \begin{align}
% \{ V_{g}\}_{D}:=& g_s^{-1} \langle V_{g}^{-1,-1} c_0^-\Bbb{P} |B\rangle_{NS}\,,\\
% =&-\frac{1}{2} \int d^{p+1}x g_s^{-1} g_c \mathcal{N}_B \epsilon_g^{\mu\nu}S_{\mu\nu}\,.
% \end{align}
% To reproduce the linearized DBI action, the normalization constant $\mathcal{N}_B$ is fixed to be
% \begin{equation}
% \mathcal{N}_B=-4\pi T_p\,,
% \end{equation}
% where $T_p$ is the Dp-brane tension of the DBI action
% \begin{equation}
% S_{DBI}=-T_p \int d^{p+1}x e^{-\phi}\sqrt{-g}=-T_p\int e^{-\phi}(1-\kappa h^\mu_\mu+\dots)\,.
% \end{equation}

In the R sector, the boundary state has the following massless state contribution
\begin{align}
\Bbb{P}|B\rangle_R=&\mathcal{N}_{B}'(\Bbb{P}\delta^{(d-p-1)}(q^i-y^i)) \frac{c_0+\bar{c}_0}{2}c_1\bar{c}_1 \nonumber\\
&\times\Biggl[cs(\gamma_0\bar{\beta}_0) (\Gamma^{0\dots p})^\alpha_{~\beta} e^{-\phi/2}\Sigma_\alpha e^{-3\bar{\phi}/2}\overline{\Sigma}^\beta+sh(\gamma_0\bar{\beta}_0) (\Gamma^{0\dots p})_\alpha^{~\beta} e^{-\phi/2}\Sigma^\alpha e^{-3\bar{\phi}/2}\overline{\Sigma}_\beta\nonumber\\
&\quad+cs (-\bar{\gamma}_0\beta_0) (\Gamma^{0\dots p})_{\alpha}^{~\beta} e^{-3\phi/2}\Sigma^\alpha e^{-\bar{\phi}/2}\overline{\Sigma}_\beta-sh(\bar{\gamma}_0\beta_0)(\Gamma^{0\dots p})^\alpha_{~\beta} e^{-3\phi/2}\Sigma_\alpha e^{-\bar{\phi}/2} \overline{\Sigma}^\beta\Biggr]|0\rangle\,,
\end{align}
where we used the shorthand notations $cs:=\cosh$ and $sh:=\sinh.$ Note that the boundary state above is in the mixed picture. Its contribution to the SFEOM to be discussed later will be in the picture $(-1/2,-1/2)$ as was prescribed in \cite{FarooghMoosavian:2019yke}. The prescription requires the computation of the following
\begin{equation}\label{eqn:diskBracket}
([]_D)_0:=\frac{1}{g_s}{\bf{P}} \tilde{\mathcal{G}} {\Bbb{P}}|B\rangle_R\,,
\end{equation}
where $\bf{P}$ is the projection operator onto $(-1/2,-1/2)$ picture and
\begin{equation}
\tilde{\mathcal{G}}:= \frac{1}{2}(\mathcal{X}_0+\bar{\mathcal{X}}_0)\,.
\end{equation}
Going to the $(-1/2,-1/2)$ picture will also make the appearance of the GSO even states manifest. 

We shall first apply $\mathcal{X}_0$ to the (-3/2,-1/2) picture boundary state
\begin{align}\label{eqn:RR boundary res1}
\mathcal{X}_0\Bbb{P} |B\rangle_R^{-\frac{3}{2},-\frac{1}{2}}=&-i\frac{\sqrt{\alpha'}\mathcal{N}'_B}{2}\partial^q_A (\Bbb{P}\delta^{(d-p-1)}) \frac{c_0+\bar{c}_0}{2}c_1\bar{c}_1 cs (\bar{\gamma}_0\beta_0) (\Gamma^{0\dots p} \Gamma^A)^{\alpha\beta} e^{-\phi/2} \Sigma_\alpha e^{-\bar{\phi}/2}\overline{\Sigma}_\beta |0\rangle\nonumber\\
&+i\frac{\sqrt{\alpha'}\mathcal{N}_B'}{2}\partial_A^q (\Bbb{P}\delta^{(d-p-1)})\frac{c_0+\bar{c}_0}{2}c_1\bar{c}_1sh(\bar{\gamma}_0\beta_0) (\Gamma^{0\dots p}\Gamma^A)_{\alpha\beta} e^{-\phi/2}\Sigma^\alpha e^{-\bar{\phi}/2} \overline{\Sigma}^\beta|0\rangle
\nonumber\\
&-\frac{\mathcal{N}_B'}{2}(\Bbb{P}\delta^{(d-p-1)})\eta_{-1}c_1\bar{c}_1cs(\bar{\gamma}_0\beta_0) (\Gamma^{0\dots p})_{\alpha}^{~\beta} e^{\phi/2}\Sigma^\alpha e^{-\bar{\phi}/2}\overline{\Sigma}_\beta|0\rangle\nonumber\\
&+\frac{\mathcal{N}_B'}{2}(\Bbb{P}\delta^{(d-p-1)})\eta_{-1}c_1\bar{c}_1sh(\bar{\gamma}_0\beta_0) (\Gamma^{0\dots p})^{\alpha}_{~\beta} e^{\phi/2}\Sigma_\alpha e^{-\bar{\phi}/2}\overline{\Sigma}^\beta|0\rangle\,.
\end{align}
The first two lines vanish since $\Bbb{P}\delta^{(d-p-1)}$ is a constant, but we keep them for the formality of the expression. By using the identities \cite{deLacroix:2017lif}
\begin{equation}
\beta_n e^{p\phi}|0\rangle=0\quad\text{for}\quad n\geq -p-\frac{1}{2}\,, \quad \bar{\beta}_n e^{p\bar{\phi}}|0\rangle=0\quad\text{for}\quad n\geq -p-\frac{1}{2}\,, \quad 
\end{equation}
we simplify \eqref{eqn:RR boundary res1} into
\begin{align}
\mathcal{X}_0\Bbb{P}|B\rangle^{-\frac{3}{2},-\frac{1}{2}}=&-i\frac{\sqrt{\alpha'}\mathcal{N}'_B}{2}\partial^q_A (\Bbb{P}\delta^{(d-p-1)}) \frac{c_0+\bar{c}_0}{2}c_1\bar{c}_1  (\Gamma^{0\dots p} \Gamma^A)^{\alpha\beta} e^{-\phi/2} \Sigma_\alpha e^{-\bar{\phi}/2}\overline{\Sigma}_\beta |0\rangle\nonumber\\
&-\frac{\mathcal{N}_B'}{2}(\Bbb{P}\delta^{(d-p-1)})\eta_{-1}c_1\bar{c}_1 (\Gamma^{0\dots p})_{\alpha}^{~\beta} e^{\phi/2}\Sigma^\alpha e^{-\bar{\phi}/2}\overline{\Sigma}_\beta|0\rangle\,.
\end{align}
Similarly, we find
\begin{align}
\overline{\mathcal{X}}_0 \Bbb{P}|B\rangle_R^{-\frac{1}{2},-\frac{3}{2}}=&-i\frac{\sqrt{\alpha'}\mathcal{N}'_B}{2}\partial^q_A (\Bbb{P}\delta^{(d-p-1)}) \frac{c_0+\bar{c}_0}{2}c_1\bar{c}_1 (\Gamma^{0\dots p} \Gamma^A)^{\alpha\beta} e^{-\phi/2} \Sigma_\alpha e^{-\bar{\phi}/2}\overline{\Sigma}_\beta |0\rangle\nonumber\\
&-\frac{\mathcal{N}_B'}{2}(\Bbb{P}\delta^{(d-p-1)}) \bar{\eta}_{-1}c_1\bar{c}_1 (\Gamma^{0\dots p})^{\alpha}_{~\beta} e^{-\phi/2}\Sigma_\alpha e^{\bar{\phi}/2} \overline{\Sigma}^\beta|0\rangle\nonumber\\
\end{align}
As a result, we find
\begin{align}
g_s([]_D)_0=&-i\frac{\sqrt{\alpha'}\mathcal{N}'_B}{2}\partial^q_A (\Bbb{P}\delta^{(d-p-1)}) \frac{c_0+\bar{c}_0}{2}c_1\bar{c}_1  (\Gamma^{0\dots p} \Gamma^A)^{\alpha\beta} e^{-\phi/2} \Sigma_\alpha e^{-\bar{\phi}/2}\overline{\Sigma}_\beta |0\rangle\nonumber\\
&-\frac{\mathcal{N}_B'}{4}(\Bbb{P}\delta^{(d-p-1)})\eta_{-1}c_1\bar{c}_1(\Gamma^{0\dots p})_{\alpha}^{~\beta} e^{\phi/2}\Sigma^\alpha e^{-\bar{\phi}/2}\overline{\Sigma}_\beta|0\rangle\nonumber\\
&-\frac{\mathcal{N}_B'}{4}(\Bbb{P}\delta^{(d-p-1)}) \bar{\eta}_{-1}c_1\bar{c}_1 (\Gamma^{0\dots p})^{\alpha}_{~\beta} e^{-\phi/2}\Sigma_\alpha e^{\bar{\phi}/2} \overline{\Sigma}^\beta|0\rangle\,.
\end{align}
For later use, we can insert factors of $P^{10}_\pm:=(1\pm \Gamma^{(10)})/2$ without affecting the answer
\begin{align}
g_s([]_D)_0=&-i\frac{\sqrt{\alpha'}\mathcal{N}'_B}{2}\partial^q_A (\Bbb{P}\delta^{(d-p-1)}) \frac{c_0+\bar{c}_0}{2}c_1\bar{c}_1  (P_+^{10}\Gamma^{0\dots p} \Gamma^A)^{\alpha\beta} e^{-\phi/2} \Sigma_\alpha e^{-\bar{\phi}/2}\overline{\Sigma}_\beta |0\rangle\nonumber\\
&-\frac{\mathcal{N}_B'}{4}(\Bbb{P}\delta^{(d-p-1)})\eta_{-1}c_1\bar{c}_1(P_-^{10}\Gamma^{0\dots p})_{\alpha}^{~\beta} e^{\phi/2}\Sigma^\alpha e^{-\bar{\phi}/2}\overline{\Sigma}_\beta|0\rangle\nonumber\\
&-\frac{\mathcal{N}_B'}{4}(\Bbb{P}\delta^{(d-p-1)}) \bar{\eta}_{-1}c_1\bar{c}_1 (P_+^{10}\Gamma^{0\dots p})^{\alpha}_{~\beta} e^{-\phi/2}\Sigma_\alpha e^{\bar{\phi}/2} \overline{\Sigma}^\beta|0\rangle\,.
\end{align}

To fix the normalization, we shall compute the RR one-point function on the disk
\begin{align}
\{ V_{RR}\}_{D}:=&g_s^{-1} \langle V_{R,R}^{-\frac{3}{2},-\frac{3}{2}} c_0^-{\bf{P}}\tilde{\mathcal{G}} \Bbb{P}|B\rangle_R\\
=&-\int d^{p+1} x \frac{1}{16\sqrt{2}\pi} \mathcal{N}_B' \sum_{k}\frac{1}{(2k)!} C_{1\dots 2k} \text{Tr}\left( \Gamma^{1\dots 2k}\Gamma^{0\dots p}\right)\,,
\end{align}
which gives, for $p+1$ form field, 
\begin{equation}
\{V_{RR}\}_{D^2} =-\frac{1}{\pi}\int d^{p+1}x  \mathcal{N}_B' C_{0\dots p}\,.
\end{equation} 
As a result the normalization constant $\mathcal{N}_B'$ is fixed to be
\begin{equation}
\mathcal{N}_B'=\pm\sqrt{2}\pi T_p\,,
\end{equation}
where different sign choices lead to D-branes or anti-D-branes.
\subsection{Toroidal compactifications}\label{subsection:toruscft}
In this section, we study the worldsheet CFT convention for the toroidal orientifold compactification we discussed in \S\ref{subsection:explicitExample}. We will first study the original NSNS background around which we will perturb by including fluxes. The original background can be, therefore, described by the conventional RNS formalism. Because the toroidal manifold has no non-trivial curvature, most of the worldsheet conventions for the flat spacetime carry over. There are several important differences, however. First, the momentum along the compact directions is quantized. Second, strings can wind, which gives rise to the winding states. Third, there are moduli degrees of freedom, some of which will be correlated with the axio-dilaton.  We shall carefully deal with those issues in this section.

Even though we already laid out the conventions for the supergravity background, we will reiterate a few important ingredients here. In the NSNS background, the targe spacetime is $\Bbb{R}^{1,3}\times T^6,$ whose metric is given as
\begin{equation}
ds^2=G_{AB}dX^A dX^B=g_{\mu\nu} dX^\mu dX^\nu +g_{ab} dY^a dY^b\,,
\end{equation} 
where the Greek indices range from $0$ to $3,$ and the Latin indices range from $1$ to $6.$ $Y^i$ are the coordinates along the compact directions, whose fundamental domain is $[0,1].$  We choose the metric for the non-compact directions to be 
\begin{equation}
g_{\mu\nu}=\eta_{\mu\nu}=\text{diag}(-1,1,1,1)\,.
\end{equation}
As in the previous section, we decompose $T^6$ into $T^2_1\times T^2_2\times T^2_3,$ and denote the coordinates of $T^2_i$ by $Y^{2i-1}$ and $Y^{2i}.$ We define the metric of $T_i^2$ to be
\begin{equation}
ds^2= \frac{\im t_i}{\im u_i} \left( (dY^{2i-1})^2+2\re u_i dY^{2i-1} dY^{2i}+|u_i|^2 (dY^{2i})^2\right)\,,
\end{equation}
and the complex coordinates 
\begin{equation}
Z^i=Y^{2i-1}+u_i Y^{2i}\,.
\end{equation}
For simplicity, we shall restrict $u_i$ to $u_i=i\im u_i,$ whenever needed. Note also that we chose a somewhat unconventional orientation following the convention of \cite{Kachru:2002he}, in which the volume form is given as
\begin{equation}
d \text{Vol}_{T_i^2}=-\im t_i dY^{2i-1}\wedge dY^{2i}=\frac{\im t_i}{2i\im u_i}dZ^i\wedge d\overline{Z}^i\,.
\end{equation}

We write the matter part of the action of the worldsheet CFT in the flat background
\begin{equation}
S_{matter}=\frac{1}{2\pi} \int d^2z G_{AB} \left(\frac{1}{\alpha'}\partial X^A \bar{\partial}X^B+\frac{1}{2}\psi^A \bar{\partial} \psi^B+\frac{1}{2} \bar{\psi}^A \partial \bar{\psi}^B\right)\,,
\end{equation}
and the ghost part of the action as
\begin{equation}
S_{ghost}=\frac{1}{2\pi} \int d^2z [b\bar{\partial}c+\bar{b}\partial \bar{c}+\beta\bar{\partial}\gamma+\bar{\beta}\partial \bar{\gamma}]\,.
\end{equation}
For simplicity, we shall set the real part of the axio-dilaton, complex structure moduli $u_i,$ and K\"{a}hler moduli $t_i$ to zero, and introduce an orthonormal frame 
\begin{equation}
G_{AB}dX^A dX^B= \eta_{AB} d\tilde{X}^A d\tilde{X}^B\,,
\end{equation} 
where we defined the vielbein
\begin{equation}
\tilde{X}^A:= e^{A}_B X^B\,,
\end{equation}
such that
\begin{equation}
G_{AB}=\eta_{CD} e^C_A e^D_B\,.
\end{equation}
Similarly, we introduce new worldsheet fermions in the orthonormal frame
\begin{equation}
\lambda^A :=e^A_B \psi^B\,,\quad \bar{\lambda}^A:= e^A_B \bar{\psi}^B\,.
\end{equation}

The OPEs of the matter fields are given as
\begin{align}
X^A (z,\bar{z})X^B (0,0)\sim& -\frac{\alpha'}{2}G^{AB}\ln |z|^2\,,\quad\psi^A(z)\psi^B(0)\sim \frac{G^{AB}}{z}\,,\quad\bar{\psi}^A (z)\bar{\psi}^B(0)\simeq \frac{G^{AB}}{\bar{z}}\,.
\end{align}
The OPEs involving the ghost fields are the same as that of the worldsheet theory, whose target spacetime is $\Bbb{R}^{1,9}.$ The OPEs involving the matter fields in the orthonormal frame are given as
\begin{align}
\tilde{X}^A (z,\bar{z})\tilde{X}^B (0,0)\sim& -\frac{\alpha'}{2}\eta^{AB}\ln |z|^2\,,\quad\lambda^A(z)\lambda^B(0)\sim \frac{\eta^{AB}}{z}\,,\quad\bar{\lambda}^A (z)\bar{\lambda}^B(0)\simeq \frac{\eta^{AB}}{\bar{z}}\,.
\end{align}
We write the mode expansions of the matter fields
\begin{equation}
\partial \tilde{X}^A=-i\sqrt{\frac{\alpha'}{2}} \sum_{m=-\infty}^\infty \frac{\tilde{\alpha}^A}{z^{m+1}}\,,\quad \bar{\partial}\tilde{X}^A=-i\sqrt{\frac{\alpha'}{2}} \sum_{m=-\infty}^\infty \frac{\bar{\tilde{\alpha}}^A}{\bar{z}^{m+1}}\,,
\end{equation}
\begin{equation}
\lambda^A=\sum_{n\in\Bbb{Z}+\nu} \frac{\lambda^A_n}{z^{n+\frac{1}{2}}}\,,\quad \bar{\lambda}^A=\sum_{n\in\Bbb{Z}+\nu}\frac{\bar{\lambda}^A_n}{\bar{z}^{n+\frac{1}{2}}}\,.
\end{equation}
We then have the following (anti)-commutation relations
\begin{equation}
[\tilde{\alpha}^A_m,\tilde{\alpha}^B_n]=[\bar{\tilde{\alpha}}^A_m,\bar{\tilde{\alpha}}^B_n]=m\eta^{AB} \delta_{m,-n}\,,
\end{equation}
\begin{equation}
\{\lambda^A_m,\lambda^B_n\}=\{\bar{\lambda}^A_m,\bar{\lambda}^B_n\}=\eta^{AB}\delta_{m,-n}\,.
\end{equation}

Now, let us study the spin fields. Identical to the previous section, we denote the 16 component spinors by $\Sigma^{\alpha^{(10)}}$ and $\Sigma_{\alpha^{(10)}}.$ Note that the chirality of the 4d spinor determines the chirality of the 6 dimensional spinor; we need not specify the chirality of the 6d spinor unless necessary. We shall assume that $e^{-\phi/2}\Sigma_{\alpha^{(10)}}$ and $e^{-3\phi/2}\Sigma^{\alpha^{(10)}}$ and their anti-holomorphic counterparts are GSO even. Upon decomposing the 10d spinor into the 4d spinor, we find
\begin{equation}
\Sigma_{\alpha^{10}}=S_{\alpha}\otimes \eta_{\alpha^{(6)}}+S_{\dot{\alpha}} \otimes \eta_{\alpha^{(6)}}\,,
\end{equation}
and
\begin{equation}
\Sigma^{\alpha^{10}}=S^{\alpha}\otimes \eta^{\alpha^{(6)}}+S^{\dot{\alpha}} \otimes \eta^{\alpha^{(6)}}\,,
\end{equation}
where $S$ is the 4d spinor, and $\eta$ is the 6d spinor. For the details of the spinors, see \S\ref{app:spin}. 

We conclude this section with the conventions for the vertex operators corresponding to unit quantized NSNS flux. To illustrate the convention, we choose 
\begin{equation}
H= H_{123} dY^1\wedge dY^2\wedge dY^3\,.
\end{equation}
This NSNS threeform flux should be properly quantized, as required. The aforementioned $H$ field can be reproduced by the following $B_2$ field
\begin{equation}
B_2=H_{123}Y^1dY^2\wedge dY^3\,.
\end{equation}
or, equivalently, its fully anti-symmetrized form
\begin{equation}
B_2'=\frac{1}{2}B_{AB}dX^A\wedge dX^B=\frac{1}{3} H_{123} \left( Y^1 dY^2 \wedge dY^3-Y^2 dY^1 \wedge dY^3+Y^3dY^1\wedge dY^2\right)\,,
\end{equation}
Note that $B_2'$ and $B_2$ are related by a gauge transformation
\begin{equation}
B_2'=B_2+d\chi\,,
\end{equation}
where
\begin{equation}
\chi=\frac{1}{3}Y^1Y^3dY^2-\frac{1}{3} Y^1Y^2dY^3\,.
\end{equation}
The corresponding NSNS vertex operator in $(-1,-1)$ picture is
\begin{align}
\mathcal{O}_{NSNS}=&\frac{1}{4\pi}c\bar{c} B_{AB} e^{-\phi} \psi^A e^{-\bar{\phi}}\bar{\psi}^B\,,\\
=& \frac{1}{4\pi}c\bar{c} H_{123} Y^1 e^{-\phi}\psi^2e^{-\bar{\phi}}\bar{\psi}^3\,.\label{eq:H3vertOp}
\end{align}
Or, equivalently, one can use the fully anti-symmetrized version. 

Note that \eqref{eq:H3vertOp} is written in the local chart of the torus with coordinates $Y^i$ and $B$-field profile as written is not periodic under $Y^i\rightarrow Y^i+1$. We expect that effects of such local charts to be absent when computing physical observables, which in particular implies that only $H_{123}$ will appear in the final results. For the physical quantities involving only the momentum modes, this is evidenced by the fact that constant $B$-field shift does not affect the momentum mode spectrum. Moreover, under $Y^i\rightarrow Y^i+1$, $B$-field shifts by an integer amount which is one of the duality actions for torus CFTs in constant $B$-field background. Therefore, even though SFT action may be subject to change, the resulting SFT action should still produce the same physics as the original SFT action.

In any case, since K\"{a}hler moduli remain as moduli in our background, the use of local chart $Y^i$ should pose no problem as we take the large volume limit.\footnote{We thank Xi Yin for his comments on the local charts and the large volume limit. It will be interesting to see how the computations involving winding states are (in)dependent on the local charts.} The computation involving $\mathcal{O}_{NSNS}$ is evidently simpler in the $B_2$ gauge. $\mathcal{O}_{NSNS}$ is even under the orientifold action and thus survives as an operator of the orientifold theory.

\section{String field theory for flux compactifications}\label{sec:SFT}
In this section, we demonstrate how SFT provides a systematic framework to study the explicit example discussed in section \ref{subsection:explicitExample}, where the only expansion parameter is effectively $g_s$. We begin by briefly reviewing the result of the recently constructed BV master action for open-closed-unoriented super-SFT in \cite{FarooghMoosavian:2019yke}.

\subsection{Review of SFT}\label{subsection:sft review}
We will mostly follow \cite{deLacroix:2017lif,FarooghMoosavian:2019yke}, where the full details of the discussion can be found. The main idea of SFT is that the consideration of generic off-shell states (as opposed to considering the usual $Q_B$-closed states only) allows one to write a gauge-invariant action from which off-shell computations can be performed. The SFEOM is obtained by varying the action, and the action can be expanded around a solution of SFEOM to provide the usual Feynman rules.

We start with a worldsheet BCFT for a pure NSNS IIB background in the context of RNS formalism. In our specific example of toroidal orientifold, this will be the usual free boson and fermion CFT (with proper orbifolding) together with $bc$ and $\beta\gamma$ ghosts. Typically, in the on-shell worldsheet computations, only $Q_B$-closed states of the worldsheet BCFT are of interest. In contrast, generic states of the worldsheet BCFT play important roles in the construction of SFT.

We define $H_{m,n}$ to be the set of GSO-even bulk CFT states $|s\rangle$ of picture number $(m,n)$ satisfying\footnote{We will be working in the small Hilbert space where states are annihilated by $\eta_0$ and $\tilde{\eta}_0$.}
\begin{equation}
b_0^-|s\rangle=L_0^-|s\rangle=0\,,
\end{equation}
where
\begin{equation}
b_0^\pm:=b_0\pm \bar{b}_0\,,
\end{equation}
and
\begin{equation}
L_0^\pm:=L_0\pm\bar{L}_0\,.
\end{equation}
Similarly, we define $H_n$ to be the set of boundary CFT states with picture number $n$. The subspaces of these Hilbert spaces required for the construction of a BV master action for SFT are given by
\begin{align}
H^c:=& H_{-1,-1}\oplus H_{-1/2,-1}\oplus H_{-1,-1/2}\oplus H_{-1/2,-1/2}\,,\\
\tilde{H}^c:=& H_{-1,-1}\oplus H_{-3/2,-1}\oplus H_{-1,-3/2}\oplus H_{-3/2,-3/2}\,,\\
H^o:=& H_{-1}\oplus H_{-1/2}\,,\\
\tilde{H}^o:=&H_{-1}\oplus H_{-3/2}\,,
\end{align}
We denote by $\Psi^c,\tilde{\Psi}^c,\Psi^o,$ and $\tilde{\Psi}^o$ general elements of $H^c,\tilde{H}^c,H^o,$ and $\tilde{H}^o$ respectively. $\Psi^c$ can be expanded in a basis of $H^c$ with coefficients, and these coefficients are `string fields.' For example, if a basis of $H^c$ is given by states $|v^c_i\rangle$, then we have the expansion $|\Psi\rangle=\sum_i w_c^i|v^c_i\rangle$, where $w_c^i$ are string fields. They in particular include the gravi-dilaton and $B-$ `field' $e_{\mu\nu}(k)$
\begin{equation}
\Psi^c\sim\int d^dke_{\mu\nu}(k)c\bar{c}e^{-\phi}\Psi^\mu e^{-\bar{\phi}}\bar{\Psi}^\nu e^{ik\cdot X}+....
\end{equation}
On-shell condition requires $k^\mu e_{\mu\nu}(k)=k^\nu e_{\mu\nu}(k)=k^2=0$, but such a condition is not required to be satisfied for generic string fields that are off-shell, and we consider general $e_{\mu\nu}(k)$ for example.

Since the states carrying off-shell string fields still belong to the worldsheet BCFT Hilbert space, we can compute their worldsheet correlators, which are then functionals of string fields such as $e_{\mu\nu}(k)$. Unlike the on-shell correlators, though, such off-shell correlators are not in general invariant under the change of local coordinate charts around the vertex operator insertions and locations of PCOs. Nonetheless, it has been shown \cite{Hata:1993gf} that such off-shell effects can be absorbed into string field redefinitions, and any on-shell observables are well-defined.

The off-shell correlators integrated over some part of the worldsheet moduli space (with a specific choice of local charts and PCO locations) provide the Feynman vertices of the quantum BV master action for SFT. The three choices here, which part of the worldsheet moduli space is integrated, what the local chart choices are as a function of the worldsheet moduli, and where the PCOs are placed as a function of the worldsheet moduli, are constrained by the requirement of gauge invariance of the action. Consistent choices of such string vertices to all orders in string perturbation theory have been explicitly found \cite{Zwiebach:1990nh,Zwiebach:1990qj,Zwiebach:1992bw,Headrick:2018ncs,Headrick:2018dlw,Costello:2019fuh,Cho:2019anu}, while one can always work with simpler choices of string vertices at low orders in string perturbation theory.

For superstrings, PCO plays an essential role in the construction of a consistent SFT action. We define $\mathcal{G}$ by
\begin{equation}
\mathcal{G}|s^o\rangle=\begin{cases}
|s^o\rangle& \text{if }|s^o\rangle\in H_{-1}\\
\frac{1}{2}(\mathcal{X}_0+\overline{\mathcal{X}}_0)|s^o\rangle& \text{if }|s^o\rangle\in H_{-3/2}
\end{cases}\,,
\end{equation}
\begin{equation}
\mathcal{G}|s^c\rangle=\begin{cases}
|s^c\rangle&\text{if }|s^c\rangle \in H_{-1,-1}\\
\mathcal{X}_0 |s^c\rangle &\text{if }|s^c\rangle\in H_{-3/2,-1}\\
\overline{\mathcal{X}}_0|s^c\rangle&\text{if }|s^c\rangle\in H_{-1,-3/2}\\
\mathcal{X}_0\overline{\mathcal{X}}_0|s^c\rangle&\text{if }|s^c\rangle\in H_{-3/2,-3/2}
\end{cases}\,,
\end{equation}
where $\mathcal{X}_0$ and $\overline{\mathcal{X}}_0$ are zero modes of left-moving and right-moving PCOs.

We are ready to introduce the 1PI effective action of open-closed-unoriented SFT. In a widely used convention \cite{FarooghMoosavian:2019yke}, the action is given by
\begin{align}\label{eqn:SFT1PIaction0}
S_{1PI}=&-\frac{1}{2g_s^2} \langle\tilde{\Psi}^c|c_0^-Q_B \mathcal{G}|\tilde{\Psi}^c\rangle+\frac{1}{g_s^2} \langle \tilde{\Psi}^c|c_0^-Q_B|\Psi^c\rangle -\frac{1}{2g_s} \langle \tilde{\Psi}^o|Q_B\mathcal{G}|\tilde{\Psi}^o\rangle+\frac{1}{g_s}\langle \tilde{\Psi}^o|Q_B|\Psi^o\rangle\nonumber\\
&+\frac{Z_{D^2}}{g_s}+\frac{Z_{\Bbb{RP}^2}}{g_s}+\{\tilde{\Psi}^c\}_D+\{\tilde{\Psi}^c\}_{\Bbb{RP}^2}+\sum_{N\geq0,M\geq0}\frac{1}{N!M!}\{(\Psi^c)^N;(\Psi^o)^M\}\,,
\end{align}
where $c_0^-=(c_0-\bar{c}_0)/2$. This convention is convenient for proving general results such as gauge invariance of the action and the main identity. In this work, we shall instead use the string field theory action in the following convention to make the comparison with supergravity action more convenient
\begin{align}\label{eqn:SFT1PIaction}
S_{1PI}=&\frac{4}{\alpha' g_c^2}\left[-\frac{1}{2} \langle\tilde{\Psi}^c|c_0^-Q_B \mathcal{G}|\tilde{\Psi}^c\rangle+ \langle \tilde{\Psi}^c|c_0^-Q_B|\Psi^c\rangle \right]-\frac{1}{2g_s} \langle \tilde{\Psi}^o|Q_B\mathcal{G}|\tilde{\Psi}^o\rangle+\frac{1}{g_s}\langle \tilde{\Psi}^o|Q_B|\Psi^o\rangle\nonumber\\
&+\frac{Z_{D^2}}{g_s}+\frac{Z_{\Bbb{RP}^2}}{g_s}+\{\tilde{\Psi}^c\}_D+\{\tilde{\Psi}^c\}_{\Bbb{RP}^2}+\sum_{N\geq0,M\geq0}\frac{1}{N!M!}\{(\Psi^c)^N;(\Psi^o)^M\}\,,
\end{align}
such that string field for a canonically normalized closed string state in the CFT has the canonical kinetic term. For example, string field of the form
\begin{equation}
    |\Psi\rangle= g_c\int\frac{d^{10}k}{(2\pi)^{10}} h_{ab}(k) c\bar{c} e^{-\phi} \psi^a e^{-\bar{\phi}}\bar{\psi}^b e^{ik\cdot X}|0\rangle 
\end{equation}
has the kinetic term
\begin{equation}
    \frac{2}{\alpha' g_c^2}\langle \Psi|c_0^-Q_B|\Psi\rangle=-\frac{1}{2}\int \frac{d^{10}k}{(2\pi)^{10}} h_{ab}(-k)k^2 h^{ab}(k)\,.
\end{equation}

The second line of \eqref{eqn:SFT1PIaction} consists of disk and $\Bbb{RP}^2$ partition functions together with the interactions, where we separated out the closed string 1-point vertices on the disk and $\Bbb{RP}^2$. These are defined by
\begin{equation}
\{\tilde{A}^c\}_D:= g_s^{-1}\langle \tilde{A}^c|c_0^-\hat{\mathcal{G}}e^{-\beta (L_0+\bar{L}_0)}|B\rangle\,,
\end{equation}
where $\hat{\mathcal{G}}$ is 1 when acting on $H_{-1,-1}$ and $\frac{1}{2}\left( \mathcal{X}_0+\overline{\mathcal{X}}_0 \right)$ when acting on $H_{-3/2,-3/2}$, $|B\rangle$ is the matter-ghost boundary state of BCFT, and $\beta$ is a positive number related to the choice of the local chart around the bulk operator insertion on the disk $D$. We similarly defined $\{\tilde{A}^c\}_{\Bbb{RP}^2}$. Note that when ${\tilde A}^c$ is the graviton or massless RR vertex operator, this definition for $\{\cdot\}_D$ agrees with the same notation we introduced in section \ref{sec:bdrystate}. 

Even though the disk and $\Bbb{RP}^2$ partition functions do not depend on the string fields and thus contribute to the action only by constant terms, they are required by the background independence of string field theory as explained for example in \cite{Zwiebach:1997fe}.

$\{(\Psi^c)^N;(\Psi^o)^M\}$ is obtained by summing over all 1PI Feynman diagrams with $N$ closed string fields and $M$ open string fields. Again, disk and $\Bbb{RP}^2$ 1-point closed string vertices defined above are excluded in the definition of $\{\Psi^c;\}$. $\{(\Psi^c)^N;(\Psi^o)^M\}$, in particular, includes summing over all string vertices with $N$ closed string fields and $M$ open string fields of all possible worldsheet topologies and also lower point string vertices connected together by string field propagators or higher point string vertices connected to itself by propagators. Such 1PI diagrams are well-defined even in the presence of tadpoles. At any fixed finite order in string perturbation theory, only a finite number of 1PI diagrams will contribute to $\{(\Psi^c)^N;(\Psi^o)^M\}$. It can be shown that as long as the original choice of string vertices were consistent in the sense of gauge invariance, $S_{1PI}$ satisfies the classical BV master equation.

In this work, the only relevant closed string interaction vertex at genus zero is the sphere cubic vertex. We choose the normalization such that insertion of three closed string states are computed in the usual convention of \cite{Polchinski:1988jq}. For generic string vertices, appropriate normalizations should be introduced such that the factorization relations hold.\footnote{We thank Ashoke Sen for explaining to us the importance of correct normalizations for string vertices.} 

As already emphasized, $S_{1PI}$ is a functional of string fields which are coefficient fields in the expansion of $\Psi^c,\tilde{\Psi}^c,\Psi^o,$ and $\tilde{\Psi}^o$ into basis elements. There are gauge transformations leaving $S_{1PI}$ invariant, which are often partially fixed using the Seigel gauge.

We are interested in SFEOM, which can be obtained by varying $S_{1PI}$. Therefore, It is natural to introduce `string brackets' obtained by varying 1PI interaction terms in $S_{1PI}$. For the disk 1-point vertex, we introduce
\begin{equation}
[]_D:=g_s^{-1}{\bf{P}}\hat{\mathcal{G}}e^{-\beta(L_0+\bar{L}_0)}|B\rangle\,,
\end{equation}
where $\bf{P}$ is a projection operator onto $H^c$. We similarly define the string bracket $[]_{\Bbb{RP}^2}$ for $\Bbb{RP}^2$.

For $\{(\Psi^c)^N;(\Psi^o)^M\}$, we define the corresponding string brackets $[A_1^c... A_N^c;A_1^o...A_M^o]^c\in\tilde{H}^c$ and $[A_1^c... A_N^c;A_1^o...A_M^o]^o\in\tilde{H}^o$ via
\begin{equation}\label{eqn:stringbracketDEF}
\langle A_0^c|c_0^-|[A_1^c... A_N^c;A_1^o...A_M^o]^c\rangle=\{A_o^cA_1^c... A_N^c;A_1^o...A_M^o \},~~~\forall|A^c_0\rangle\in H^c,
\end{equation}
and
\begin{equation}
\langle A_0^o|c_0^-|[A_1^c... A_N^c;A_1^o...A_M^o]^o\rangle=\{A_1^c... A_N^c;A_0^oA_1^o...A_M^o \},~~~\forall|A^o_0\rangle\in H^o.
\end{equation}

By varying $S_{1PI}$, we obtain SFEOM given by
\begin{align}\label{eqn:eom01}
&\frac{4}{\alpha' g_c^2}Q_B\left( |\Psi^c\rangle-\mathcal{G}|\tilde{\Psi}^c\rangle\right) +|[]_D\rangle+|[]_{\Bbb{RP}^2}\rangle=0\,,\\\label{eqn:eom02}
&\frac{4}{\alpha' g_c^2}Q_B|\tilde{\Psi}^c\rangle +\sum_{N=1}^\infty \sum_{M=0}^\infty \frac{1}{(N-1)!M!}\left[ (\Psi^c)^{N-1};(\Psi^o)^M\right]^c=0\,,\\\label{eqn:eom03}
& Q_B\left( |\Psi^o\rangle-\mathcal{G}|\tilde{\Psi}^o\rangle\right)=0\,,\\\label{eqn:eom04}
&Q_B|\tilde{\Psi}^o\rangle+g_s\sum_{N=0}^\infty \sum_{M=0}^\infty \frac{1}{N!(M-1)!}\left[(\Psi^c)^N;(\Psi^o)^{M-1}\right]^o=0\,.
\end{align}
One can multiply $\mathcal{G}$ to the second and the fourth equations and combine them with the first and the third equations to obtain
\begin{align}\label{eqn:eom1}
&\frac{4}{\alpha' g_c^2}Q_B|\Psi^c\rangle +\sum_{N=1}^\infty \sum_{M=0}^\infty \frac{1}{(N-1)!M!}\mathcal{G}\left[ (\Psi^c)^{N-1};(\Psi^o)^{M}\right]^c+|[]_D\rangle+|[]_{\Bbb{RP}^2}=0\,,\\\label{eqn:eom2}
&Q_B|\Psi^o\rangle+g_s\sum_{N=0}^\infty \sum_{M=1}^\infty\frac{1}{N!(M-1)!}\mathcal{G}\left[(\Psi^c)^N;(\Psi^o)^{M-1}\right]^o=0\,. 
\end{align}

The solution corresponding to the original background represented by a pure NSNS worldsheet CFT is given by $|\Psi^c\rangle=|\Psi^o\rangle=0$. The presence of tadpoles at subleading orders in $g_s$ necessarily requires nontrivial string field configurations to be turned on. If one is interested in obtaining a background that requires turning on nontrivial field configurations from the original worldsheet CFT background, such field configurations should enter as string fields in the above SFEOM and further solve it.

For the equations of motion \eqref{eqn:eom1} and \eqref{eqn:eom2} to have solutions, non-$Q_B$ terms in \eqref{eqn:eom1} and \eqref{eqn:eom2} must be BRST exact. Once the solution $|\Psi^c\rangle$ and $|\Psi^o\rangle$ is obtained, one can obtain $\tilde{\Psi}^c$ and $\tilde{\Psi}^o$ accordingly. Note that there is a freedom to add any $Q_B$-closed term to $\tilde{\Psi}^c$ and $\tilde{\Psi}^o$, meaning that they represent free string fields that decouple from the interacting string field degrees of freedom.

\subsection{$\epsilon$ expansion}\label{subsection:epsilon expansion}
Now, we shall study the expansion scheme for the solution to SFEOM \eqref{eqn:eom1} and \eqref{eqn:eom2}, which corresponds to the example introduced in section \ref{subsection:explicitExample}. The pure NSNS background which serves as the starting point for SFT is given by an exact worldsheet BCFT whose matter part consists of four non-compact free bosons $X^\mu$ ($\mu=0,1,2,3$), six compact free bosons $Y^i$ ($i=1,2,3,4,5,6$) for toroidal orientifold $T^6/\mathbb{Z}_2$, and their fermionic partners $\psi^\mu$ and $\psi^i$ as discussed in section \ref{sec:WS}. We also have 4 D3-branes, which fill in the four non-compact spacetime and 64 O3-planes as discussed in section \ref{subsection:explicitExample}. This worldsheet CFT provides the Hilbert spaces $H^c,\tilde{H}^c,H^o,$ and $\tilde{H}^o$ which are building blocks of SFT, and SFEOM \eqref{eqn:eom1} and \eqref{eqn:eom2} are written in terms of the coefficient string fields when states in $H^c$ and $H^o$ are expanded in a basis.

To define a consistent expansion scheme, we shall introduce a small parameter $\epsilon.$ We will then take a large complex structure limit such that the imaginary part of the complex structure moduli will be treated as a large parameter 
\begin{equation}
\mathcal{O}(\im u_i)=\mathcal{O}(\epsilon^{-1})\,.
\end{equation}
At the same time, we shall treat string coupling as a small parameter such that 
\begin{equation}
\mathcal{O}(\im \tau)=\mathcal{O}(\epsilon^{-1}).
\end{equation}
Therefore, we shall treat the complex structure moduli and the axio-dilaton modulus on the same footing.

Because the threeform fluxes $F$ and $H$ generate potential for the moduli, the moduli vev will be severely constrained by the SFEOM. To make sure that the $\epsilon$ expansion scheme is self-consistent, the scaling 
\begin{equation}\label{eqn:epsilon expansion}
\mathcal{O}(\im u_i)=\mathcal{O}(\im \tau)=\mathcal{O}(\epsilon^{-1})\,,
\end{equation}
shall be justified with the solutions to SFEOM. We shall do so in the next section, and show that the $\epsilon$ expansion scheme is self-consistent as the tadpole cancellation of SFEOM at order $\mathcal{O}(\epsilon)$ will lead to
\begin{equation}\label{eqn:ui tau in sfeom}
 u^i=p^i\tau\,,   
\end{equation}
 in agreement with the low energy supergravity. Note, in particular, that due to the non-renormalization theorem \cite{Dine:1986vd,Burgess:2005jx}, we expect the relation $u^i=p^u\tau$ to hold up to all orders in $g_s$ and hence $\epsilon$ perturbatively. Once the identification \eqref{eqn:ui tau in sfeom} is made at $\mathcal{O}(\epsilon),$ we can effectively trade $\epsilon$ with $g_s,$ thereby establishing $g_s^{1/2}$ expansion scheme.

The scaling \eqref{eqn:epsilon expansion} inevitably introduces explicit factors of $\epsilon$ in the evaluation of OPE and correlators of the CFT $T^6/\mathbb{Z}_2$. A systematic expansion in $\epsilon$ can be obtained by employing the vielbein introduced in section \ref{sec:WS}. Then, OPE and correlators of fields in the orthonormal frame do not involve explicit factors of $\epsilon$, and the vielbein can be regarded as an expansion parameter carrying specific powers of $\epsilon$. We will use the shorthand notation $Y:=\Bbb{R}^{1,3}\times T^6$ and $\tilde{Y}:=\Bbb{R}^{1,3}\times T^6/\Bbb{Z}_2.$

For example, the NSNS 3-form flux corresponds to the following vertex operator of the worldsheet CFT:
\begin{equation}\label{eqn:H3vertop}
\mathcal{O}_{NSNS}=\frac{1}{4\pi}c\bar{c} B_{ij}e^{-\phi}\psi^ie^{-\bar{\phi}}\bar{\psi}^j,
\end{equation}
where locally, we have
\begin{equation}
dB=H_3\,,
\end{equation}
and
\begin{equation}
B=\frac{1}{3!} H_{ijk} Y^i dY^j\wedge dY^k\,.
\end{equation}
As explained in \eqref{eq:H3vertOp}, $\mathcal{O}_{NSNS}$ is expressed in the local chart of the torus with coordinates $Y^i$. In order to discuss $g_s$-dependence, we should discuss the picture-raised vertex operator, which includes the unambiguous field strength $H_3$ (as opposed to $B$)
\begin{equation}
\mathcal{X}_0\mathcal{O}_{NSNS}\sim c\bar{c} H_{ijk}\psi^i\psi^j e^{-\bar{\phi}}\bar{\psi}^k+...
\end{equation}
Introducing the vielbeins,
\begin{equation}
c\bar{c} H_{ijk}\psi^i\psi^j e^{-\bar{\phi}}\bar{\psi}^k=c\bar{c} \left(e^{-1}\right)_{i'}^i \left(e^{-1}\right)_{j'}^j \left(e^{-1}\right)_{k'}^k H_{ijk}\lambda^{i'}\lambda^{j'}e^{-\bar{\phi}}\bar{\lambda}^{k'} \,,
\end{equation}
where primed indices $i',j',$ and $k'$ stand for the orthonormal frame. From the metric of $T^6$ which in particular depends on $u_i$, it is straightforward to read off the scaling of vielbeins in terms of $\epsilon$
\begin{equation}
e^{2i'}_{2i}\sim \epsilon^{-\frac{1}{2}} \delta^{i'}_i,~~~e^{2i'-1}_{2i-1}\sim \epsilon^{\frac{1}{2}}\delta^{i'}_i,~~~i=1,2,3.
\end{equation}
Each $H_{ijk}$ given in \eqref{eqn:H3explicit} are order 1 numbers, and $H_{ijk}\neq0$ only when one of $\{i,j,k\}$ is odd and the other two are even. Therefore, we conclude that
\begin{equation}
\left(e^{-1}\right)_{i'}^i \left(e^{-1}\right)_{j'}^j \left(e^{-1}\right)_{k'}^k H_{ijk}\sim \epsilon^{\frac{1}{2}}.
\end{equation}
Therefore, the NSNS part of the string field solution $\Psi^c$ to SFEOM \eqref{eqn:eom1} corresponding to $H_3$-flux term should start at order $\epsilon^{\frac{1}{2}}$.

The story is similar for the RR 3-form flux, except that there is an extra factor of $g_s$ in the SFT convention where NSNS and RR string fields are normalized in the same way unlike the supergravity convention \cite{Alexandrov:2021shf,Alexandrov:2021dyl}. The vertex operator for the Ramond-Ramond field in $(-1/2,-1/2)$ picture is then given as
\begin{equation}\label{eqn:F3vertop}
\mathcal{O}_{RR}=\frac{i}{3!}\frac{g_s \sqrt{\alpha'}}{16\sqrt{2}\pi}c\bar{c} F_{ijk}e^{-\phi/2}\Sigma_{\alpha} (\Gamma^{ijk})^{\alpha\beta}e^{-\bar{\phi}/2}\bar{\Sigma}_{\beta}\,,
\end{equation}
Each $F_{ijk}$ given in \eqref{eqn:F3explicit} are order 1 numbers, and $F_{ijk}\neq0$ only when one of $\{i,j,k\}$ is even and the other two are odd. Therefore, we conclude that the RR part of the string field solution $\Psi^c$ to SFEOM \eqref{eqn:eom1} corresponding to $F_3$-flux term should start at order $\epsilon^{\frac{1}{2}}$.

In contrast, the solution \eqref{eqn:supergravitysolp} implies that the corresponding string field solutions for the metric and $F_5$ start at order $\epsilon$. We, therefore, take the following expansion scheme for the string field solutions $\Psi^c=\Psi^c_0$ and $\Psi^o=\Psi^o_0$:
\begin{equation}
\Psi^c_0=\sum_{n=1}^\infty \epsilon^{n/2}\left(V_{NS,n}^{-1,-1}+V_{R,n}^{-\frac{1}{2},-\frac{1}{2}}\right),~~~~V_{NS,n}^{-1,-1}\in H_{-1,-1},~V_{R,n}^{-\frac{1}{2},-\frac{1}{2}}\in H_{-1/2,-1/2},
\end{equation}
\begin{equation}
\Psi^o_0=\sum_{n=1}^\infty \epsilon^{n/2}v_{NS,n}^{-1},~~~~ v_{NS,n}^{-1}\in H_{-1}.
\end{equation}
Note here that we also introduce a nontrivial open string field profile given by $v_{NS,n}^{-1}$. This is because the change in the bulk geometry back-reacts to D-branes, which then requires nontrivial open string field profiles. Since the closed string field solution only has NSNS and RR field profiles, the back-reaction only induces NS open string fields. As disk diagrams are suppressed by a factor of $g_s$ compared to sphere diagrams, we will not need to obtain $\Psi^o_0$ if we are only interested in closed string dynamics up to the second order in the expansion parameter $\epsilon^{1/2}$.

Finally, we shall discuss the gauge we will impose on the solutions of SFEOM. We introduce the projection operator $\Bbb{P}$ which projects to states on which $L_0+\bar{L}_0$ acts nilpotently. We impose Siegel gauge condition on $(1-\Bbb{P})\Psi^c_0$: $b_0^+(1-\Bbb{P})\Psi^c_0=0$. For $\mathbb{P}\Psi^c_0$, we impose Siegel gauge condition except for the subspace of states spanned by
\begin{equation}
    (\partial c+\bar{\partial}\bar{c})c\bar{c}e^{-\phi}\psi_Ae^{-2\bar\phi}\bar{\partial}\bar{\xi}~~\text{and}~~(\partial c+\bar{\partial}\bar{c})c\bar{c}e^{-\bar\phi}\bar{\psi}_Ae^{-2\phi}{\partial}{\xi},
\end{equation}
on which we impose
\begin{equation}
    (\partial c+\bar{\partial}\bar{c})c\bar{c}\left(e^{-\phi}\psi_Ae^{-2\bar\phi}\bar{\partial}\bar{\xi}-e^{-\bar\phi}\bar{\psi}_Ae^{-2\phi}{\partial}{\xi}\right)=0.
\end{equation}
Then, the states which survive the gauge condition in the subspace are
\begin{equation}
    (\partial c+\bar{\partial}\bar{c})c\bar{c}\left(e^{-\phi}\psi_Ae^{-2\bar\phi}\bar{\partial}\bar{\xi}+e^{-\bar\phi}\bar{\psi}_Ae^{-2\phi}{\partial}{\xi}\right).
\end{equation}
These states are the auxiliary fields which have played an important role in establishing the field redefinition between the massless string fields and the low energy gravity fields \cite{Hull:2009mi}.

\subsection{Background solution - order $\epsilon^{1/2}$}\label{sec:1stbgd}
We are ready to study the background solution to the SFEOM at the first two orders in $\epsilon^{1/2}$-expansion. At the leading order in $\epsilon^{1/2}$, the SFEOM \eqref{eqn:eom1} reduces to
\begin{equation}
Q_B V_{NS,1}^{-1,-1}=0,
\end{equation}
\begin{equation}
Q_B V_{R,1}^{-\frac{1}{2},-\frac{1}{2}}=0.
\end{equation}
The solution of interest to us is given by the 3-form fluxes
\begin{equation}
\epsilon^{1/2}V_{NS,1}^{-1,-1}=\mathcal{O}_{NSNS},~~\epsilon^{1/2}V_{R,1}^{-\frac{1}{2},-\frac{1}{2}}=\mathcal{O}_{RR},
\end{equation}
which were discussed in \eqref{eqn:H3vertop} and \eqref{eqn:F3vertop}. It is straightforward to check that $\mathcal{O}_{NSNS}$ and $\mathcal{O}_{RR}$ are $Q_B$-closed.

\subsection{Background solution - order $\epsilon$}\label{sec:2ndbgd}
At the second order in $\epsilon^{1/2},$ the string field equations of motion are
\begin{align}\label{eqn:sec pert NS} 
&\frac{4}{\alpha' g_c^2}Q_B V_{NS,2}^{-1,-1}+\frac{1}{2}\mathcal{G}\left[ V_{NS,1}^{-1,-1}\otimes V_{NS,1}^{-1,-1}+V_{R,1}^{-\frac{1}{2},-\frac{1}{2}}\otimes V_{R,1}^{-\frac{1}{2},-\frac{1}{2}} \right]^c_{S^2}+\epsilon^{-1}|[]_{D+\Bbb{RP}^2}\rangle_{NSNS}=0\,,\\\label{eqn:sec pert R}
&\frac{4}{\alpha' g_c^2}Q_B V_{R,2}^{-\frac{1}{2},-\frac{1}{2}}+ \mathcal{G} \left[ V_{NS,1}^{-1,-1}\otimes V_{R,1}^{-\frac{1}{2},-\frac{1}{2}}\right]^c_{S^2} +\epsilon^{-1} |[]_{D+\Bbb{RP}^2}\rangle_{RR}=0\,.
\end{align}
Here, $[...]_{S^2}^c$ is the string bracket obtained by varying the sphere string vertices, $[]_{D+\Bbb{RP}^2}=[]_D+[]_{\Bbb{RP}^2}$, and $|[]_{D+\Bbb{RP}^2}\rangle_{NSNS}$ and $|[]_{D+\Bbb{RP}^2}\rangle_{RR}$ are NSNS and RR sector projection of the state $|[]_{D+\Bbb{RP}^2}\rangle$ respectively.

Note that the sphere amplitudes contain an overall normalization constant
\begin{equation}
C_{S^2}=\frac{8\pi}{\alpha' g_c^2}=\frac{2^5\pi^3}{\alpha'\kappa_{10}^2 g_s^2}\,,
\end{equation}
and the boundary states contain $g_s^{-1}$ in the overall normalization. Therefore, all of the terms in the second-order equations of motion are of the same order in $\epsilon$ as they should be.

Following \cite{deLacroix:2017lif,Cho:2018nfn}, to solve \eqref{eqn:sec pert NS} and \eqref{eqn:sec pert R}, we shall split the equations of motion into two parts using the projection $\Bbb{P}$ introduced above. We split
\begin{equation}
V_{NS,2}^{-1,-1}=W_{NS}^{(2)}+X_{NS}^{(2)}\,,
\end{equation}
and
\begin{equation}
V_{R,2}^{-\frac{1}{2},-\frac{1}{2}}=W_{R}^{(2)}+X_{R}^{(2)}\,,
\end{equation}
such that $X_{NS}^{(2)}$ and $X_R^{(2)}$ satisfy
\begin{equation}
\Bbb{P} X_{NS}^{(2)}=\Bbb{P}X_{R}^{(2)}=0\,,
\end{equation}
and $W_{NS}^{(2)}$ and $W_R^{(2)}$ satisfy
\begin{equation}
\Bbb{P} W_{NS}^{(2)}=W_{NS}^{(2)},~~\Bbb{P}W_{R}^{(2)}=W_{R}^{(2)}\,.
\end{equation}
Using $\{ Q_B,b_0\}=L_0\,$, we can solve $X_{NS}^{(2)}$ and $X_{R}^{(2)}$ rather easily
\begin{align}\label{eqn:XNS sol}
\frac{4}{\alpha' g_c^2}X_{NS}^{(2)}=&-\frac{1}{2(L_0+\bar{L}_0)}(b_0+\bar{b}_0)(1-\Bbb{P})\mathcal{G}\left[ V_{NS,1}^{-1,-1}\otimes V_{NS,1}^{-1,-1}+V_{R,1}^{-\frac{1}{2},-\frac{1}{2}}\otimes V_{R,1}^{-\frac{1}{2},-\frac{1}{2}} \right]^c_{S^2}\nonumber\\
&-\frac{\epsilon^{-1}}{L_0+\bar{L}_0}(b_0+\bar{b}_0)(1-\Bbb{P})|[]_{D+\Bbb{RP}^2}\rangle_{NSNS}\,,
\end{align}
and
\begin{align}\label{eqn:XR sol}
\frac{4}{\alpha' g_c^2}X_{R}^{(2)}=&-\frac{1}{(L_0+\bar{L}_0)}(b_0+\bar{b}_0)(1-\Bbb{P})\mathcal{G}\left[ V_{NS,1}^{-1,-1} V_{R,1}^{-\frac{1}{2},-\frac{1}{2}} \right]^c_{S^2}\nonumber\\
&-\frac{\epsilon^{-1}}{L_0+\bar{L}_0}(b_0+\bar{b}_0)(1-\Bbb{P})|[]_{D+\Bbb{RP}^2}\rangle_{RR}\,.
\end{align}

$\Bbb{P}$-parts of the equations of motion are more involved. We write
\begin{align}\label{eqn:PprojNS}
&\frac{4}{\alpha' g_c^2}Q_B W_{NS}^{(2)}+\frac{1}{2}\Bbb{P}\mathcal{G}\left[ V_{NS,1}^{-1,-1}\otimes V_{NS,1}^{-1,-1}+V_{R,1}^{-\frac{1}{2},-\frac{1}{2}}\otimes V_{R,1}^{-\frac{1}{2},-\frac{1}{2}} \right]^c_{S^2}+\epsilon^{-1} \Bbb{P}|[]_{D^2+\Bbb{RP}^2}\rangle_{NSNS}=0\,,
\end{align}
and
\begin{equation}
\frac{4}{\alpha' g_c^2}Q_B W_{R}^{(2)}+\Bbb{P} \mathcal{G} \left[ V_{NS,1}^{-1,-1}\otimes V_{R,1}^{-\frac{1}{2},-\frac{1}{2}}\right]^c_{S^2} +\epsilon^{-1}\Bbb{P} |[]_{D+\Bbb{RP}^2}\rangle_{RR}=0\,.
\end{equation}
To find the solutions $W_{NS}^{(2)}$ and $W_R^{(2)}$, we shall first compute the string brackets appearing in the above.

\subsubsection{Equations of motion of the $L_0^+$ nilpotent NSNS fields}\label{subsec:NS eom}
In this section, we shall study the following equation
\begin{align}\label{eqn:L0 nil NSNS}
&\frac{4}{\alpha' g_c^2}Q_B W_{NS}^{(2)}+\frac{1}{2}\Bbb{P}\mathcal{G}\left[ V_{NS,1}^{-1,-1}\otimes V_{NS,1}^{-1,-1}+V_{R,1}^{-\frac{1}{2},-\frac{1}{2}}\otimes V_{R,1}^{-\frac{1}{2},-\frac{1}{2}} \right]^c_{S^2}+\epsilon^{-1} \Bbb{P}|[]_{D^2+\Bbb{RP}^2}\rangle_{NSNS}=0\,,
\end{align}
First, we will compute the closed string source terms in \eqref{eqn:L0 nil NSNS} i.e. string brackets. In order for the equation \eqref{eqn:L0 nil NSNS} to admit a solution to $W_{NS}^{(2)},$ the source term must be BRST-exact. 

\begin{comment}
We will show that the BRST-exactness implies that the tadpole due to the source terms in \eqref{eqn:L0 nil NSNS} should vanish, and furthermore we will show how the tadpole cancellation condition is a string field theory analog of
\begin{equation}
\partial_{\varphi}\left(\sqrt{-G}\mathcal{V}_{eff}\right)=0
\end{equation}
of the low-energy supergravity, where $\varphi$ is a supergravity field and $V$ is the effective potential.

\end{comment}

To compute 
\begin{equation}
\frac{1}{2}\Bbb{P}\mathcal{G} \left[ V_{NS,1}^{-1,-1}\otimes V_{NS,1}^{-1,-1}+V_{R,1}^{-\frac{1}{2},-\frac{1}{2}}\otimes V_{R,1}^{-\frac{1}{2},-\frac{1}{2}} \right]^c_{S^2} \,,
\end{equation}
we can compute an overlap between a generic state $\phi$ satisfying $\Bbb{P}\phi=\phi$ with 
\begin{equation}
V_{NS,1}^{-1,-1}\otimes V_{NS,1}^{-1,-1}+V_{R,1}^{-\frac{1}{2},-\frac{1}{2}}\otimes V_{R,1}^{-\frac{1}{2},-\frac{1}{2}} \,,
\end{equation}
which will produce the sphere cubic string vertex 
\begin{equation}
\{\phi,V_{NS,1}^{-1,-1},V_{NS,1}^{-1,-1}\}_{S^2}+\{\phi,V_{R,1}^{-\frac{1}{2},-\frac{1}{2}},V_{R,1}^{-\frac{1}{2},-\frac{1}{2}}\}_{S^2}\,.
\end{equation}

Let us first study the contribution from $V_{NS}^{-1,-1}\otimes V_{NS}^{-1,-1}.$ The corresponding string vertex $\{\phi,V_{NS,1}^{-1,-1},V_{NS,1}^{-1,-1}\}_{S^2}$, which is nonzero only when $\phi$ is in the NSNS sector, is defined not only by the choice of local charts around three punctures on the sphere but also by the location of left- and right-moving PCOs. It is straightforward to see that this string vertex can be nonzero only if $\phi$ takes the form
\begin{equation}
\phi= c\bar{c}e^{-\phi}\psi^i e^{-\bar{\phi}}\psi^j f(X^A),
\end{equation}
where $i$ and $j$ are along the compact directions and $f(X^A)$ should lie in the $L_0^+$-nilpotent space to survive the $\Bbb{P}$ projection. Since $V_{NS,1}^{-1,-1}$ has no nontrivial free boson zero modes along the non-compact direction, $f(X^A)$ cannot have nontrivial excitations along the compact directions since it will not survive the $\Bbb{P}$ projection in that case. $Q_B\phi$ is then proportional to derivatives of $f(X^A)$ with respect to the non-compact free boson zero modes, meaning that $\{\phi,V_{NS,1}^{-1,-1},V_{NS,1}^{-1,-1}\}_{S^2}$ is independent of the choice of PCO locations. Similarly, it is independent of the choice of local charts since $V_{NS,1}^{-1,-1}$ is on-shell, and local chart effects on $\phi$ result in derivatives of $f(X^A)$ with respect to the non-compact free boson zero modes.\footnote{In case the sphere three-point vertex of interest involves more generic operator insertions, we should specify the local charts and the PCO locations. One can place three NSNS vertex operator insertions at $z=0,1,\infty$ on the sphere and place PCO at $z=e^{\pm i\pi/3}$, and average over cyclic permutations of the three NSNS vertex operator insertions. This ensures that the sphere cubic vertex is symmetric under the permutation of string fields \cite{Sen:2019jpm}.}

We find
\begin{align}
\frac{1}{2}\Bbb{P}\left[V_{NS,1}^{-1,-1}\otimes V_{NS,1}^{-1,-1}\right]^c_{S^2}=&-  C_{S^2}\frac{\epsilon^{-1}}{2(4\pi)^2}\frac{\alpha'}{2}(\partial c+\bar{\partial}\bar{c})c\bar{c} H_{acd}H_{bef}G^{ce}G^{df}  e^{-\phi}\psi^a e^{-\bar{\phi}}\bar{\psi}^b\,.
\end{align}
Using $C_{S^2}=2^5\pi^3/(\alpha'\kappa_{10}^2g_s^2),$ we find
\begin{align}
\frac{1}{2}\Bbb{P}\left[V_{NS,1}^{-1,-1}\otimes V_{NS,1}^{-1,-1}\right]^c_{S^2}=&-\frac{\pi\epsilon^{-1}}{2\kappa_{10}^2g_s^2}(\partial c+\bar{\partial}\bar{c})c\bar{c} H_{acd}H_{bef}G^{ce}G^{df}e^{-\phi}\psi^a e^{-\bar{\phi}}\bar{\psi}^b\,.
\end{align}
Similarly, we find
\begin{equation}
\frac{1}{2}\Bbb{P}\left[V_{R,1}^{-\frac{1}{2},-\frac{1}{2}}\otimes V_{R,1}^{-\frac{1}{2},-\frac{1}{2}}\right]^c_{S^2}=-\frac{\pi }{2\epsilon\kappa_{10}^2 }(\partial c+\bar{\partial}\bar{c})c\bar{c} \left(F_{Acd}F_{Bef}G^{ce}G^{df}-\frac{G_{AB} }{3!}|F|^2\right) e^{-\phi}\psi^Ae^{-\bar{\phi}}\bar{\psi}^B\,.
\end{equation}
Finally, we have
\begin{equation}\label{eqn:boundary state sec4}
\frac{1}{\epsilon}\Bbb{P}|[]_{D^2}\rangle_{NSNS}=\sum_{y_{D3}}\frac{2\pi T_3 }{g_s\epsilon} (\Bbb{P}\delta^{(6)}(y-y_{D3}))\left[\frac{\partial c+\bar{\partial}\bar{c}}{2}c\bar{c}\left( e^{-\phi} \psi^A S_{AB}e^{-\bar{\phi}}\bar{\psi}^B - (\eta\bar{\partial}\bar{\xi} e^{-2\bar{\phi}}-\partial\xi\bar{\eta}e^{-2\phi}) \right)\right]\,,
\end{equation}
\begin{equation}
\frac{1}{\epsilon}\Bbb{P}|[]_{\Bbb{RP}^2}\rangle_{NSNS}=-\sum_{y_{O3}}\frac{\pi T_3 }{2g_s\epsilon} (\Bbb{P}\delta^{(6)}(y-y_{O3}))\left[\frac{\partial c+\bar{\partial}\bar{c}}{2}c\bar{c}\left( e^{-\phi} \psi^A S_{AB}e^{-\bar{\phi}}\bar{\psi}^B - (\eta\bar{\partial}\bar{\xi} e^{-2\bar{\phi}}-\partial\xi\bar{\eta}e^{-2\phi}) \right)\right]\,.
\end{equation}
Note that we included an additional factor of 2 to the sphere diagram to properly take into account that the original sphere diagram was properly normalized for $T^6,$ not $T^6/\Bbb{Z}_2.$ This is because the volume integral over $T^6$ is twice that of $T^6/\Bbb{Z}_2.$ Furthermore, as studied in \cite{FarooghMoosavian:2019yke}, for each closed string vertex operator, we shall multiply an additional factor of $1/\sqrt{2}$ to correctly take into account the orientifolding. These factors cancel out for the closed string source terms studied above. We also note that such an additional factor is not needed as we normalized the boundary states such that two D3-branes at $y_i$ in $T^6$ correspond to a single D3-brane at $y_i$ in $T^6/\Bbb{Z}_2.$ This implies, for example, that the boundary state for a single D3-brane on an O-plane should be half of \eqref{eqn:boundary state sec4}. 

Now, we shall solve \eqref{eqn:L0 nil NSNS} by two steps. First, we will show that the matter parts of the source terms in \eqref{eqn:L0 nil NSNS} cancel provided that moduli vevs are tuned to solve supergravity equations of motion at $\mathcal{O}(\epsilon).$ Second, with this simplification in hand, we will find an explicit form of $W_{NS}^{(2)}$ that solves \eqref{eqn:L0 nil NSNS}.

A few important remarks are in order. It should be and it is possible to directly solve the equations of motion in string field theory without making any reference to the low-energy supergravity. Hence, it may seem unnecessary to use the low-energy supergravity as an input.\footnote{Note furthermore that the non-trivial field redefinitions between supergravity fields and string fields may obscure this ad-hoc procedure \cite{Hull:2009mi}.} We nevertheless proceed to use the low-energy supergravity as an input in this section, as the input from the low-energy supergravity can be used to greatly simplify the SFEOM. The reader may wonder if this simplification can be achieved by invoking the tadpole cancellation in string field theory. As we will discuss in \S\ref{section:tadpole}, the tadpole cancellation in SFT leads to the equivalent conclusions, but there exist subtleties that need to be carefully studied. Therefore, to avoid technicalities, as promised, we shall use the supergravity as an input, and revisit the issue of the tadpole cancellation in \S\ref{section:tadpole}.

The matter part of the source terms in \eqref{eqn:L0 nil NSNS} takes the following form
\begin{equation}
    \mathcal{A}_{AB} (\partial c+\bar{\partial}\bar{c})c\bar{c} e^{-\phi}\psi^Ae^{-\bar{\phi}}\bar{\psi}^B\,.
\end{equation}
Therefore, by computing an overlap against a state generated by the following test vertex operator
\begin{equation}
    V_{test}^{-1,-1}=\epsilon_{AB} c\bar{c} e^{-\phi} \psi^A e^{-\bar{\phi}}\bar{\psi}^B\,,
\end{equation}
one can study if $\mathcal{A}_{AB}$ vanishes at the F-term locus. In this section, we shall illustrate that absence of $\epsilon^{AB}\mathcal{A}_{AB}$ is guaranteed by the F-term conditions for all choices of the polarization $\epsilon_{AB}.$ We shall refer $\epsilon^{AB}\mathcal{A}_{AB}$ as $\epsilon^{AB}$ tadpole.

Let us first study the case where $V_{test}^{-1,-1}$ is the dilaton vertex operator at zero momentum
\begin{equation}
V_{D}^{-1,-1}(k,n)=-\frac{g_c}{\sqrt{8}}c\bar{c}%\left(G_{AB}e^{-\phi}\psi^Ae^{-\bar{\phi}}\bar{\psi}^B-(\eta{\bar{\partial}}{\bar{\xi}}e^{-2\bar{\phi}}-\partial\xi{\bar{\eta}}e^{-2\phi}) \right).
\left(G_{AB}-k_A\bar{k}_B-k_B\bar{k}_A\right)e^{-\phi}\psi^Ae^{-\bar{\phi}}\bar{\psi}^B\,,
\end{equation}
where we used the primary form of the dilaton vertex operator. Note that $\bar{k}_A$ is defined as
\begin{equation}
\bar{k}_A:=\frac{n_A}{n\cdot k}\,,
\end{equation}
where $n_A$ is chosen as a generic null vector \cite{Blumenhagen:2013fgp}, and $k_A$ is a vector that is chosen to be transverse to the worldvolume of the spacetime filling D3-branes and O3-planes. Note that $k_A$ shouldn't be understood as the momentum of the dilaton state, as $k_A$ and $\bar{k}_A$ are auxiliary vectors. Furthermore, we shall average over different choices of $k_A$ so that there is no preference over a particular direction, such that
\begin{equation}
V_{D}^{-1,-1}=\frac{1}{6} \sum_i V_D^{-1,-1}(k_i,n_i)\,,
\end{equation}
where $k_i$ is a unit vector along $i$-th direction, meaning $k_i^2G^{ii}=1$, where $i$ is not summed over, and $n_i$ is a null vector that is defined as
\begin{equation}
n_i:= n_0+k_i\,,\quad n_0=(1,0,0,0,0,0,0,0,0,0)\,.
\end{equation}
One can check that the dilaton vertex operator has a vanishing one-point function on the disk and $\Bbb{RP}^2.$ Therefore, the dilaton correlation function receives only contributions from the closed string sources. From the NSNS fluxes, we obtain
\begin{equation}\label{eqn:dilaton tadpole NSNS}
\frac{1}{2}\{ V_{D}^{-1,-1}, V_{NS,1}^{-1,-1}, V_{NS,1}^{-1,-1} \}_{S^2}=-\frac{1}{\epsilon}\int_{\tilde{Y}} d^{10}X\frac{g_c}{\sqrt{8}}\frac{\pi}{3 \kappa_{10}^2 g_s^2} H_{acd}H_{bef}G^{ab}G^{ce}G^{df}\,.
\end{equation}
From the RR fluxes, we obtain
\begin{equation}\label{eqn:dilaton tadpole RR}
\frac{1}{2}\{ V_{D}^{-1,-1}, V_{R,1}^{-\frac{1}{2},-\frac{1}{2}}, V_{R,1}^{-\frac{1}{2},-\frac{1}{2}} \}_{S^2}=\frac{1}{\epsilon}\int_{\tilde{Y}} d^{10}X\frac{g_c}{\sqrt{8}}\frac{\pi }{3\kappa_{10}^2} F_{acd}F_{bef} G^{ab}G^{ce}G^{df}\,. 
\end{equation}
Combining \eqref{eqn:dilaton tadpole NSNS} and \eqref{eqn:dilaton tadpole RR}, we find that the overlap between the source term and the dilaton state vanishes at the F-term locus
\begin{equation}\label{eqn:dilaton tadpole full}
\boxed{\frac{1}{g_s\epsilon}\frac{g_c}{\sqrt{8}}\frac{2\pi}{\kappa_{10}^2} \int_{\tilde{Y}} d^{10}X \frac{\partial}{\partial\log g_s}\left(\frac{1}{\im\tau} G_3\cdot\overline{G}_3\right)= \frac{1}{g_s\epsilon}\frac{\pi g_c}{\sqrt{8}}\int_{\Bbb{R}^{1,3}} d^4X \frac{\partial}{\partial\log g_s}\mathcal{V}_1=0\,,}
\end{equation}

Note that the $C_0$ contribution in $G_3$ is neglected as we are working in the string perturbation theory with $C_0=0$ background. As a result, up to a numerical factor, we find that the dilaton tadpole is exactly the derivative of the effective potential with respect to the dilaton.\footnote{Similarly, the complex structure tadpole results in the analogous result. Because the constraints due to the complex structure tadpole are redundant to that of the dilaton tadpole, we will not write them here.} 

\begin{comment}    
Therefore, the existence of the solution requires
\begin{equation}
|H_3|^2=g_s^2|F_3|^2\,,
\end{equation}
which is equivalent to
\begin{equation}
G_+\cdot G_-=0\,,
\end{equation}
at the leading order in $g_s.$ It is very important to note that the ISD condition $G_-=0$ of the low energy supergravity requires $|H_3|^2=g_s^2|F_3|^2,$ which is in exact agreement with the condition we find here. 

Note that one shouldn't be surprised that the condition 
\begin{equation}
\frac{\partial}{\partial\log g_s}\mathcal{V}_1=0\,,
\end{equation}
implies not only the F-term condition for the axio-dilaton
\begin{equation}
\int_{T^6/\Bbb{Z}_2} G_3\wedge \Omega=0\,,
\end{equation}
but also the F-term conditions for the complex structure moduli
\begin{equation}
\int_{T^6/\Bbb{Z}_2} G_3\wedge\chi_a=0\,.
\end{equation}
This is because the derivative of the F-term potential w.r.t. the axio-dilaton modulus $\tau$ involves the F-term of both the complex structure moduli and the axio-dilaton, c.f. \cite{Marsh:2011aa}, 
\begin{equation}
\partial_\tau V=e^K\left(\mathcal{D}_\tau  (D_aW)(D^a\overline{W})\right)=0\,,
\end{equation}
where $\mathcal{D}_\tau$ denotes the K\"{a}hler and geometrically covariant derivative, $a$ runs for the axio-dilaton and the complex structure moduli, and we took into account the no-scale structure.
\end{comment}

Next, we shall choose the following test vertex operator
\begin{equation}
V_{test}^{-1,-1}=V_{h}^{-1,-1}:=\frac{1}{4\pi} h_{\mu\nu} c\bar{c} e^{-\phi}\psi^\mu e^{-\bar{\phi}}\bar{\psi}^\nu\,,
\end{equation}
where $h_{\mu\nu}$ is a symmetric rank two tensor along the non-compact directions. We compute
\begin{equation}\label{eqn:grav trace H}
\frac{1}{2}\{V_{h}^{-1,-1},V_{NS,1}^{-1,-1},V_{NS,1}^{-1,-1} \}=0\,,
\end{equation}
\begin{equation}\label{eqn:grav trace F}
\frac{1}{2}\{ V_{h}^{-1,-1}, V_{R,1}^{-\frac{1}{2},-\frac{1}{2}}, V_{R,1}^{-\frac{1}{2},-\frac{1}{2}} \}_{S^2}=-\frac{1}{24\kappa_{10}^2\epsilon}\int_{\tilde{Y}} d^{10}X \frac{1}{2}h^\mu_\mu|F|^2 \,,
\end{equation}
\begin{equation}\label{eqn:grav trace DBI}
\{V_{h}^{-1,-1}\}_{D^2+\Bbb{RP}^2}=-\frac{1}{2g_s\epsilon}\int_{\tilde{Y}} d^{10}X \frac{1}{2}h^\mu_\mu\mu_3\rho_{D3}^{loc} \,.
\end{equation}
We combine \eqref{eqn:grav trace H}, \eqref{eqn:grav trace F}, and \eqref{eqn:grav trace DBI} to obtain the $h$ tadpole
\begin{equation}
-\frac{1}{g_s\epsilon}\int_{\tilde{Y}} d^{10}X \frac{1}{2}h^\mu_\mu\left(-\frac{g_s}{4\kappa_{10}^2}\frac{1}{3!}|F_3|^2-\frac{1}{2}\mu_3\rho_{D3}^{loc}\right) \,,
\end{equation}
Using the fact that we have the ISD flux, we find that the above expression is proportional to the tadpole cancellation condition we imposed
\begin{equation}\label{eqn:external graviton tadpole}
\boxed{\int_{\tilde{Y}} d^{10}X\left(\frac{1}{2\kappa_{10}^2} [ H_3 \wedge F_3]+\mu_3\rho_{D3}^{loc}\right)=0\,.}
\end{equation}
Again, we found that the matter sector tadpole for this particular polarization vanishes. 

Finally, we study the tadpole associated with the following vertex operator 
\begin{equation}\label{eqn:v dg vertex}
V_{test}^{-1,-1}=V_{\delta g}^{-1,-1}=\frac{1}{4\pi}c\bar{c}\delta g_{ab} e^{-\phi}\psi^a e^{-\bar{\phi}}\bar{\psi}^b\,,
\end{equation}
where $\delta g_{ab}$ is a symmetric rank 2 tensor along compact directions. We compute the string brackets
\begin{align}\label{eqn:app d tadpole NSNS}
\frac{1}{2}\{ V_{\delta g}^{-1,-1}(k_i,n_i)\otimes V_{NS,1}^{-1,-1}\otimes V_{NS,1}^{-1,-1}\}_{S^2}=&-\frac{\pi g_c }{8\pi g_s^2\epsilon\kappa_{10}^2 }\int_{\tilde{Y}} d^{10}X \delta g^{ab} H_{aij}H_{b}^{~ij}\,.
\end{align}
Similarly, we compute
\begin{align}\label{eqn:app d tadpole RR}
\frac{1}{2}\{ V_{\delta g}^{-1,-1}(k_i,n_i)\otimes V_{R,1}^{-1,-1}\otimes V_R^{-1,-1}\}_{S^2}=&-\frac{\pi g_c }{8\pi\epsilon\kappa_{10}^2 }\int_{\tilde{Y}} d^{10}X \left[ \delta g^{ab}F_{aij}F_b^{~ij}-\frac{\delta g^{ab} G_{ab}}{3!} |F|^2\right]\,,
\end{align}
\begin{equation}\label{eqn:app d tadpole DBI}
\{V_{\delta g}^{-1,-1}\}_{D^2+\Bbb{RP}^2}=\frac{1}{2g_s\epsilon}\int_{\tilde{Y}} d^{10}X \frac{1}{2}\delta g^a_a\mu_3\rho_{D3}^{loc} \,.
\end{equation}

We can combine \eqref{eqn:app d tadpole NSNS}, \eqref{eqn:app d tadpole RR}, \eqref{eqn:app d tadpole DBI} and to find the combined tadpole
\begin{equation}
-\frac{\pi g_c}{8\pi g_s^2\epsilon\kappa_{10}^2}\int_{\tilde{Y}} d^{10}X\left[ (H_{iab}H^{iab}+g_s^2F_{iab}F^{iab}-\frac{1}{3!} \left(g_s^2|F|^2+|H|^2\right)\right]\,.
\end{equation}
Note that we used the Bianchi identity to rewrite the D3-brane charge in terms of the NSNS 3-form flux. As one can check, the above equation is equivalent to the variation of the effective potential w.r.t. the internal graviton in Einstein-frame
\begin{equation}
\boxed{-\frac{\pi g_c }{4\pi g_s\epsilon\kappa_{10}^2} \int_{\tilde{Y}} d^{10}X \delta^{ab}\frac{\partial}{\partial G^{ab}}\left(\sqrt{-G}\frac{1}{\im\tau} G_3\cdot\overline{G}_3\right)=0}\,,
\end{equation}
which vanishes at the F-term locus. This concludes that the matter part of the source terms in the NSNS sector of the SFEOM cancel at the F-term locus.

As a result, the choice of fluxes for the explicit example we consider here are such that the NSNS source terms in \eqref{eqn:L0 nil NSNS} completely cancel each other except for the ghost-dilaton term:
\begin{equation}\label{eqn:WNSgh}
    \frac{4}{\alpha'g_c^2}Q_B W^{(2)}_{NS}=-\sum_i \rho_i\frac{\pi T_3}{g_s\epsilon}\left(\Bbb{P}\delta^{(6)}(y-y_i)\right)(\partial c+\bar{\partial}\bar{c})(\eta\bar{\partial}\bar{\xi} e^{-2\bar{\phi}}-\partial\xi\bar{\eta}e^{-2\phi})\,,
\end{equation}
where $\rho_i=1$ for a D3-brane and $\rho_i=-1/4$ for an O3-plane. This equation can be solved by
\begin{align}\label{eqn:NS sol zero}
    \frac{4}{\alpha' g_c^2}W_{NS}^{(2)}=&-\frac{\pi}{18\kappa_{10}^2g_s^2\epsilon} c\bar{c}\bigg(B_{ab}B^{ab}(\eta\bar{\partial}\bar{\xi} e^{-2\bar{\phi}}-\partial\xi\bar{\eta}e^{-2\phi})-2B_{ac}B^{cb}c\bar{c}e^{-\phi}\psi^ae^{-\bar\phi}\bar{\psi}_b\nonumber
    \\
    &-2i\sqrt{\frac{\alpha'}{2}}B_{ab}H^{abc}(\partial c+\bar{\partial}\bar{c})\left(e^{-\phi}\psi_ce^{-2\bar\phi}\bar{\partial}\bar{\xi}+e^{-\bar\phi}\bar{\psi}_ce^{-2\phi}{\partial}{\xi}\right)    \bigg).
\end{align}
The solution $W_{NS}^{(2)}$ presents nontrivial field profiles for the ghost-dilaton (the first term in the first line), metric along $T^6$ (the second term in the first line), and the auxiliary field (the second line). However, this does not necessarily imply that these fields carry nontrivial field profiles in the supergravity frame, since string fields and supergravity fields at this order are related to each other via nonlinear field redefinitions \cite{Hull:2009mi}. For example, a nontrivial supergravity $B$-field at the first order induces a nontrivial ghost-dilaton string field at the second order proportional to $B_{AB}B^{AB}$, as in the above profile of $W_{NS}^{(2)}$, meaning that a nonzero ghost-dilaton string field profile may be equivalent to a zero ghost-dilaton supergravity field profile.

In terms of the mode expansion, the Dirac delta function can be written as
\begin{equation}
\delta^{(6)}(Y)=\frac{2}{\im t_1\im t_2\im t_3}\sum_{\vec{n}\in \Bbb{Z}^6} e^{2\pi i \vec{n}\cdot Y}\,.
\end{equation}
Note that the volume of the orientifold is $\im t_1\im t_2\im t_3/2.$ As the $\Bbb{P}$ projection eliminates a non-trivial mode, we find, therefore, 
\begin{equation}
\Bbb{P} \delta^{(6)}(Y)=\frac{2}{\im t_1\im t_2\im t_3}=\frac{1}{\text{Vol}_{T_6/\Bbb{Z}_2}}\,.
\end{equation}
This implies that the Green's function $\mathcal{G}$ defined in \eqref{eq:green} is trivial under the $\Bbb{P}$ projection
\begin{equation}
\Bbb{P}\mathcal{G}(X;Y)=0\,.
\end{equation}
Therefore, we expect that $W_{NS}^{(2)}$ becomes trivial as we implement the field redefinition mapping the string fields to the supergravity fields. Since on-shell physical observables are independent of such field redefinitions, we will proceed with the solution \eqref{eqn:NS sol zero}.

In contrast, since the first order solution had nontrivial profile only for the $\Bbb{P}$-projected sector, we expect that $(1-\Bbb{P})$ sector of the second order solution to agree with the supergravity description for the momenta modes along the compact directions. It is useful to recall that the mode expansion of the Green's function reads
\begin{equation}
\mathcal{G}^{(6)}(X;0)=-\sum_{\vec{n}\in\Bbb{Z}^6}(1-\delta_{\vec{n},0})\frac{e^{2\pi i \vec{n}\cdot X}}{4\pi^2 |n|^2\text{Vol}_{T^6/\Bbb{Z}_2}}=-\sum_{\vec{n}\in\Bbb{Z}^6\backslash0} \frac{\alpha'}{L_0+\bar{L}_0} \frac{e^{2\pi i\vec{n}\cdot X}}{\text{Vol}_{T^6/\Bbb{Z}_2}}\,.
\end{equation}
As a result, for the states projected out by $\Bbb{P}$, the Green's function acts as \cite{deLacroix:2017lif}
\begin{equation}\label{eqn:green}
-\frac{1}{L_0+\bar{L}_0}(b_0+\bar{b}_0)\,.
\end{equation}
Note that \eqref{eqn:green} is precisely the information contained in $X_{NS}^{(2)}$ we found in \eqref{eqn:XNS sol} previously
\begin{align}
\frac{4}{\alpha'g_c^2}X_{NS}^{(2)}=&-\frac{1}{2(L_0+\bar{L}_0)}(b_0+\bar{b}_0)(1-\Bbb{P})\mathcal{G}\left[ V_{NS,1}^{-1,-1}\otimes V_{NS,1}^{-1,-1}+V_{R,1}^{-\frac{1}{2},-\frac{1}{2}}\otimes V_{R,1}^{-\frac{1}{2},-\frac{1}{2}} \right]^c_{S^2}\nonumber\\
&-\frac{\epsilon^{-1}}{L_0+\bar{L}_0}(b_0+\bar{b}_0)(1-\Bbb{P})|[]_{D+\Bbb{RP}^2}\rangle_{NSNS}\,,
\end{align}
which agrees with the expectation from the supergravity description.

\subsubsection{Equations of motion of the $L_0^+$ nilpotent RR fields}
Now let us study the zero modes sector of the RR sector of the string field equations of motion
\begin{equation}\label{eqn:sft eom rr2}
\frac{4}{\alpha' g_c^2}Q_B W_{R}^{(2)}+\Bbb{P} \mathcal{G} \left[ V_{NS,1}^{-1,-1}\otimes V_{R,1}^{-\frac{1}{2},-\frac{1}{2}}\right]^c_{S^2} +\frac{1}{\epsilon}\Bbb{P} |[]_{D+\Bbb{RP}^2}\rangle_{RR}=0\,.
\end{equation}
We first compute the following string bracket
\begin{align}
\Bbb{P}\mathcal{G} \left[ V_{NS,1}^{-1,-1}\otimes V_{R,1}^{-\frac{1}{2},-\frac{1}{2}}\right]^c_{S^2} =&\frac{g_s}{\epsilon}\Bbb{P} \mathcal{X}_0\bar{\mathcal{X}}_0 \left[\frac{1}{4\pi} c\bar{c}B_{ab}e^{-\phi}\psi^a e^{-\bar{\phi}}\bar{\psi}^b\right.\nonumber\\
&\otimes\left. \frac{i}{3!}\frac{\sqrt{\alpha'}}{16\sqrt{2}\pi } c\bar{c} F_{cde} e^{-\phi/2}\Sigma_\alpha \left(\frac{1+\Gamma^{10}}{2}\Gamma^{cde}\right)^{\alpha\beta}e^{-\bar{\phi}/2}\bar{\Sigma}_{\beta}\right]_{S^2}^c\,.
\end{align}
Note here that we inserted 
\begin{equation}
\frac{1+\Gamma^{10}}{2}\,,
\end{equation}
without affecting the result. One is free to do so because $\Sigma_\alpha=\Sigma_\beta (\Gamma^{10})^{\beta}_{~\alpha}.$ We can freely pass $\Bbb{P}$ through $\mathcal{X}_0$ and $\bar{\mathcal{X}}_0$ as they are of weight zero. We obtain
\begin{align}
\Bbb{P}\left[ V_{NS,1}^{-1,-1}\otimes V_{R,1}^{-\frac{1}{2},-\frac{1}{2}}\right]^c_{S^2} =&-i\frac{ g_s C_{S^2}}{3!\epsilon} \frac{\sqrt{\alpha'}}{64\sqrt{2}\pi^2} (\partial c+\bar{\partial}\bar{c}) c\bar{c} B_{ab} F_{cde}  e^{-3\phi/2} \Sigma^{\gamma} e^{-3\bar{\phi}/2}\overline{\Sigma}^\delta\nonumber \\ &\qquad  \times  (\Gamma^a)_{\gamma\alpha} \left(\frac{1+\Gamma^{10}}{2}\Gamma^{cde}\right)^{\alpha\beta} (\Gamma^b)_{\beta\delta}\,.
\end{align}
Note that we inserted an additional factor of 2 to take into account the orientifolding. Using the identity
\begin{equation}
C_{S^2}=\frac{2^5\pi^3}{\alpha'\kappa_{10}^2g_s^2}\,,
\end{equation}
we write
\begin{align}\label{eqn:source eom rr1 1}
\Bbb{P}\left[ V_{NS,1}^{-1,-1}\otimes V_{R,1}^{-\frac{1}{2},-\frac{1}{2}}\right]^c_{S^2}=&-\frac{i}{3!\epsilon}\frac{\pi}{2\kappa_{10}^2g_s\sqrt{2\alpha'}}(\partial c+\bar{\partial}\bar{c}) c\bar{c} B_{ab} F_{cde}  e^{-3\phi/2} \Sigma^{\gamma} e^{-3\bar{\phi}/2}\overline{\Sigma}^\delta\nonumber\\ &\qquad \times (\Gamma^a)_{\gamma\alpha} \left(\frac{1+\Gamma^{10}}{2}\Gamma^{cde}\right)^{\alpha\beta} (\Gamma^b)_{\beta\delta} \,.
\end{align}

To compute
\begin{equation}
\mathcal{X}_0\bar{\mathcal{X}}_0\mathcal{S}_{RR}^{-\frac{3}{2},-\frac{3}{2}}:=\mathcal{G}\Bbb{P} \left[ V_{NS,1}^{-1,-1}\otimes V_{R,1}^{-\frac{1}{2},-\frac{1}{2}}\right]^c_{S^2}\,,
\end{equation}
we shall now act the zero mode of PCOs on \eqref{eqn:source eom rr1 1}. As a result, we obtain
\begin{align}
g_s\mathcal{X}_0\mathcal{S}_{RR}^{-\frac{3}{2},-\frac{3}{2}}=&\frac{\pi}{24\sqrt{2} \epsilon\kappa_{10}^2} (\partial c+\bar{\partial}\bar{c})c\bar{c}\partial_a B_{bc} F_{def} e^{-\phi/2}  \Sigma_\alpha (\Gamma^{ab}) (P_+^{10}\Gamma^{def}) (\Gamma^c)e^{-3\bar{\phi}/2}\overline{\Sigma}^\beta\nonumber\\
&+\frac{\pi i}{12 \epsilon\kappa_{10}^2\sqrt{2\alpha'}} \eta c\bar{c}B_{ab}F_{cde} e^{\phi/2}\Sigma^\gamma (\Gamma^a)(P_+^{10}\Gamma^{cde})(\Gamma^b) e^{-3\bar{\phi}/2}\overline{\Sigma}^\delta\,.
\end{align}
Note that we used the following identity
\begin{align}
\mathcal{C}:=&-\frac{1}{2\pi i}\oint \frac{dz}{z} \left(\partial\eta b e^{2\phi}(z)+\partial\left(\eta b e^{2\phi}\right)(z)\right) \left(\frac{\partial c+\bar{\partial}\bar{c}}{2}c\bar{c} e^{-3\phi/2}\Sigma^\gamma\right)(0)\\
=&  -\frac{1}{2\pi i}\oint \frac{dz}{z} \frac{1}{2} \eta(z)  c(0)\bar{c}(0)  \left( e^{\phi/2}\Sigma^\gamma\right)(0)+\dots\\
=&-\frac{1}{2}\eta c\bar{c} e^{\phi/2}\Sigma^\gamma\,.
\end{align}
Also, we dropped the contribution from
\begin{equation}
\mathcal{T}:=-\frac{1}{2\pi i}\oint dz \left( z \partial \eta(z)\right)e^{\phi/2}(0)\,,
\end{equation}
because one can treat the above integral as
\begin{align}
\mathcal{T}=&\frac{1}{2\pi i}\oint \frac{dz}{z^{1/2}}\left( \eta(z) e^{\phi}(z) \right) e^{-\phi/2}(0)\,\\
=&\gamma_1 \cdot e^{-\phi/2}\,,
\end{align}
which should vanish in the RR vacuum. We compute
\begin{align}
g_s\overline{\mathcal{X}}_0\left(\mathcal{X}_0 \mathcal{S}_{RR}^{-\frac{3}{2},-\frac{3}{2}}\right)_1:=& \overline{\mathcal{X}}_0\left[\frac{\pi}{24\sqrt{2}\epsilon\kappa_{10}^2} (\partial c+\bar{\partial}\bar{c})c\bar{c}\partial_a B_{bc} F_{def} e^{-\phi/2}  \Sigma_\alpha (\Gamma^{ab}) (P_+^{10}\Gamma^{def}) (\Gamma^c)e^{-3\bar{\phi}/2}\overline{\Sigma}^\beta\right]\,,\\
=&-\frac{\pi}{24\sqrt{2}\epsilon\kappa_{10}^2}\bar{\eta}c\bar{c}\partial_a B_{bc}F_{def} e^{-\phi/2}\Sigma_\alpha (P_+^{10}\Gamma^{ab})(\Gamma^{def})(\Gamma^c) e^{\bar{\phi}/2}\overline{\Sigma}^\beta
\end{align}
where we used the following identity
\begin{align}
\overline{\mathcal{C}}:=& -\frac{1}{2\pi i}\oint \frac{d\bar{z}}{\bar{z}} \bar{\partial}\left(\bar{\eta} \bar{b}e^{2\bar{\phi}}\right)(\bar{z})\left( \frac{\partial c+\bar{\partial}\bar{c}}{2}c\bar{c} e^{-3\bar{\phi}/2}\overline{\Sigma}^\delta\right)(0)\\
=&-\frac{1}{2}\bar{\eta}c\bar{c} e^{\bar{\phi}/2}\overline{\Sigma}^\delta\,.
\end{align}
Next, we compute 
\begin{align}
g_s\overline{\mathcal{X}}_0\left(\mathcal{X}_0 \mathcal{S}_{RR}^{-\frac{3}{2},-\frac{3}{2}}\right)_2:=&\overline{\mathcal{X}}_0\left[\frac{\pi i}{12\epsilon\kappa_{10}^2\sqrt{2\alpha'}} \eta c\bar{c}B_{ab}F_{cde} e^{\phi/2}\Sigma^\gamma (\Gamma^a)(P_+^{10}\Gamma^{cde})(\Gamma^b) e^{-3\bar{\phi}/2}\overline{\Sigma}^\delta\right]\,\\
=&-\frac{\pi}{48\sqrt{2}\epsilon\kappa_{10}^2}\eta c\bar{c} \partial_a B_{bc} F_{def} e^{\phi/2}\Sigma^\gamma (\Gamma^b)(P_+^{10}\Gamma^{def})(\Gamma^{ca}) e^{-\bar{\phi}/2}\overline{\Sigma}_\delta
\end{align}
As a result, we find that the source term coming from the closed string vertex operators is
\begin{align}
\mathcal{X}_0\overline{\mathcal{X}}_0\mathcal{S}_{RR}^{-\frac{3}{2},-\frac{3}{2}}= -\frac{\pi}{24\sqrt{2}g_s\epsilon\kappa_{10}^2}\partial_a B_{bc}F_{def} \Bigl(&\bar{\eta}c\bar{c} e^{-\phi/2}\Sigma_\alpha (\Gamma^{ab})(P_+^{10}\Gamma^{def})(\Gamma^c)e^{\bar{\phi}/2}\overline{\Sigma}^\beta\nonumber\\
&+\eta c\bar{c} e^{\phi/2}\Sigma^\alpha (\Gamma^a)(P_+^{10}\Gamma^{def})(\Gamma^{bc}) e^{-\bar{\phi}/2}\overline{\Sigma}_\beta\Bigr)\,.
\end{align}
As we already saw, the ISD condition forces 
\begin{equation}
H_3\cdot F_3=0\,.
\end{equation}
Furthermore, we have $\partial^a B_{ab}=0.$ Therefore, we only need to consider the fully anti-symmetrized Gamma matrices, which leads to
\begin{align}\label{eqn:closed source in RR}
\mathcal{S}_{RR}^{-\frac{1}{2},-\frac{1}{2}}=&-\frac{\pi}{4\sqrt{2}g_s\epsilon\kappa_{10}^2}\frac{1}{3!}\frac{1}{3!} H_{abc}F_{def} c\bar{c} \biggl(\eta e^{\phi/2}\Sigma^\alpha P_-^{10}\Gamma^{abc} \Gamma^{def} e^{-\bar{\phi}/2}\overline{\Sigma}_\beta\nonumber\\ &\qquad\qquad\qquad\qquad\qquad\qquad-\bar{\eta} e^{-\phi/2}\Sigma_\alpha P_+^{10} \Gamma^{abc}\Gamma^{def}  e^{\bar{\phi}/2} \overline{\Sigma}^\beta\biggr)\,,
\end{align}
where
\begin{equation}
\mathcal{S}_{RR}^{-\frac{1}{2},-\frac{1}{2}}:=g_s^2\Bbb{P}\mathcal{G}[V_{NS,1}^{-1,-1}\otimes V_{R,1}^{-\frac{1}{2},-\frac{1}{2}}]_{S^2}^c\,.
\end{equation}
Using the fact that the orientation of the internal manifold is chosen to be
\begin{equation}
-dY^1\wedge dY^2\wedge dY^3\wedge dY^4\wedge dY^5\wedge dY^6\,,
\end{equation}
\footnote{This unconventional choice of the orientation is equivalent to the conventional choice of the conventional orientation with the flipped sign for the Chern-Simons action.} and
\begin{equation}
-iP_\pm^{10} \tilde{\Gamma}=\pm P_\pm^{10}\Gamma^{0123}\,,
\end{equation}
we rewrite \eqref{eqn:closed source in RR} as
\begin{align}
\mathcal{S}_{RR}^{-\frac{1}{2},-\frac{1}{2}}=-\frac{\pi}{4\sqrt{2}g_s\epsilon\kappa_{10}^2}[H_3\wedge F_3]c\bar{c}\biggl(&\eta e^{\phi/2}\Sigma^\alpha (P_-^{10} \Gamma^{0123})_\alpha^{~\beta}e^{-\bar{\phi}/2}\overline{\Sigma}_\beta\nonumber\\
&+\bar{\eta} e^{-\phi/2}\Sigma_\alpha (P_+^{10}\Gamma^{0123})^\alpha_{~\beta} e^{\bar{\phi}/2}\overline{\Sigma}^\beta\biggr)\,.
\end{align}

We also include the zero-mode contribution from the boundary states
\begin{align}
\frac{1}{\epsilon}([]_{D^2+\Bbb{RP}^2})_0=\frac{\pi \mu_3}{\sqrt{2}g_s\epsilon}\sum_{i}\rho_i\Biggl[&i\sqrt{\alpha'}\partial^q_A (\Bbb{P}\delta^{(6)}(q_i-y_i)) \frac{\partial c+\bar{\partial}\bar{c}}{2}c\bar{c}  (P_+^{10}\Gamma^{0\dots 3} \Gamma^A)^{\alpha\beta} e^{-\phi/2} \Sigma_\alpha e^{-\bar{\phi}/2}\overline{\Sigma}_\beta\nonumber\\
&-\frac{1}{2}(\Bbb{P}\delta^{(6)}(q_i-y_i))\eta c\bar{c}(P_-^{10}\Gamma^{0\dots 3})_{\alpha}^{~\beta} e^{\phi/2}\Sigma^\alpha e^{-\bar{\phi}/2}\overline{\Sigma}_\beta|0\rangle\nonumber\\
&-\frac{1}{2}(\Bbb{P}\delta^{(6)}(q_i-y_i)) \bar{\eta}c\bar{c} (P_+^{10}\Gamma^{0\dots 3})^{\alpha}_{~\beta} e^{-\phi/2}\Sigma_\alpha e^{\bar{\phi}/2} \overline{\Sigma}^\beta|0\rangle\Biggr]\,,
\end{align}
where $\rho_i=1$ for D3-branes and $\rho_i=-1/4$ for O3-planes. Note that the first line is identically zero due to $\Bbb{P}$-projection.

\begin{comment}
The non-trivial RR tadpole is given as, up to an overall factor,
\begin{equation}
\int_{\tilde{Y}} d^{10}X \left(\frac{1}{2\kappa_{10}^2}[H_3\wedge F_3]+\mu_3 \rho_{D3}^{loc}\right)\,,
\end{equation}
which again vanishes for our choice of fluxes.
\end{comment}

It is straightforward to check that the source terms combine to give the $\Bbb{P}$-projected Bianchi identity. Therefore, for the explicit example we consider here, the sum of the source terms in \eqref{eqn:sft eom rr2} identically vanish. We hence take the solution to be
\begin{equation}
W_R^{(2)}=0
\end{equation}
Similar to the case of NSNS sector solution, the non-trivial spacetime profile of the RR fields are contained in the massive RR states $X_R^{(2)}$, which in particular contains the Green's function.

\subsection{Massless tadpoles, ISD, and integrated Bianchi identity}\label{section:tadpole}
In the previous sections, we obtained the solutions to SFEOM to the second order in $\epsilon^{1/2}$-expansion. In doing so, we worked on the F-term locus for the explicit example under consideration, which was essential in several cancellations among the source terms of the SFEOM at the second order.

Now, in this section, we shall investigate if the solvability of SFEOM reproduces the ISD condition and the integrated Bianchi identity that the RHS of \eqref{eqn:Bianchi} integrates to zero.

To investigate this question, we shall proceed as follows. To construct open-closed-unoriented string field theory, we choose a well defined worldsheet BCFT, for a purely geometric background, e.g., a toroidal or a Calabi-Yau orientifold compactification. Crucially, the defining data of the BCFT includes a D-brane configuration which we take to be a certain number of D3-branes localized on the compact directions.\footnote{To construct SFT action, one does not necessarily need to start with a D-brane configuration that cancels the Ramond-Ramond tadpole. The Ramond-Ramond tadpole cancelation will be enforced by the solvability of SFEOM at a higher order.} We also place O3-planes at the fixed planes of the orientifold action. With this open-closed-unoriented SFT, we can now let the complex structure to be large of order $g_s^{-1}$ and turn on a generic choice of quantized $H_3$ and $F_3$ fluxes. Within this set-up, we will now study the solvability of SFEOM perturbatively in $\epsilon$ expansion.

%let us assume that we are still working with the large complex structure moduli of order $g_s^{-1}$ but away from the F-term locus together with a generic spacetime filling D3 and O3 configuration, and a generic choice of constant $H_3$ and $F_3$ fluxes. The question we investigate in this subsection is, does the solvability of SFEOM reproduce ISD condition and the integrated Bianchi identity that the RHS of \eqref{eqn:Bianchi} integrates to zero?

Solvability of SFEOM such as \eqref{eqn:L0 nil NSNS} or \eqref{eqn:sft eom rr2} requires that non-$Q_B$ terms add up to a $Q_B$-exact expression. Typically in the SFT literature (e.g. \cite{Sen:2014dqa,deLacroix:2017lif}), this condition is shown to be equivalent to the requirement that there are no massless tadpoles. It is natural to expect that such a requirement leads to the ISD condition and the integrated Bianchi identity, resulting in constraints on the number of D3-branes and fluxes. However, there is a subtlety in the relation between the solvability of SFEOM and the absence of massless tadpoles when there are boundary terms, as we address now.

$\Bbb{P}$-projected part of SFEOM takes the following general form
\begin{equation}\label{eqn:QBtadpole}
    Q_B W = \Bbb{P}s,
\end{equation}
where $s$ is a linear combination of appropriate string brackets. Upon taking the inner product against a state $\langle v|$ which is $Q_B$-closed and in $\Bbb{P}$ sector, the RHS gives the expression for the massless tadpole
\begin{equation}
    \phi(v)=\langle v| c_0^-|s\rangle,
\end{equation}
while the LHS vanishes \emph{if} the $Q_B$-contour can be deformed from $W$ to $v$ without introducing the boundary term. Such a boundary term is typically dropped in SFT literature and $\phi(v)=0$ is considered equivalent to the existence of the solution $W$. However, the boundary contribution arises ubiquitously when the space of string fields includes nontrivial functions of the target spacetime coordinates.

$Q_B$ acts as the Laplacian $\nabla^2$ on the functions of spacetime coordinates. Therefore, the inner product for the LHS of \eqref{eqn:QBtadpole} is proportional to $\int d^DXf(X)\nabla^2w(X)$ where $f(X)$ is the wavefunction part of $v$ satisfying $\nabla^2f(X)=0$ and $w(X)$ is the wavefunction part of $W$. Even though $\nabla^2f(X)=0$, this does not imply that $\int d^DXf(X)\nabla^2w(X)=0$ since integration by parts may introduce total derivative terms which do not vanish generically.

In our example, the solution $W_{NS}^{(2)}$ is quadratic in the torus coordinates $Y^a$ and $\int d^DXf(X)\nabla^2w(X)$ is clearly nonzero. Therefore, $\phi(v)$ does not necessarily need to vanish. This is the reason why we were able to solve for $W_{NS}^{(2)}$ even when $\phi(v)$ for \eqref{eqn:WNSgh} is nonzero with $v$ being the zero-momentum ghost-dilaton state.\footnote{Similar nonzero massless tadpole can be found in \cite{Cho:2018nfn} for the pp-wave solution with 5-form flux in their equation (3.3), where $v=c\tilde{c}e^{-\phi}e^{-\tilde\phi}\psi^+\tilde{\psi}^+$ produces nonzero $\phi(v)$.}

Nonetheless, there are still $Q_B$-closed $\Bbb{P}$-projected states $v$'s such that $\phi(v)=0$ is required for the SFEOM to be solvable. The inner product for the LHS of \eqref{eqn:QBtadpole}, in addition to the overlap of the wavefunctions, is also proportional to the overlap of the oscillators of $v$ and $c_0^-Q_BW$. Therefore, for $v$'s such that the oscillator overlap is zero, we should still require that $\phi(v)=0$ for \eqref{eqn:QBtadpole} to be solvable.

In general, to obtain massless tadpoles which are required to vanish, one should proceed as follows. The form of the solution $W$ should be assumed first, which in particular requires which oscillators it may contain and which spacetime coordinates its wavefunctions depend on. Then for any $Q_B$-closed $\Bbb{P}$-projected $v$ such that its oscillators does not have an overlap with $c_0^-Q_BW$, we require that $\phi(v)=0$. If this requirement cannot be satisfied, then the solution $W$ with the assumed properties does not exist.

For example, based on the expectation from supergravity and the four dimensional Poincare invariance, we can propose an ansatz that $W_{NS}^{(2)}$ has no oscillators along the non-compact directions and its wavefunction does not depend on the non-compact directions. Then $\phi(v)=0$ for \eqref{eqn:PprojNS} is required for any $Q_B$-closed $\Bbb{P}$-projected $v$ whose oscillators are entirely along the non-compact directions. We take $v$ to be the zero momentum 4d graviton state $v_g=c\bar{c}e^{-\phi}\psi_{\mu}e^{-\bar\phi}\psi_{\nu}$ which leads to 
\begin{equation}\label{eqn:4dgravtad}
    \phi(v_g)\sim \eta_{\mu\nu}\int_{\tilde{Y}}d^{10}X\left(|F|^2+12 \kappa_{10}^2 g_s^{-1}\mu_3\rho_{D3}^{loc}\right)=0.
\end{equation}

For $W_R^{(2)}$, similarly, we can propose an ansatz such that it does not have an overlap with $C_4$ field along the non-compact directions. Then this ansatz leads to the tadpole cancellation condition for $C_4$
\begin{equation}\label{eqn:integBianchi}
\int_{\tilde{Y}} d^{10}X \left(\frac{1}{2\kappa_{10}^2}[H_3\wedge F_3]+\mu_3 \rho_{D3}^{loc}\right)\,=0,
\end{equation}
which is exactly the integrated Bianchi constraint. By combining \eqref{eqn:4dgravtad} and \eqref{eqn:integBianchi}, we obtain
\begin{equation}
    \int_{\tilde Y}F_3\wedge (G_3+i*_6G_3)=0,
\end{equation}
which leads to the ISD condition. Therefore, the absence of appropriate massless tadpoles lead to the constraints which we encountered in supergravity analysis.

\subsection{Concluding remarks}
It is worth noting that even though our choice of the fluxes imply that the F-term potential is minimized, it does not imply that we have checked that the SFT background solution we obtained is supersymmetric. Because the four-dimensional low-energy supergravity analysis predicts that the background must have at least $\mathcal{N}=1$ supersymmetry, it would be important to check if the string field theory analysis indeed confirms that the background preserves supersymmetry. There are several ways to study if the solution is supersymmetric. The most straightforward way is to explicitly construct the supercharges. In SFT, (super-)isometries correspond to gauge transformations which preserve the background solution. Among them, supercharges in particular belong to the R-NS and NS-R sectors. It is well known how to compute them order by order in $g_s$ in SFT literature (see e.g. \cite{Sen:2015uoa,deLacroix:2017lif}), which is very similar to solving for the string spectrum around a background solution.

Another important point worth emphasizing is that the tadpoles due to localized sources in SFT are smeared. In SFT, we have formulated the tadpole as the obstruction to solving SFEOM. As the components of SFEOM which are projected out by $\Bbb{P}$ can be solved always, the only obstruction to solving SFEOM arises in the $L_0^+$-nilpotent sector of SFEOM i.e. sector surviving the $\Bbb{P}$-projection. Because, the $\Bbb{P}$-projection picks out the ``constant" mode of the source terms in SFEOM, for example,
\begin{equation}
    \Bbb{P}\delta^{(6)}(X-Y)=\frac{1}{\text{Vol}}\,,
\end{equation}
the localized sources enter the $L_0^+$-nilpotent components of SFEOM in smeared forms. Hence, the relevant physical tadpole in SFT is the smeared tadpole. 

Finally, we remark that the form of the solutions $X_{NS}^{(2)},$ $W_{NS}^{(2)},$ $X_{R}^{(2)},$ and $W_{R}^{(2)}$ are the fully string theoretic version of the supergravity solutions \eqref{eqn:supergravitysolm} and \eqref{eqn:supergravitysolp}, where the massless string field profiles are expected to agree with the supergravity under nontrivial field redefinitions \cite{Hull:2009mi}. We can also solve for $\tilde{\Psi}^c$ using \eqref{eqn:eom01}, and we denote the corresponding solution by $\tilde{\Psi}^c_0$. We can now expand the 1PI effective action \eqref{eqn:SFT1PIaction} around the order $\epsilon$ background solution $\Psi_0^c$ and $\tilde{\Psi}_0^c$ we obtained here by taking $\Psi^c=\Psi^c_0+\varphi^c$, where $\varphi^c$ now stands for the dynamical string fields around the background given by $\Psi_0^c$:
\begin{equation}
S_{1PI}[\Psi^c,\tilde{\Psi}^c,\Psi^o,\tilde{\Psi}^o]=S_{1PI}[\Psi^c_0+\varphi^c,\tilde{\Psi}^c_0+\tilde{\varphi}^c,\Psi^o,\tilde{\Psi}^o]=S'_{1PI}[\varphi^c,\tilde{\varphi}^c,\Psi^c_0,\tilde{\Psi}^c_0,\Psi^o,\tilde{\Psi}^o].
\end{equation}
The new 1PI effective action $S'_{1PI}[\varphi^c,\tilde{\varphi}^c,\Psi^c_0,\tilde{\Psi}^c_0,\Psi^o,\tilde{\Psi}^o]$ is now free from tadpoles of $\varphi^c$ and $\tilde{\varphi}^c$ to order $\epsilon$. Discussion on the consistency of $S'_{1PI}[\varphi^c,\tilde{\varphi}^c,\Psi^c_0,\tilde{\Psi}^c_0,\Psi^o,\tilde{\Psi}^o]$ as a classical BV master action can be found for example in \cite{deLacroix:2017lif}. Solutions to the linearized equations of motions for $S'_{1PI}$ correspond to the string spectrum to order $\epsilon$. Furthermore, $S'_{1PI}$ provides Feynman rules for the background described by $\Psi^c_0$ and $\tilde{\Psi}^c_0$ up to order $\epsilon$, from which observables such as string amplitudes may be obtained. We emphasize that $S'_{1PI}$ is the ``worldsheet" description of the new background and produces genuinely stringy results. Explicit computation of such physical quantities will be carried out in the future publications.

\section{Conclusions}\label{sec:conclusions}
In summary, we have demonstrated that SFT provides a suitable framework for investigating certain flux compactification backgrounds. In the context of our particular example, inverse of complex structure moduli can be treated as expansion parameters equivalent to the string coupling, and we detailed the systematic expansion process within SFT. The construction of the background solution order by order in string coupling highlights how SFT is not much different from string perturbation theory in practice when a good expansion parameter aligns with it.

This work points towards several promising directions for future research. First, it will be important to obtain higher-order solutions in our specific example. The anticipation of an SFT solution existing to all orders in string coupling is inherent, and its form is expected to resemble that of the supergravity solution closely. The pursuit of this higher-order solution becomes essential when aiming to compute physical observables at those orders. In parallel, there arises the related task of computing the supercharges associated with the background at these higher orders. The connection between the presence of supersymmetry and the anticipation of a solution persisting to all orders in string coupling is an intriguing avenue to explore. It will be interesting to investigate the consequences of imposing the requirement that the background solution respects specific supersymmetries at all orders.\footnote{In \cite{Sen:2015uoa}, the problem on massless tadpoles and the supersymmetry restoration to the first few orders in $g_s$ has been studied. An explicit example of SO(32) heterotic string theory on Calabi-Yau 3-folds was considered.} Additionally, it is worth noting that the open string field profile becomes nontrivial at higher orders, offering insight into how D-branes back-react to the bulk geometry.

One of the most important tasks is the study of physical observables associated with the flux compactification backgrounds. While we have described the background solution in this work, we have not delved into its physical quantities, such as spectra and scattering processes. SFT offers a rigorous and consistent platform for conducting such computations, unveiling genuinely stringy effects that remain inaccessible through low-energy effective descriptions.

One notable physical observable that can be investigated using the background solution presented in this study pertains to the order $g_s$ correction to the string spectrum. While all massive modes are anticipated to decay owing to their interactions with massless modes, the emergence of an imaginary part in the pole of the propagator, as a result of decay effects, occurs at order $g_s^2$. Consequently, it becomes meaningful to determine the order $g_s$ correction to the masses of various excitations, which will also be needed to study decay rates at higher orders. In practice, beyond the contributions from closed string background solution profiles, additional contributions to the mass matrix arise from the disk and $\Bbb{RP}^2$, introducing a novel facet specific to the flux compactification example under consideration here.

An intriguing and relatively straightforward class of excitations to explore involves the leading Regge trajectory of the non-compact four dimensions. These excitations exhibit trivial string brackets with both the first-order NSNS and RR 3-form fluxes, as well as the second-order RR 5-form flux. Consequently, the analysis of the mass matrix becomes particularly simple. Additionally, it will be interesting to study how specific linear combinations of complex structure and axio-dilaton moduli are lifted, aligning with the insights from the superpotential analysis.

Another significant aspect worth investigating is the scattering of massless modes. SFT lets one compute corrections to the conventional Virasoro-Shapiro amplitude. For example, the computation of the Virasoro-Shapiro amplitude involving moduli fields determines the curvature of the moduli space, or equivalently, in $\mathcal{N}=1$ compactifications, the K\"{a}hler potential of the moduli space. Despite the importance of the computation of the K\"{a}hler potential at higher order in $g_s$ and $\alpha',$ such computations were only carried out in fluxless compactifications at the leading order in the backreaction caused by D-branes \cite{Kiritsis:1994ta,Antoniadis:1997eg,Berg:2005ja,Berg:2011ij,Berg:2014ama,Haack:2015pbv,Haack:2018ufg,Kim:2023sfs,Kim:2023eut}. The challenge to generalize such computations to non-trivial RR flux backgrounds can be overcome with the machinery built in this paper. It would be very interesting to compute $\alpha',$ $g_s,$ and flux corrections to the K\"{a}hler potential of the moduli space in flux compactifications.\footnote{To carry out such computations in string field theory, one may need string vertices for higher vertices and genera. Interesting developments on the related issues happened recently, see, for example, \cite{Firat:2021ukc,Erbin:2022rgx,Firat:2023glo,Erbin:2023hcs,Firat:2023suh}.}

We would also like to remark that the string field solution in this work was expressed in a local chart of the torus. As far as we are aware, it has not been explored if SFT allows for a systematic patch-by-patch description like general relativity. Nonetheless, we expect that physical observables are independent of the choice of local charts of the target spacetime. In the follow-up works, we plan to explicitly compute physical observables metioned above and check this independence.

While our discussion in this work primarily focused on string perturbation theory, it is important to note that SFT has demonstrated remarkable power in the recent exploration of non-perturbative D-instanton effects \cite{Sen:2019qqg,Sen:2020cef,Sen:2020eck,Sen:2021qdk,Sen:2021tpp,Sen:2021jbr,Eniceicu:2022nay,Alexandrov:2022mmy,Agmon:2022vdj,Eniceicu:2022dru,Chakravarty:2022cgj,Eniceicu:2022xvk,Alexandrov:2021shf,Alexandrov:2021dyl}. The suitability of SFT for D-instantons arises from its ability to systematically integrate out open string fields on the D-instanton, resulting in the effective field theory for the dynamic closed string fields. This process relies on crucial insights from field theory.

As an illustration of the potential of SFT in this context, in \cite{Alexandrov:2022mmy} the overall numerical factor of the non-perturbative superpotential was fixed by regulating the IR divergence arising from the zero modes of Euclidean D-branes appearing in fluxless orientifold compactifications. Shortly after, the massive state contribution to the one-loop pfaffian was shown to be determined by open-unoriented topological string partition function at one-loop \cite{Kim:2023cbh}. Extending this line of inquiry to examine how Euclidean D-branes contribute to the superpotential in non-trivial flux backgrounds holds the promise of shedding light on the intricate roles played by non-perturbative effects in moduli stabilization.

We have investigated one of the simplest instances of flux compactification amenable to the direct application of SFT. It is essential to highlight that there appear to be no immediate obstructions to extending the same methodologies to other examples such as Calabi-Yau orientifold compactifications. While it might be necessary to introduce additional expansion parameters, such as $\alpha’$, alongside the string coupling, this double expansion can also be systematically explored within the framework of SFT. One may also introduce D7-branes and O7-planes, provided their configurations are such that the dilaton profile does not develop singularity. It is important to acknowledge that while SFT may not be universally applicable to all cases of interest, it unquestionably opens up new perspectives for the examination of flux compactifications and more general string backgrounds.

One additional speculative consideration pertains to the potential use of SFT in exploring time-dependent backgrounds. In this context, one would primarily seek time-dependent solutions to SFEOM. However, it becomes imperative to ensure that these solutions allow for a meaningful expansion in string coupling, which may itself be time-dependent. Equally important is the determination of whether the relevant physical observables can be unambiguously computed using the ingredients of SFT.

Notably, SFT has already demonstrated its efficacy in elucidating time-dependent backgrounds in the case of open string rolling tachyons \cite{Sen:2002nu,Sen:2002in,Sen:2004zm,Sen:2004cq,Sen:2004yv,Sen:2004nf,Cho:2023khj}. In this case, open-closed SFT provides an explanation for how rigid gauge transformations of the closed string background correspond to symmetry transformations on the open string fields, thus validating the energy of the rolling tachyon solution originally proposed in \cite{Sen:2002nu}. Nevertheless, the applicability of such success to other intriguing time-dependent closed string backgrounds remains a subject of future investigation.

\section*{Acknowledgements}
The work of MC is supported by the Sam B. Treiman fellowship at the Princeton Center for Theoretical Science. The work of MK was supported by a Pappalardo fellowship. We would like to especially thank Barton Zwiebach for extended discussions on the background independence and the ghost-dilaton theorem in string field theory, and Ashoke Sen for extended discussions on the ghost-dilaton coupling and field redefinitions in string field theory. We thank Atakan Hilmi F\i rat, Carlo Maccaferri, and Marcus Berg for useful discussions on related topics. We thank Jakob Moritz, Richard Nally, Michael Haack, Ashoke Sen, Barton Zwiebach, and Xi Yin for their comments on the draft.
\appendix
\newpage
\section{Toroidal compactifications}\label{app:tor}
In this section, we present a simple toroidal compactification with the explicit complex structure dependence of the metric. Let us consider type IIB compactification on $T^6.$ Most of the discussion here will also apply to toroidal orbifold compactifications. We shall focus on a rather special sublocus of the moduli space such that one can treat $T^6$ as $T^2\times T^2\times T^2\,,$ with each of $T^2$ has a complex structure modulus and a K\"{a}hler modulus. 

We shall take the coordinates of $T^2_i$ to be $Y^{2i-1}$ and $Y^{2i}$ and let $Y^j$ to take values in $[0,1].$ Having a fixed domain for the torus will make it easier to discuss moduli fields and their associated vertex operators. We take the line element of a torus $T^2_i$ to be, in the \emph{string-frame}, 
\begin{equation}
ds^2=\frac{\im t_i}{\im u_i}\left((dY^{2i-1})^2+2 \re u_i dY^{2i-1}dY^{2i}+|u_i|^2 (dY^{2i})^2\right)\,,
\end{equation}
and we define the complex coordinate $Z^i$ as
\begin{equation}
dZ^i=dY^{2i-1}+u_i dY^{2i}\,.
\end{equation}
We shall choose the orientation of the torus such that the volume form is written as
\begin{equation}
-\im t_i dY^{2i-1}\wedge dY^{2i}\,. 
\end{equation}
This choice of orientation is equivalent to the convention of \cite{Kachru:2002he}. We have the following relations
\begin{equation}
dY^{2i}= \frac{1}{2i \im u_i}(dZ^i-d\bar{Z}^i)\,,
\end{equation}
\begin{equation}
dY^{2i-1}=\frac{1}{2i\im u_i}\left( u_i d\bar{Z}^i-\bar{u}_i dZ^i\right)\,,
\end{equation}
where the index $i$ is not summed over. Also, note that we have
\begin{equation}
dY^{2i-1}\wedge dZ^i=-\frac{u_i}{2i\im u_i}dZ^i\wedge d\overline{Z}^i\,,
\end{equation}
and
\begin{equation}
dY^{2i}\wedge dZ^i=\frac{1}{2i\im u_i}dZ^i\wedge d\overline{Z}^i\,.
\end{equation}
Therefore, we can write the volume form as
\begin{equation}
d\text{Vol}_{T^2_i}=-\im t_i dY^{2i-1}\wedge dY^{2i}=\frac{\im t_i}{2i\im u_i} dZ^i \wedge d\bar{Z}^i\,.
\end{equation}
Phrased differently, we define the K\"{a}hler form $J^i$ as
\begin{equation}
J^i:= \frac{\im t_i}{2i \im u_i}dZ^i\wedge d\overline{Z}^i\,.
\end{equation}
Similarly, we define the holomorphic threeform of the six torus as
\begin{equation}
\Omega:=d Z^1\wedge dZ^2\wedge dZ^3\,,
\end{equation}
which results in
\begin{equation}
-\frac{i}{8}\int_{T^6} \Omega\wedge\overline{\Omega}= \im u_1\im u_2\im u_3\,.
\end{equation}

It is instructive to consider the inner product between differential forms. Let us consider a one-form $F=d Y^{2i-1}.$ Then, its norm is
\begin{equation}
|F|^2=\frac{\im u_i}{\im t_i}\,.
\end{equation} 
Similarly, for a one-form $H=dY^{2i},$ the norm is
\begin{equation}
|H|^2=\frac{\im u_i}{|u_i|^2\im t_i}\,.
\end{equation}
So, we have
\begin{equation}
\int_{T_i^2} F\wedge \star_{2i}F=\im u_i\,,\int_{T_i^2}F\wedge H=1\,,\int_{T_i^2}H\wedge\star_{2i}H=\frac{\im u_i}{|u_i|^2}\,.
\end{equation}

\section{Spinor conventions for toroidal compactifications}\label{app:spin}
In this section, shall set the conventions for the spinor fields and the gamma matrices. The spinor group decomes into $Spin(1,3)\times Spin(6),$ where $Spin(6)\equiv SU(4).$ We shall decompose the 16 components spinor into 
\begin{equation}
16=(2,4)\oplus (\bar{2},\bar{4})\,,
\end{equation}
where $2$ is an irrep of $Spin(1,3)$ and $4$ is an irrep of $SU(4).$ For the four-dimensional spinor indices, we shall use $\alpha$ and $\dot{\alpha}$ and for the 6 and 10 dimensional spinor indices we shall use $\alpha^{(6)}$ and $\alpha^{(10)},$ respectively. We write a six-dimensional Gamma matrix
\begin{equation}
\tilde{\Gamma}^i\,,
\end{equation}
where $i=1,\dots,6.$ We additionally define
\begin{equation}
\tilde{\Gamma}=i \prod_{i=1}^6 \tilde{\Gamma}^i\,,
\end{equation}
which satisfies
\begin{equation}
\tilde{\Gamma}^2=I_4\,,\quad \{\tilde{\Gamma},\tilde{\Gamma}^i\}=0\,,\quad \{\tilde{\Gamma}^a,\tilde{\Gamma}^b\}=g^{ab} I_4\,,
\end{equation}
where $I_4$ is an eight-dimensional identity matrix. We then now write the ten-dimensional Gamma matrices
\begin{equation}
\Gamma^\mu= \gamma^\mu\otimes \tilde{\Gamma}\,,\quad \Gamma^a=I_2\otimes \tilde{\Gamma}^a\,,
\end{equation}
where $\gamma^\mu$ is a four-dimensional Gamma matrix that satisfies
\begin{equation}
\{\gamma^\mu,\gamma^\nu\}=\eta^{\mu\nu}I_2\,.
\end{equation}

We introduce auxiliary matrices $\overline{\Gamma}^i,$ which can be thought of as an $8\times 8$ component 6d gamma matrix. We define $\overline{\Gamma}^i$ as
\begin{equation}
\overline{\Gamma}^1= \sigma_1\otimes \sigma_3\otimes I_2 \,,\quad \overline{\Gamma}^2=\sigma_2\otimes \sigma_3\otimes I_2\,, \quad\overline{\Gamma}^3=I_2\otimes \sigma_1\otimes \sigma_3\,,
\end{equation}
\begin{equation}
\overline{\Gamma}^4=I_2\otimes \sigma_2\otimes \sigma_3\,,\quad \overline{\Gamma}^5=\sigma_3\otimes I_2\otimes \sigma_1\,,\quad\overline{\Gamma}^6=\sigma_3\otimes I_2\otimes \sigma_2\,,
\end{equation}
and
\begin{equation}
\overline{\Gamma}:=i \prod_{i=1}^6\overline{\Gamma}^i=-\sigma_3\otimes\sigma_3\otimes \sigma_3\,,
\end{equation}
where $\sigma_i$ are Pauli matrices. Then, we have
\begin{equation}
e^j_i(\tilde{\Gamma}^i)_{\alpha^{(6)}}^{~\dot{\beta}^{(6)}}=\overline{P}_+ \overline{\Gamma}^j \overline{P}_-\,,\quad e^j_i(\tilde{\Gamma}^i)_{\dot{\alpha}^{(6)}}^{~\beta^{(6)}}=\overline{P}_- \overline{\Gamma}^j \overline{P}_+\,,
\end{equation}
\begin{equation}
\overline{P}_{\pm}\overline{\Gamma}^i \overline{P}_{\pm}=\overline{P}_{\pm}\overline{P}_{\mp}\overline{\Gamma}^i=0\,,
\end{equation}
where we defined
\begin{equation}
\overline{P}_\pm=\frac{I_8+\overline{\Gamma}}{2}\,.
\end{equation}
Similarly, we define $\Gamma^{10}$ as
\begin{equation}
\Gamma^{10}:=-i\Gamma^0\dots \Gamma^3\overline{\Gamma}\,,
\end{equation}
and $P_\pm^{10}$ as
\begin{equation}
P_\pm^{10}:=\frac{1+\Gamma^{10}}{2}\,.
\end{equation}

We also introduce complex Gamma matrices
\begin{equation}
\tilde{\Gamma}^a_\Bbb{C}=\tilde{\Gamma}^{2a-1}+u_a\tilde{\Gamma}^{2a}\,,\quad \tilde{\Gamma}_\Bbb{C}^{\bar{a}}=\tilde{\Gamma}^{2a-1}+\bar{u}_a\tilde{\Gamma}^{2a}\,.
\end{equation}
Then, 
\begin{equation}
\tilde{\Gamma}^a_+:=\sqrt{\im t_a}\tilde{\Gamma}^a_\Bbb{C}\,,
\end{equation}
and
\begin{equation}
\tilde{\Gamma}^a_-:=\sqrt{\im t_a}\tilde{\Gamma}^{\bar{a}}_\Bbb{C}\,,
\end{equation}
act as lowering and raising operators for the 6d spinors. 

We can now construct irreps of the internal spinors. Because of the representation of the gamma matrices we chose, it is convenient to consider a subgroup $SU(3)\times U(1)\subset SU(4),$ and represent the irreducible representations of $SU(4)$ as a linear combination of irreducible representations of $SU(3)\times U(1).$ The representation under $SU(3)$ will be classified how the spinor transforms under the natural $SU(3)$ rotation, and the $U(1)$ charge of the fermion will be given by
\begin{equation}
Q=2 \sum_a\left(\tilde{\Gamma}_+^a\tilde{\Gamma}_-^a-1/2\right)\,.
\end{equation}

Let us start by constructing the lowest weight representation $\eta_-$ that satisfies
\begin{equation}
\tilde{\Gamma}^a_-\eta_-=0\,,
\end{equation}
for all $a.$ $\eta_-$ is $1_{-3}$ under $SU(3)_{U(1)}.$ Acting one raising operator, we find 
\begin{equation}
v_a\tilde{\Gamma}^a_+\eta_-\,,
\end{equation}
which is in $3_{-1}$ under $SU(3)_{U(1)}.$ Acting the raising operators twice, we find
\begin{equation}
\frac{1}{2!}f_{abc} \tilde{\Gamma}^b_+\tilde{\Gamma}^c_+\eta_-\,,
\end{equation}
which is in $\bar{3}_{1}$ under $SU(3)_{U(1)}.$ Finally, we find
\begin{equation}
\eta_+:= \frac{1}{3!}\epsilon_{abc} \tilde{\Gamma}^a_+\tilde{\Gamma}^b_+\tilde{\Gamma}^c_+\eta_-\,,
\end{equation}
which is in $1_3$ under $SU(3)_{U(1)}.$ Then, we have
\begin{equation}
4=1_{3}\oplus 3_{-1}\,,\quad \bar{4}=1_{-3}\oplus \bar{3}_1\,. 
\end{equation}
We then find
\begin{equation}
\Omega_{abc}=\im u_1\im u_2\im u_3\overline{\eta}_+\tilde{\Gamma}^a_+\tilde{\Gamma}^b_+\tilde{\Gamma}^c_+\eta_-\,,
\end{equation}
for the covariantly constant spinor
\begin{equation}
\overline{\eta}_\pm\eta_{\pm}=1\,.
\end{equation}
Note that we defined
\begin{equation}
\overline{\eta}_\pm:=\eta^\dagger\,.
\end{equation}
\newpage
\bibliographystyle{JHEP}
\bibliography{refs}

\providecommand{\href}[2]{#2}\begingroup\raggedright\begin{thebibliography}{100}

\bibitem{SupernovaSearchTeam:1998fmf}
{\scshape Supernova Search Team} collaboration, A.~G. Riess et~al.,
  \emph{{Observational evidence from supernovae for an accelerating universe
  and a cosmological constant}},
  \href{http://dx.doi.org/10.1086/300499}{\emph{Astron. J.} {\bf 116} (1998)
  1009--1038}, [\href{http://arxiv.org/abs/astro-ph/9805201}{{\tt
  astro-ph/9805201}}].

\bibitem{SupernovaCosmologyProject:1998vns}
{\scshape Supernova Cosmology Project} collaboration, S.~Perlmutter et~al.,
  \emph{{Measurements of $\Omega$ and $\Lambda$ from 42 high redshift
  supernovae}}, \href{http://dx.doi.org/10.1086/307221}{\emph{Astrophys. J.}
  {\bf 517} (1999) 565--586},
  [\href{http://arxiv.org/abs/astro-ph/9812133}{{\tt astro-ph/9812133}}].

\bibitem{Matorras:2022xel}
P.~Matorras, \emph{{Supersymmetry searches in ATLAS and CMS}},
  \href{http://dx.doi.org/10.22323/1.406.0076}{\emph{PoS} {\bf CORFU2021}
  (2022) 076}.

\bibitem{Maldacena:1997re}
J.~M. Maldacena, \emph{{The Large N limit of superconformal field theories and
  supergravity}},
  \href{http://dx.doi.org/10.4310/ATMP.1998.v2.n2.a1}{\emph{Adv. Theor. Math.
  Phys.} {\bf 2} (1998) 231--252},
  [\href{http://arxiv.org/abs/hep-th/9711200}{{\tt hep-th/9711200}}].

\bibitem{Aharony:1999ti}
O.~Aharony, S.~S. Gubser, J.~M. Maldacena, H.~Ooguri and Y.~Oz, \emph{{Large N
  field theories, string theory and gravity}},
  \href{http://dx.doi.org/10.1016/S0370-1573(99)00083-6}{\emph{Phys. Rept.}
  {\bf 323} (2000) 183--386}, [\href{http://arxiv.org/abs/hep-th/9905111}{{\tt
  hep-th/9905111}}].

\bibitem{Kachru:2003aw}
S.~Kachru, R.~Kallosh, A.~D. Linde and S.~P. Trivedi, \emph{{De Sitter vacua in
  string theory}},
  \href{http://dx.doi.org/10.1103/PhysRevD.68.046005}{\emph{Phys. Rev. D} {\bf
  68} (2003) 046005}, [\href{http://arxiv.org/abs/hep-th/0301240}{{\tt
  hep-th/0301240}}].

\bibitem{Balasubramanian:2005zx}
V.~Balasubramanian, P.~Berglund, J.~P. Conlon and F.~Quevedo,
  \emph{{Systematics of moduli stabilisation in Calabi-Yau flux
  compactifications}},
  \href{http://dx.doi.org/10.1088/1126-6708/2005/03/007}{\emph{JHEP} {\bf 03}
  (2005) 007}, [\href{http://arxiv.org/abs/hep-th/0502058}{{\tt
  hep-th/0502058}}].

\bibitem{DeWolfe:2005uu}
O.~DeWolfe, A.~Giryavets, S.~Kachru and W.~Taylor, \emph{{Type IIA moduli
  stabilization}},
  \href{http://dx.doi.org/10.1088/1126-6708/2005/07/066}{\emph{JHEP} {\bf 07}
  (2005) 066}, [\href{http://arxiv.org/abs/hep-th/0505160}{{\tt
  hep-th/0505160}}].

\bibitem{Silverstein:2004id}
E.~Silverstein, \emph{{TASI / PiTP / ISS lectures on moduli and microphysics}},
   in \emph{{Theoretical Advanced Study Institute in Elementary Particle
  Physics (TASI 2003): Recent Trends in String Theory}}, pp.~381--415, 5, 2004.
\newblock \href{http://arxiv.org/abs/hep-th/0405068}{{\tt hep-th/0405068}}.
\newblock \href{http://dx.doi.org/10.1142/9789812775108_0004}{DOI}.

\bibitem{Grana:2005jc}
M.~Grana, \emph{{Flux compactifications in string theory: A Comprehensive
  review}}, \href{http://dx.doi.org/10.1016/j.physrep.2005.10.008}{\emph{Phys.
  Rept.} {\bf 423} (2006) 91--158},
  [\href{http://arxiv.org/abs/hep-th/0509003}{{\tt hep-th/0509003}}].

\bibitem{Douglas:2006es}
M.~R. Douglas and S.~Kachru, \emph{{Flux compactification}},
  \href{http://dx.doi.org/10.1103/RevModPhys.79.733}{\emph{Rev. Mod. Phys.}
  {\bf 79} (2007) 733--796}, [\href{http://arxiv.org/abs/hep-th/0610102}{{\tt
  hep-th/0610102}}].

\bibitem{Friedan:1985ge}
D.~Friedan, E.~J. Martinec and S.~H. Shenker, \emph{{Conformal Invariance,
  Supersymmetry and String Theory}},
  \href{http://dx.doi.org/10.1016/0550-3213(86)90356-1}{\emph{Nucl. Phys. B}
  {\bf 271} (1986) 93--165}.

\bibitem{Friedan:1985ey}
D.~Friedan, S.~H. Shenker and E.~J. Martinec, \emph{{Covariant Quantization of
  Superstrings}},
  \href{http://dx.doi.org/10.1016/0370-2693(85)91466-2}{\emph{Phys. Lett. B}
  {\bf 160} (1985) 55--61}.

\bibitem{Witten:2012bh}
E.~Witten, \emph{{Superstring Perturbation Theory Revisited}},
  \href{http://arxiv.org/abs/1209.5461}{{\tt 1209.5461}}.

\bibitem{Witten:2013cia}
E.~Witten, \emph{{More On Superstring Perturbation Theory: An Overview Of
  Superstring Perturbation Theory Via Super Riemann Surfaces}},
  \href{http://arxiv.org/abs/1304.2832}{{\tt 1304.2832}}.

\bibitem{Sen:2015hia}
A.~Sen and E.~Witten, \emph{{Filling the gaps with PCO\textquoteright{}s}},
  \href{http://dx.doi.org/10.1007/JHEP09(2015)004}{\emph{JHEP} {\bf 09} (2015)
  004}, [\href{http://arxiv.org/abs/1504.00609}{{\tt 1504.00609}}].

\bibitem{Berenstein:1999jq}
D.~Berenstein and R.~G. Leigh, \emph{{Superstring perturbation theory and
  Ramond-Ramond backgrounds}},
  \href{http://dx.doi.org/10.1103/PhysRevD.60.106002}{\emph{Phys. Rev. D} {\bf
  60} (1999) 106002}, [\href{http://arxiv.org/abs/hep-th/9904104}{{\tt
  hep-th/9904104}}].

\bibitem{Berenstein:1999ip}
D.~Berenstein and R.~G. Leigh, \emph{{Quantization of superstrings in
  Ramond-Ramond backgrounds}},
  \href{http://dx.doi.org/10.1103/PhysRevD.63.026004}{\emph{Phys. Rev. D} {\bf
  63} (2001) 026004}, [\href{http://arxiv.org/abs/hep-th/9910145}{{\tt
  hep-th/9910145}}].

\bibitem{Berkovits:1999im}
N.~Berkovits, C.~Vafa and E.~Witten, \emph{{Conformal field theory of AdS
  background with Ramond-Ramond flux}},
  \href{http://dx.doi.org/10.1088/1126-6708/1999/03/018}{\emph{JHEP} {\bf 03}
  (1999) 018}, [\href{http://arxiv.org/abs/hep-th/9902098}{{\tt
  hep-th/9902098}}].

\bibitem{Berkovits:1999in}
N.~Berkovits, \emph{{Quantization of the superstring with manifest U(5)
  superPoincare invariance}},
  \href{http://dx.doi.org/10.1016/S0370-2693(99)00548-1}{\emph{Phys. Lett. B}
  {\bf 457} (1999) 94--100}, [\href{http://arxiv.org/abs/hep-th/9902099}{{\tt
  hep-th/9902099}}].

\bibitem{Berkovits:2000fe}
N.~Berkovits, \emph{{Super Poincare covariant quantization of the
  superstring}},
  \href{http://dx.doi.org/10.1088/1126-6708/2000/04/018}{\emph{JHEP} {\bf 04}
  (2000) 018}, [\href{http://arxiv.org/abs/hep-th/0001035}{{\tt
  hep-th/0001035}}].

\bibitem{Berkovits:2000ph}
N.~Berkovits and B.~C. Vallilo, \emph{{Consistency of superPoincare covariant
  superstring tree amplitudes}},
  \href{http://dx.doi.org/10.1088/1126-6708/2000/07/015}{\emph{JHEP} {\bf 07}
  (2000) 015}, [\href{http://arxiv.org/abs/hep-th/0004171}{{\tt
  hep-th/0004171}}].

\bibitem{Berkovits:2000nn}
N.~Berkovits, \emph{{Cohomology in the pure spinor formalism for the
  superstring}},
  \href{http://dx.doi.org/10.1088/1126-6708/2000/09/046}{\emph{JHEP} {\bf 09}
  (2000) 046}, [\href{http://arxiv.org/abs/hep-th/0006003}{{\tt
  hep-th/0006003}}].

\bibitem{Berkovits:2001us}
N.~Berkovits, \emph{{Relating the RNS and pure spinor formalisms for the
  superstring}},
  \href{http://dx.doi.org/10.1088/1126-6708/2001/08/026}{\emph{JHEP} {\bf 08}
  (2001) 026}, [\href{http://arxiv.org/abs/hep-th/0104247}{{\tt
  hep-th/0104247}}].

\bibitem{Berkovits:1994wr}
N.~Berkovits, \emph{{Covariant quantization of the Green-Schwarz superstring in
  a Calabi-Yau background}},
  \href{http://dx.doi.org/10.1016/0550-3213(94)90106-6}{\emph{Nucl. Phys. B}
  {\bf 431} (1994) 258--272}, [\href{http://arxiv.org/abs/hep-th/9404162}{{\tt
  hep-th/9404162}}].

\bibitem{Berkovits:2001tg}
N.~Berkovits, S.~Gukov and B.~C. Vallilo, \emph{{Superstrings in 2-D
  backgrounds with RR flux and new extremal black holes}},
  \href{http://dx.doi.org/10.1016/S0550-3213(01)00413-8}{\emph{Nucl. Phys. B}
  {\bf 614} (2001) 195--232}, [\href{http://arxiv.org/abs/hep-th/0107140}{{\tt
  hep-th/0107140}}].

\bibitem{Linch:2006ig}
W.~D. Linch, III and B.~C. Vallilo, \emph{{Hybrid formalism, supersymmetry
  reduction, and Ramond-Ramond fluxes}},
  \href{http://dx.doi.org/10.1088/1126-6708/2007/01/099}{\emph{JHEP} {\bf 01}
  (2007) 099}, [\href{http://arxiv.org/abs/hep-th/0607122}{{\tt
  hep-th/0607122}}].

\bibitem{Kappeli:2006fj}
J.~Kappeli, S.~Theisen and P.~Vanhove, \emph{{Hybrid formalism and topological
  amplitudes}},  in \emph{{15th International Congress on Mathematical
  Physics}}, 7, 2006.
\newblock \href{http://arxiv.org/abs/hep-th/0607021}{{\tt hep-th/0607021}}.

\bibitem{Linch:2008rw}
W.~D. Linch, III, J.~McOrist and B.~C. Vallilo, \emph{{Type IIB Flux Vacua from
  the String Worldsheet}},
  \href{http://dx.doi.org/10.1088/1126-6708/2008/09/042}{\emph{JHEP} {\bf 09}
  (2008) 042}, [\href{http://arxiv.org/abs/0804.0613}{{\tt 0804.0613}}].

\bibitem{Beisert:2010jr}
N.~Beisert et~al., \emph{{Review of AdS/CFT Integrability: An Overview}},
  \href{http://dx.doi.org/10.1007/s11005-011-0529-2}{\emph{Lett. Math. Phys.}
  {\bf 99} (2012) 3--32}, [\href{http://arxiv.org/abs/1012.3982}{{\tt
  1012.3982}}].

\bibitem{SEN1990551}
A.~Sen, \emph{On the background independence of string field theory},
  \href{http://dx.doi.org/https://doi.org/10.1016/0550-3213(90)90400-8}{\emph{Nuclear
  Physics B} {\bf 345} (1990) 551--583}.

\bibitem{Sen:1993kb}
A.~Sen and B.~Zwiebach, \emph{{Quantum background independence of closed string
  field theory}},
  \href{http://dx.doi.org/10.1016/0550-3213(94)90145-7}{\emph{Nucl. Phys. B}
  {\bf 423} (1994) 580--630}, [\href{http://arxiv.org/abs/hep-th/9311009}{{\tt
  hep-th/9311009}}].

\bibitem{Bergman:1994qq}
O.~Bergman and B.~Zwiebach, \emph{{The Dilaton theorem and closed string
  backgrounds}},
  \href{http://dx.doi.org/10.1016/0550-3213(95)00022-K}{\emph{Nucl. Phys. B}
  {\bf 441} (1995) 76--118}, [\href{http://arxiv.org/abs/hep-th/9411047}{{\tt
  hep-th/9411047}}].

\bibitem{Belopolsky:1995vi}
A.~Belopolsky and B.~Zwiebach, \emph{{Who changes the string coupling?}},
  \href{http://dx.doi.org/10.1016/0550-3213(96)00203-9}{\emph{Nucl. Phys. B}
  {\bf 472} (1996) 109--138}, [\href{http://arxiv.org/abs/hep-th/9511077}{{\tt
  hep-th/9511077}}].

\bibitem{Sen:2017szq}
A.~Sen, \emph{{Background Independence of Closed Superstring Field Theory}},
  \href{http://dx.doi.org/10.1007/JHEP02(2018)155}{\emph{JHEP} {\bf 02} (2018)
  155}, [\href{http://arxiv.org/abs/1711.08468}{{\tt 1711.08468}}].

\bibitem{Sen:2014dqa}
A.~Sen, \emph{{Gauge Invariant 1PI Effective Action for Superstring Field
  Theory}}, \href{http://dx.doi.org/10.1007/JHEP06(2015)022}{\emph{JHEP} {\bf
  06} (2015) 022}, [\href{http://arxiv.org/abs/1411.7478}{{\tt 1411.7478}}].

\bibitem{Sen:2015hha}
A.~Sen, \emph{{Gauge Invariant 1PI Effective Superstring Field Theory:
  Inclusion of the Ramond Sector}},
  \href{http://dx.doi.org/10.1007/JHEP08(2015)025}{\emph{JHEP} {\bf 08} (2015)
  025}, [\href{http://arxiv.org/abs/1501.00988}{{\tt 1501.00988}}].

\bibitem{Sen:2015uaa}
A.~Sen, \emph{{BV Master Action for Heterotic and Type II String Field
  Theories}}, \href{http://dx.doi.org/10.1007/JHEP02(2016)087}{\emph{JHEP} {\bf
  02} (2016) 087}, [\href{http://arxiv.org/abs/1508.05387}{{\tt 1508.05387}}].

\bibitem{deLacroix:2017lif}
C.~de~Lacroix, H.~Erbin, S.~P. Kashyap, A.~Sen and M.~Verma, \emph{{Closed
  Superstring Field Theory and its Applications}},
  \href{http://dx.doi.org/10.1142/S0217751X17300216}{\emph{Int. J. Mod. Phys.
  A} {\bf 32} (2017) 1730021}, [\href{http://arxiv.org/abs/1703.06410}{{\tt
  1703.06410}}].

\bibitem{FarooghMoosavian:2019yke}
S.~Faroogh~Moosavian, A.~Sen and M.~Verma, \emph{{Superstring Field Theory with
  Open and Closed Strings}},
  \href{http://dx.doi.org/10.1007/JHEP01(2020)183}{\emph{JHEP} {\bf 01} (2020)
  183}, [\href{http://arxiv.org/abs/1907.10632}{{\tt 1907.10632}}].

\bibitem{Cho:2018nfn}
M.~Cho, S.~Collier and X.~Yin, \emph{{Strings in Ramond-Ramond Backgrounds from
  the Neveu-Schwarz-Ramond Formalism}},
  \href{http://dx.doi.org/10.1007/JHEP12(2020)123}{\emph{JHEP} {\bf 12} (2020)
  123}, [\href{http://arxiv.org/abs/1811.00032}{{\tt 1811.00032}}].

\bibitem{Sen:2015uoa}
A.~Sen, \emph{{Supersymmetry Restoration in Superstring Perturbation Theory}},
  \href{http://dx.doi.org/10.1007/JHEP12(2015)075}{\emph{JHEP} {\bf 12} (2015)
  075}, [\href{http://arxiv.org/abs/1508.02481}{{\tt 1508.02481}}].

\bibitem{Giddings:2001yu}
S.~B. Giddings, S.~Kachru and J.~Polchinski, \emph{{Hierarchies from fluxes in
  string compactifications}},
  \href{http://dx.doi.org/10.1103/PhysRevD.66.106006}{\emph{Phys. Rev. D} {\bf
  66} (2002) 106006}, [\href{http://arxiv.org/abs/hep-th/0105097}{{\tt
  hep-th/0105097}}].

\bibitem{Gukov:1999ya}
S.~Gukov, C.~Vafa and E.~Witten, \emph{{CFT's from Calabi-Yau four folds}},
  \href{http://dx.doi.org/10.1016/S0550-3213(00)00373-4}{\emph{Nucl. Phys. B}
  {\bf 584} (2000) 69--108}, [\href{http://arxiv.org/abs/hep-th/9906070}{{\tt
  hep-th/9906070}}].

\bibitem{Becker:2002nn}
K.~Becker, M.~Becker, M.~Haack and J.~Louis, \emph{{Supersymmetry breaking and
  alpha-prime corrections to flux induced potentials}},
  \href{http://dx.doi.org/10.1088/1126-6708/2002/06/060}{\emph{JHEP} {\bf 06}
  (2002) 060}, [\href{http://arxiv.org/abs/hep-th/0204254}{{\tt
  hep-th/0204254}}].

\bibitem{Denef:2004ze}
F.~Denef and M.~R. Douglas, \emph{{Distributions of flux vacua}},
  \href{http://dx.doi.org/10.1088/1126-6708/2004/05/072}{\emph{JHEP} {\bf 05}
  (2004) 072}, [\href{http://arxiv.org/abs/hep-th/0404116}{{\tt
  hep-th/0404116}}].

\bibitem{Demirtas:2019sip}
M.~Demirtas, M.~Kim, L.~Mcallister and J.~Moritz, \emph{{Vacua with Small Flux
  Superpotential}},
  \href{http://dx.doi.org/10.1103/PhysRevLett.124.211603}{\emph{Phys. Rev.
  Lett.} {\bf 124} (2020) 211603}, [\href{http://arxiv.org/abs/1912.10047}{{\tt
  1912.10047}}].

\bibitem{Kachru:2019dvo}
S.~Kachru, M.~Kim, L.~Mcallister and M.~Zimet, \emph{{de Sitter vacua from ten
  dimensions}}, \href{http://dx.doi.org/10.1007/JHEP12(2021)111}{\emph{JHEP}
  {\bf 12} (2021) 111}, [\href{http://arxiv.org/abs/1908.04788}{{\tt
  1908.04788}}].

\bibitem{Sethi:2017phn}
S.~Sethi, \emph{{Supersymmetry Breaking by Fluxes}},
  \href{http://dx.doi.org/10.1007/JHEP10(2018)022}{\emph{JHEP} {\bf 10} (2018)
  022}, [\href{http://arxiv.org/abs/1709.03554}{{\tt 1709.03554}}].

\bibitem{Kachru:2018aqn}
S.~Kachru and S.~P. Trivedi, \emph{{A comment on effective field theories of
  flux vacua}}, \href{http://dx.doi.org/10.1002/prop.201800086}{\emph{Fortsch.
  Phys.} {\bf 67} (2019) 1800086}, [\href{http://arxiv.org/abs/1808.08971}{{\tt
  1808.08971}}].

\bibitem{Grimm:2004uq}
T.~W. Grimm and J.~Louis, \emph{{The Effective action of N = 1 Calabi-Yau
  orientifolds}},
  \href{http://dx.doi.org/10.1016/j.nuclphysb.2004.08.005}{\emph{Nucl. Phys. B}
  {\bf 699} (2004) 387--426}, [\href{http://arxiv.org/abs/hep-th/0403067}{{\tt
  hep-th/0403067}}].

\bibitem{Hori:2003ic}
K.~Hori, S.~Katz, A.~Klemm, R.~Pandharipande, R.~Thomas, C.~Vafa et~al.,
  \emph{{Mirror symmetry}}, vol.~1 of \emph{Clay mathematics monographs}.
\newblock AMS, Providence, USA, 2003.

\bibitem{Demirtas:2021nlu}
M.~Demirtas, M.~Kim, L.~McAllister, J.~Moritz and A.~Rios-Tascon, \emph{{Small
  cosmological constants in string theory}},
  \href{http://dx.doi.org/10.1007/JHEP12(2021)136}{\emph{JHEP} {\bf 12} (2021)
  136}, [\href{http://arxiv.org/abs/2107.09064}{{\tt 2107.09064}}].

\bibitem{Dine:1986vd}
M.~Dine and N.~Seiberg, \emph{{Nonrenormalization Theorems in Superstring
  Theory}}, \href{http://dx.doi.org/10.1103/PhysRevLett.57.2625}{\emph{Phys.
  Rev. Lett.} {\bf 57} (1986) 2625}.

\bibitem{Burgess:2005jx}
C.~P. Burgess, C.~Escoda and F.~Quevedo, \emph{{Nonrenormalization of flux
  superpotentials in string theory}},
  \href{http://dx.doi.org/10.1088/1126-6708/2006/06/044}{\emph{JHEP} {\bf 06}
  (2006) 044}, [\href{http://arxiv.org/abs/hep-th/0510213}{{\tt
  hep-th/0510213}}].

\bibitem{Witten:1996bn}
E.~Witten, \emph{{Nonperturbative superpotentials in string theory}},
  \href{http://dx.doi.org/10.1016/0550-3213(96)00283-0}{\emph{Nucl. Phys. B}
  {\bf 474} (1996) 343--360}, [\href{http://arxiv.org/abs/hep-th/9604030}{{\tt
  hep-th/9604030}}].

\bibitem{Kim:2022jvv}
M.~Kim, \emph{{D-instanton superpotential in string theory}},
  \href{http://dx.doi.org/10.1007/JHEP03(2022)054}{\emph{JHEP} {\bf 03} (2022)
  054}, [\href{http://arxiv.org/abs/2201.04634}{{\tt 2201.04634}}].

\bibitem{Kallosh:2005gs}
R.~Kallosh, A.-K. Kashani-Poor and A.~Tomasiello, \emph{{Counting fermionic
  zero modes on M5 with fluxes}},
  \href{http://dx.doi.org/10.1088/1126-6708/2005/06/069}{\emph{JHEP} {\bf 06}
  (2005) 069}, [\href{http://arxiv.org/abs/hep-th/0503138}{{\tt
  hep-th/0503138}}].

\bibitem{Grimm:2011dj}
T.~W. Grimm, M.~Kerstan, E.~Palti and T.~Weigand, \emph{{On Fluxed Instantons
  and Moduli Stabilisation in IIB Orientifolds and F-theory}},
  \href{http://dx.doi.org/10.1103/PhysRevD.84.066001}{\emph{Phys. Rev. D} {\bf
  84} (2011) 066001}, [\href{http://arxiv.org/abs/1105.3193}{{\tt 1105.3193}}].

\bibitem{Bianchi:2011qh}
M.~Bianchi, A.~Collinucci and L.~Martucci, \emph{{Magnetized E3-brane
  instantons in F-theory}},
  \href{http://dx.doi.org/10.1007/JHEP12(2011)045}{\emph{JHEP} {\bf 12} (2011)
  045}, [\href{http://arxiv.org/abs/1107.3732}{{\tt 1107.3732}}].

\bibitem{Bianchi:2012kt}
M.~Bianchi, G.~Inverso and L.~Martucci, \emph{{Brane instantons and fluxes in
  F-theory}}, \href{http://dx.doi.org/10.1007/JHEP07(2013)037}{\emph{JHEP} {\bf
  07} (2013) 037}, [\href{http://arxiv.org/abs/1212.0024}{{\tt 1212.0024}}].

\bibitem{Cicoli:2022vny}
M.~Cicoli, M.~Licheri, R.~Mahanta and A.~Maharana, \emph{{Flux vacua with
  approximate flat directions}},
  \href{http://dx.doi.org/10.1007/JHEP10(2022)086}{\emph{JHEP} {\bf 10} (2022)
  086}, [\href{http://arxiv.org/abs/2209.02720}{{\tt 2209.02720}}].

\bibitem{Alexandrov:2021shf}
S.~Alexandrov, A.~Sen and B.~Stefa\'nski, \emph{{D-instantons in Type IIA
  string theory on Calabi-Yau threefolds}},
  \href{http://dx.doi.org/10.1007/JHEP11(2021)018}{\emph{JHEP} {\bf 11} (2021)
  018}, [\href{http://arxiv.org/abs/2108.04265}{{\tt 2108.04265}}].

\bibitem{Alexandrov:2021dyl}
S.~Alexandrov, A.~Sen and B.~Stefa\'nski, \emph{{Euclidean D-branes in type IIB
  string theory on Calabi-Yau threefolds}},
  \href{http://dx.doi.org/10.1007/JHEP12(2021)044}{\emph{JHEP} {\bf 12} (2021)
  044}, [\href{http://arxiv.org/abs/2110.06949}{{\tt 2110.06949}}].

\bibitem{Polchinski:1998rq}
J.~Polchinski, \emph{{String theory. Vol. 1: An introduction to the bosonic
  string}}.
\newblock Cambridge Monographs on Mathematical Physics. Cambridge University
  Press, 12, 2007,
  \href{http://dx.doi.org/10.1017/CBO9780511816079}{10.1017/CBO9780511816079}.

\bibitem{Blumenhagen:2013fgp}
R.~Blumenhagen, D.~L\"ust and S.~Theisen, \emph{{Basic concepts of string
  theory}}.
\newblock Theoretical and Mathematical Physics. Springer, Heidelberg, Germany,
  2013,
  \href{http://dx.doi.org/10.1007/978-3-642-29497-6}{10.1007/978-3-642-29497-6}.

\bibitem{Billo:1998vr}
M.~Billo, P.~Di~Vecchia, M.~Frau, A.~Lerda, I.~Pesando, R.~Russo et~al.,
  \emph{{Microscopic string analysis of the D0 - D8-brane system and dual R - R
  states}}, \href{http://dx.doi.org/10.1016/S0550-3213(98)00296-X}{\emph{Nucl.
  Phys. B} {\bf 526} (1998) 199--228},
  [\href{http://arxiv.org/abs/hep-th/9802088}{{\tt hep-th/9802088}}].

\bibitem{Callan:1987px}
C.~G. Callan, Jr., C.~Lovelace, C.~R. Nappi and S.~A. Yost, \emph{{Adding Holes
  and Crosscaps to the Superstring}},
  \href{http://dx.doi.org/10.1016/0550-3213(87)90065-4}{\emph{Nucl. Phys. B}
  {\bf 293} (1987) 83}.

\bibitem{Polchinski:1987tu}
J.~Polchinski and Y.~Cai, \emph{{Consistency of Open Superstring Theories}},
  \href{http://dx.doi.org/10.1016/0550-3213(88)90382-3}{\emph{Nucl. Phys. B}
  {\bf 296} (1988) 91--128}.

\bibitem{Billo:1997eg}
M.~Billo, P.~Di~Vecchia and D.~Cangemi, \emph{{Boundary states for moving
  D-branes}},
  \href{http://dx.doi.org/10.1016/S0370-2693(97)00329-8}{\emph{Phys. Lett. B}
  {\bf 400} (1997) 63--70}, [\href{http://arxiv.org/abs/hep-th/9701190}{{\tt
  hep-th/9701190}}].

\bibitem{DiVecchia:1997vef}
P.~Di~Vecchia, M.~Frau, I.~Pesando, S.~Sciuto, A.~Lerda and R.~Russo,
  \emph{{Classical p-branes from boundary state}},
  \href{http://dx.doi.org/10.1016/S0550-3213(97)00576-2}{\emph{Nucl. Phys. B}
  {\bf 507} (1997) 259--276}, [\href{http://arxiv.org/abs/hep-th/9707068}{{\tt
  hep-th/9707068}}].

\bibitem{DiVecchia:1999mal}
P.~Di~Vecchia and A.~Liccardo, \emph{{D Branes in String Theory, I}},
  \href{http://dx.doi.org/10.1007/978-94-011-4303-5_1}{\emph{NATO Sci. Ser. C}
  {\bf 556} (2000) 1--60}, [\href{http://arxiv.org/abs/hep-th/9912161}{{\tt
  hep-th/9912161}}].

\bibitem{Kachru:2002he}
S.~Kachru, M.~B. Schulz and S.~Trivedi, \emph{{Moduli stabilization from fluxes
  in a simple IIB orientifold}},
  \href{http://dx.doi.org/10.1088/1126-6708/2003/10/007}{\emph{JHEP} {\bf 10}
  (2003) 007}, [\href{http://arxiv.org/abs/hep-th/0201028}{{\tt
  hep-th/0201028}}].

\bibitem{Hata:1993gf}
H.~Hata and B.~Zwiebach, \emph{{Developing the covariant Batalin-Vilkovisky
  approach to string theory}},
  \href{http://dx.doi.org/10.1006/aphy.1994.1006}{\emph{Annals Phys.} {\bf 229}
  (1994) 177--216}, [\href{http://arxiv.org/abs/hep-th/9301097}{{\tt
  hep-th/9301097}}].

\bibitem{Zwiebach:1990nh}
B.~Zwiebach, \emph{{How covariant closed string theory solves a minimal area
  problem}}, \href{http://dx.doi.org/10.1007/BF02096792}{\emph{Commun. Math.
  Phys.} {\bf 136} (1991) 83--118}.

\bibitem{Zwiebach:1990qj}
B.~Zwiebach, \emph{{Quantum open string theory with manifest closed string
  factorization}},
  \href{http://dx.doi.org/10.1016/0370-2693(91)90212-9}{\emph{Phys. Lett. B}
  {\bf 256} (1991) 22--29}.

\bibitem{Zwiebach:1992bw}
B.~Zwiebach, \emph{{Interpolating string field theories}},
  \href{http://dx.doi.org/10.1142/S0217732392000951}{\emph{Mod. Phys. Lett. A}
  {\bf 7} (1992) 1079--1090}, [\href{http://arxiv.org/abs/hep-th/9202015}{{\tt
  hep-th/9202015}}].

\bibitem{Headrick:2018ncs}
M.~Headrick and B.~Zwiebach, \emph{{Convex programs for minimal-area
  problems}}, \href{http://dx.doi.org/10.1007/s00220-020-03732-1}{\emph{Commun.
  Math. Phys.} {\bf 377} (2020) 2217--2285},
  [\href{http://arxiv.org/abs/1806.00449}{{\tt 1806.00449}}].

\bibitem{Headrick:2018dlw}
M.~Headrick and B.~Zwiebach, \emph{{Minimal-area metrics on the Swiss cross and
  punctured torus}},
  \href{http://dx.doi.org/10.1007/s00220-020-03734-z}{\emph{Commun. Math.
  Phys.} {\bf 377} (2020) 2287--2343},
  [\href{http://arxiv.org/abs/1806.00450}{{\tt 1806.00450}}].

\bibitem{Costello:2019fuh}
K.~Costello and B.~Zwiebach, \emph{{Hyperbolic string vertices}},
  \href{http://dx.doi.org/10.1007/JHEP02(2022)002}{\emph{JHEP} {\bf 02} (2022)
  002}, [\href{http://arxiv.org/abs/1909.00033}{{\tt 1909.00033}}].

\bibitem{Cho:2019anu}
M.~Cho, \emph{{Open-closed Hyperbolic String Vertices}},
  \href{http://dx.doi.org/10.1007/JHEP05(2020)046}{\emph{JHEP} {\bf 05} (2020)
  046}, [\href{http://arxiv.org/abs/1912.00030}{{\tt 1912.00030}}].

\bibitem{Zwiebach:1997fe}
B.~Zwiebach, \emph{{Oriented open - closed string theory revisited}},
  \href{http://dx.doi.org/10.1006/aphy.1998.5803}{\emph{Annals Phys.} {\bf 267}
  (1998) 193--248}, [\href{http://arxiv.org/abs/hep-th/9705241}{{\tt
  hep-th/9705241}}].

\bibitem{Polchinski:1988jq}
J.~Polchinski, \emph{{Factorization of Bosonic String Amplitudes}},
  \href{http://dx.doi.org/10.1016/0550-3213(88)90522-6}{\emph{Nucl. Phys. B}
  {\bf 307} (1988) 61--92}.

\bibitem{Hull:2009mi}
C.~Hull and B.~Zwiebach, \emph{{Double Field Theory}},
  \href{http://dx.doi.org/10.1088/1126-6708/2009/09/099}{\emph{JHEP} {\bf 09}
  (2009) 099}, [\href{http://arxiv.org/abs/0904.4664}{{\tt 0904.4664}}].

\bibitem{Sen:2019jpm}
A.~Sen, \emph{{String Field Theory as World-sheet UV Regulator}},
  \href{http://dx.doi.org/10.1007/JHEP10(2019)119}{\emph{JHEP} {\bf 10} (2019)
  119}, [\href{http://arxiv.org/abs/1902.00263}{{\tt 1902.00263}}].

\bibitem{Kiritsis:1994ta}
E.~Kiritsis and C.~Kounnas, \emph{{Infrared regularization of superstring
  theory and the one loop calculation of coupling constants}},
  \href{http://dx.doi.org/10.1016/0550-3213(95)00156-M}{\emph{Nucl. Phys. B}
  {\bf 442} (1995) 472--493}, [\href{http://arxiv.org/abs/hep-th/9501020}{{\tt
  hep-th/9501020}}].

\bibitem{Antoniadis:1997eg}
I.~Antoniadis, S.~Ferrara, R.~Minasian and K.~S. Narain, \emph{{R**4 couplings
  in M and type II theories on Calabi-Yau spaces}},
  \href{http://dx.doi.org/10.1016/S0550-3213(97)00572-5}{\emph{Nucl. Phys. B}
  {\bf 507} (1997) 571--588}, [\href{http://arxiv.org/abs/hep-th/9707013}{{\tt
  hep-th/9707013}}].

\bibitem{Berg:2005ja}
M.~Berg, M.~Haack and B.~Kors, \emph{{String loop corrections to Kahler
  potentials in orientifolds}},
  \href{http://dx.doi.org/10.1088/1126-6708/2005/11/030}{\emph{JHEP} {\bf 11}
  (2005) 030}, [\href{http://arxiv.org/abs/hep-th/0508043}{{\tt
  hep-th/0508043}}].

\bibitem{Berg:2011ij}
M.~Berg, M.~Haack and J.~U. Kang, \emph{{One-Loop Kahler Metric of D-Branes at
  Angles}}, \href{http://dx.doi.org/10.1007/JHEP11(2012)091}{\emph{JHEP} {\bf
  11} (2012) 091}, [\href{http://arxiv.org/abs/1112.5156}{{\tt 1112.5156}}].

\bibitem{Berg:2014ama}
M.~Berg, M.~Haack, J.~U. Kang and S.~Sj\"ors, \emph{{Towards the one-loop
  K\"ahler metric of Calabi-Yau orientifolds}},
  \href{http://dx.doi.org/10.1007/JHEP12(2014)077}{\emph{JHEP} {\bf 12} (2014)
  077}, [\href{http://arxiv.org/abs/1407.0027}{{\tt 1407.0027}}].

\bibitem{Haack:2015pbv}
M.~Haack and J.~U. Kang, \emph{{One-loop Einstein-Hilbert term in minimally
  supersymmetric type IIB orientifolds}},
  \href{http://dx.doi.org/10.1007/JHEP02(2016)160}{\emph{JHEP} {\bf 02} (2016)
  160}, [\href{http://arxiv.org/abs/1511.03957}{{\tt 1511.03957}}].

\bibitem{Haack:2018ufg}
M.~Haack and J.~U. Kang, \emph{{Field redefinitions and K\"ahler potential in
  string theory at 1-loop}},
  \href{http://dx.doi.org/10.1007/JHEP08(2018)019}{\emph{JHEP} {\bf 08} (2018)
  019}, [\href{http://arxiv.org/abs/1805.00817}{{\tt 1805.00817}}].

\bibitem{Kim:2023sfs}
M.~Kim, \emph{{On string one-loop correction to the Einstein-Hilbert term and
  its implications on the K\"ahler potential}},
  \href{http://dx.doi.org/10.1007/JHEP07(2023)044}{\emph{JHEP} {\bf 07} (2023)
  044}, [\href{http://arxiv.org/abs/2302.12117}{{\tt 2302.12117}}].

\bibitem{Kim:2023eut}
M.~Kim, \emph{{On one-loop corrected dilaton action in string theory}},
  \href{http://arxiv.org/abs/2305.08263}{{\tt 2305.08263}}.

\bibitem{Firat:2021ukc}
A.~H. F\i{}rat, \emph{{Hyperbolic three-string vertex}},
  \href{http://dx.doi.org/10.1007/JHEP08(2021)035}{\emph{JHEP} {\bf 08} (2021)
  035}, [\href{http://arxiv.org/abs/2102.03936}{{\tt 2102.03936}}].

\bibitem{Erbin:2022rgx}
H.~Erbin and A.~H. F\i{}rat, \emph{{Characterizing 4-string contact interaction
  using machine learning}},  \href{http://arxiv.org/abs/2211.09129}{{\tt
  2211.09129}}.

\bibitem{Firat:2023glo}
A.~H. F\i{}rat, \emph{{Bootstrapping closed string field theory}},
  \href{http://dx.doi.org/10.1007/JHEP05(2023)186}{\emph{JHEP} {\bf 05} (2023)
  186}, [\href{http://arxiv.org/abs/2302.12843}{{\tt 2302.12843}}].

\bibitem{Erbin:2023hcs}
H.~Erbin and A.~H. F\i{}rat, \emph{{Open string stub as an auxiliary string
  field}},  \href{http://arxiv.org/abs/2308.08587}{{\tt 2308.08587}}.

\bibitem{Firat:2023suh}
A.~H. F\i{}rat, \emph{{Hyperbolic string tadpole}},
  \href{http://arxiv.org/abs/2306.08599}{{\tt 2306.08599}}.

\bibitem{Sen:2019qqg}
A.~Sen, \emph{{Fixing an Ambiguity in Two Dimensional String Theory Using
  String Field Theory}},
  \href{http://dx.doi.org/10.1007/JHEP03(2020)005}{\emph{JHEP} {\bf 03} (2020)
  005}, [\href{http://arxiv.org/abs/1908.02782}{{\tt 1908.02782}}].

\bibitem{Sen:2020cef}
A.~Sen, \emph{{D-instanton Perturbation Theory}},
  \href{http://dx.doi.org/10.1007/JHEP08(2020)075}{\emph{JHEP} {\bf 08} (2020)
  075}, [\href{http://arxiv.org/abs/2002.04043}{{\tt 2002.04043}}].

\bibitem{Sen:2020eck}
A.~Sen, \emph{{D-instantons, string field theory and two dimensional string
  theory}}, \href{http://dx.doi.org/10.1007/JHEP11(2021)061}{\emph{JHEP} {\bf
  11} (2021) 061}, [\href{http://arxiv.org/abs/2012.11624}{{\tt 2012.11624}}].

\bibitem{Sen:2021qdk}
A.~Sen, \emph{{Normalization of D-instanton amplitudes}},
  \href{http://dx.doi.org/10.1007/JHEP11(2021)077}{\emph{JHEP} {\bf 11} (2021)
  077}, [\href{http://arxiv.org/abs/2101.08566}{{\tt 2101.08566}}].

\bibitem{Sen:2021tpp}
A.~Sen, \emph{{Normalization of type IIB D-instanton amplitudes}},
  \href{http://dx.doi.org/10.1007/JHEP12(2021)146}{\emph{JHEP} {\bf 12} (2021)
  146}, [\href{http://arxiv.org/abs/2104.11109}{{\tt 2104.11109}}].

\bibitem{Sen:2021jbr}
A.~Sen, \emph{{Muti-instanton amplitudes in type IIB string theory}},
  \href{http://dx.doi.org/10.1007/JHEP12(2021)065}{\emph{JHEP} {\bf 12} (2021)
  065}, [\href{http://arxiv.org/abs/2104.15110}{{\tt 2104.15110}}].

\bibitem{Eniceicu:2022nay}
D.~S. Eniceicu, R.~Mahajan, C.~Murdia and A.~Sen, \emph{{Normalization of ZZ
  instanton amplitudes in minimal string theory}},
  \href{http://dx.doi.org/10.1007/JHEP07(2022)139}{\emph{JHEP} {\bf 07} (2022)
  139}, [\href{http://arxiv.org/abs/2202.03448}{{\tt 2202.03448}}].

\bibitem{Alexandrov:2022mmy}
S.~Alexandrov, A.~H. F\i{}rat, M.~Kim, A.~Sen and B.~Stefa\'nski,
  \emph{{D-instanton induced superpotential}},
  \href{http://dx.doi.org/10.1007/JHEP07(2022)090}{\emph{JHEP} {\bf 07} (2022)
  090}, [\href{http://arxiv.org/abs/2204.02981}{{\tt 2204.02981}}].

\bibitem{Agmon:2022vdj}
N.~B. Agmon, B.~Balthazar, M.~Cho, V.~A. Rodriguez and X.~Yin,
  \emph{{D-instanton Effects in Type IIB String Theory}},
  \href{http://arxiv.org/abs/2205.00609}{{\tt 2205.00609}}.

\bibitem{Eniceicu:2022dru}
D.~S. Eniceicu, R.~Mahajan, C.~Murdia and A.~Sen, \emph{{Multi-instantons in
  minimal string theory and in matrix integrals}},
  \href{http://dx.doi.org/10.1007/JHEP10(2022)065}{\emph{JHEP} {\bf 10} (2022)
  065}, [\href{http://arxiv.org/abs/2206.13531}{{\tt 2206.13531}}].

\bibitem{Chakravarty:2022cgj}
J.~Chakravarty and A.~Sen, \emph{{Normalization of D instanton amplitudes in
  two dimensional type 0B string theory}},
  \href{http://dx.doi.org/10.1007/JHEP02(2023)170}{\emph{JHEP} {\bf 02} (2023)
  170}, [\href{http://arxiv.org/abs/2207.07138}{{\tt 2207.07138}}].

\bibitem{Eniceicu:2022xvk}
D.~S. Eniceicu, R.~Mahajan, P.~Maity, C.~Murdia and A.~Sen, \emph{{The ZZ
  annulus one-point function in non-critical string theory: A string field
  theory analysis}},
  \href{http://dx.doi.org/10.1007/JHEP12(2022)151}{\emph{JHEP} {\bf 12} (2022)
  151}, [\href{http://arxiv.org/abs/2210.11473}{{\tt 2210.11473}}].

\bibitem{Kim:2023cbh}
M.~Kim, \emph{{D-instanton, threshold corrections, and topological string}},
  \href{http://dx.doi.org/10.1007/JHEP05(2023)097}{\emph{JHEP} {\bf 05} (2023)
  097}, [\href{http://arxiv.org/abs/2301.03602}{{\tt 2301.03602}}].

\bibitem{Sen:2002nu}
A.~Sen, \emph{{Rolling tachyon}},
  \href{http://dx.doi.org/10.1088/1126-6708/2002/04/048}{\emph{JHEP} {\bf 04}
  (2002) 048}, [\href{http://arxiv.org/abs/hep-th/0203211}{{\tt
  hep-th/0203211}}].

\bibitem{Sen:2002in}
A.~Sen, \emph{{Tachyon matter}},
  \href{http://dx.doi.org/10.1088/1126-6708/2002/07/065}{\emph{JHEP} {\bf 07}
  (2002) 065}, [\href{http://arxiv.org/abs/hep-th/0203265}{{\tt
  hep-th/0203265}}].

\bibitem{Sen:2004zm}
A.~Sen, \emph{{Rolling tachyon boundary state, conserved charges and
  two-dimensional string theory}},
  \href{http://dx.doi.org/10.1088/1126-6708/2004/05/076}{\emph{JHEP} {\bf 05}
  (2004) 076}, [\href{http://arxiv.org/abs/hep-th/0402157}{{\tt
  hep-th/0402157}}].

\bibitem{Sen:2004cq}
A.~Sen, \emph{{Energy momentum tensor and marginal deformations in open string
  field theory}},
  \href{http://dx.doi.org/10.1088/1126-6708/2004/08/034}{\emph{JHEP} {\bf 08}
  (2004) 034}, [\href{http://arxiv.org/abs/hep-th/0403200}{{\tt
  hep-th/0403200}}].

\bibitem{Sen:2004yv}
A.~Sen, \emph{{Symmetries, conserved charges and (black) holes in two
  dimensional string theory}},
  \href{http://dx.doi.org/10.1088/1126-6708/2004/12/053}{\emph{JHEP} {\bf 12}
  (2004) 053}, [\href{http://arxiv.org/abs/hep-th/0408064}{{\tt
  hep-th/0408064}}].

\bibitem{Sen:2004nf}
A.~Sen, \emph{{Tachyon dynamics in open string theory}},
  \href{http://dx.doi.org/10.1142/S0217751X0502519X}{\emph{Int. J. Mod. Phys.
  A} {\bf 20} (2005) 5513--5656},
  [\href{http://arxiv.org/abs/hep-th/0410103}{{\tt hep-th/0410103}}].

\bibitem{Cho:2023khj}
M.~Cho, B.~Mazel and X.~Yin, \emph{{Rolling tachyon and the Phase Space of Open
  String Field Theory}},  \href{http://arxiv.org/abs/2310.17895}{{\tt
  2310.17895}}.

\end{thebibliography}\endgroup
\end{document}